\documentstyle[12pt,newlfont,amssymb]{amsbook}

\numberwithin{equation}{subsection}

\newcommand{\ie}{i.e.\ }

\newcommand{\cf}{c.f.\ }

\newcommand{\U}{{\cal U}}
\newcommand{\Uq}{\U_q}
\newcommand{\e}{\epsilon}
\newcommand{\Ue}{\U_\e}
\newcommand{\A}{{\cal A}}
\newcommand{\UA}{\U_\A}
\newcommand{\Upq}{\U_{pq}}
\newcommand{\Uqr}{\U_{qr}}

\newcommand{\Uqf}{\U_{q\f}}
\newcommand{\Uef}{\U_{\e\f}}
\newcommand{\Ur}{\U_{r}}
\newcommand{\f}{\zeta}
\newcommand{\Uf}{\U_\f}
\newcommand{\Us}{\U_{s}}

\newcommand{\Zcentre}{{\cal Z}}

\newcommand{\Ze}{\Zcentre_\e}
\newcommand{\Zo}{\Zcentre_0}
\newcommand{\hthree}{\frak{h}_3}
\newcommand{\hfour}{\frak{h}_4}
\renewcommand{\sl}{\frak{sl}}
\newcommand{\sltwo}{\sl_2}
\newcommand{\SL}{SL}
\newcommand{\SLtwo}{\SL_2}
\newcommand{\sln}{\sl_n}
\newcommand{\gl}{\frak{gl}}
\newcommand{\GL}{GL}

\newcommand{\fun}{\mathrm{fun}}
\newcommand{\funq}{\fun_q}
\newcommand{\fune}{\fun_\e}
\newcommand{\funpq}{\fun_{pq}}
\newcommand{\g}{\frak{g}}
\newcommand{\h}{\frak{h}}
\newcommand{\n}{\frak{n}}
\newcommand{\kmn}{\hat{\frak{n}}}
\newcommand{\kmg}{\hat{\g}}
\newcommand{\kmh}{\hat{\h}}
\newcommand{\kmsltwo}{\widehat{\sl_2}}
\newcommand{\kmsln}{\widehat{\sl_n}}
\newcommand{\kmW}{\hat{W}}

\renewcommand{\H}{\frak{H}}  
\newcommand{\hatL}[1]{\widehat{L(#1)}}  
\newcommand{\deriv}{D}

\newcommand{\ev}{\operatorname{ev}}
\newcommand{\ad}{\operatorname{ad}}
\newcommand{\id}{\operatorname{id}}
\newcommand{\End}{\operatorname{End}}

\newcommand{\Ad}{\operatorname{Ad}}
\newcommand{\Hom}{\operatorname{Hom}}
\newcommand{\Aut}{\operatorname{Aut}}
\newcommand{\cs}[1]{{\cal #1}}                  
\newcommand{\alg}[1]{{\cal #1}}
\newcommand{\comb}[2]{\brackets{ \begin{array}{c} #1 \\ #2 \end{array}}}
\newcommand{\brackets}[1]{\left[ #1 \right]}      
\newcommand{\range}[1]{\brackets{#1}}             
\newcommand{\comm}[1]{\brackets{#1}}              
\newcommand{\ideal}[1]{\langle #1 \rangle}
\newcommand{\pairing}[1]{\ideal{#1}}
\newcommand{\structure}[1]{(#1)}
\newcommand{\half}{\frac{1}{2}}

\newcommand{\twomatrix}[4]{\begin{pmatrix} #1& #2\\ #3& #4\end{pmatrix}}
\newcommand{\inv}{^{-1}}
\newcommand{\czek}{^\vee}
\newcommand{\tensor}{\otimes}
\newcommand{\Tensor}{\bigotimes}
\newcommand{\dottensor}{\dot{\otimes}}
\newcommand{\directsum}{\oplus}
\newcommand{\Directsum}{\bigoplus}

\newcommand{\mapto}{\rightarrow}
\newcommand{\isomorphic}{\simeq}
\newcommand{\compose}{\circ}
\newcommand{\union}{\cup}
\newcommand{\intersect}{\cap}
\newcommand{\N}{{\Bbb N}}
\newcommand{\Z}{{\Bbb Z}}
\newcommand{\Zplus}{\Z_{>0}}
\newcommand{\Zminus}{\Z_{<0}}
\newcommand{\Zcross}{{\Z^\times}}
\newcommand{\R}{{\Bbb R}}
\newcommand{\Rplus}{\R_{>0}}
\newcommand{\Rcross}{{\R^\times}}
\newcommand{\C}{{\Bbb C}}
\newcommand{\Ccross}{{\C^\times}}

\renewcommand{\l}{\ell}
\newcommand{\ng}{N}             
\newcommand{\unity}{\boldsymbol{1}}
\newcommand{\free}[1]{\brackets{#1}}
\newcommand{\freeNC}[1]{\brackets{\brackets{#1}}}
\newcommand{\Cfree}[1]{\C\free{#1}}
\newcommand{\CfreeNC}[1]{\C\freeNC{#1}}            
\newcommand{\Zfree}[1]{\Z\free{#1}}
\newcommand{\w}{\omega}
\newcommand{\set}[1]{\{ #1 \}}
\newcommand{\modulus}[1]{\vert #1 \vert}
\newcommand{\embed}{\hookrightarrow}
\newcommand{\serc}{S.E.R.C.\ }

\theoremstyle{plain}
\newtheorem{theorem}[subsection]{Theorem}
\newtheorem{lemma}[subsection]{Lemma}
\newtheorem{proposition}[subsection]{Proposition}
\newtheorem{corollary}[subsection]{Corollary}
\theoremstyle{definition}
\newtheorem{definition}[subsection]{Definition}

\theoremstyle{remark}
\newtheorem{remark}[subsection]{Remark}
\newtheorem{example}[subsection]{Example}
\newtheorem{notation}{Notation} 
\newenvironment{proof}{\begin{pf}}{\end{pf}}
\newcounter{thmparts}

\begin{document}

\begin{flushright}
  August 1994\\
  QMW-PH/94-36\\
  hep-th/9409079\\
  AMS-PPS \#19941114001
\end{flushright}

\pagestyle{empty}

\vspace*{3cm}
\begin{center}
  A thesis presented for\\
  the Degree of Doctor of Philosophy of\\
  the University of London
\end{center}
\vspace{1cm}

\begin{center}
  \size{17}{20pt}\selectfont
  Representations\\
  at a Root of Unity of\\
  $q$-Oscillators and\\
  Quantum Kac-Moody  Algebras
\end{center}
\vspace{0.5cm}

\begin{center}
  by\\[0.5cm]
  \Large
  Jens-Ulrik Holger Petersen
\end{center}
\vspace{1cm}

\begin{center}
  Theoretical Physics Group\\
  Physics Department\\
  Queen Mary and Westfield College\\
  London E1 4NS\\
  {\small E-mail address: j.petersen@@qmw.ac.uk}
\end{center}
\vspace{1cm}

\begin{center}
  July 1994
\end{center}
\vspace{1cm}

\clearpage 

\null 
\vspace{4cm}
\begin{center}
  \Large To my wife, Noriko, and\\
  the little baby she is now carrying
\end{center}

\clearpage

\chapter*{Abstract}
\pagestyle{empty}

\vspace{-0.5cm}

The subject of this thesis is quantum groups and quantum algebras at a
root of unity.  After an introductory chapter, I set up my notation in
chapter~2. The rest of the thesis is presented in three parts.

In part~I, quantum matrix groups and quantum enveloping algebras are
discussed.  In chapter~\ref{chap:matrix-quantum-groups}, I present two
well-known $2\times2$ matrix quantum groups, including their
coaction on the quantum plane and specialisations at a root of
unity.  Chapter~\ref{chap:quantum-enveloping-algebras} develops a
quite detailed description of quantum enveloping algebras and their
specialisation at an odd root of unity. The results from this chapter
are required in part~III.

Part~II is devoted to certain deformations of the quantum mechanical
oscillator algebra: so called $q$-oscillators. In
chapter~\ref{chap:q-oscillators}, a standard $q$-oscillator and its
Fock module is described, including its specialisation at a root of
unity. In chapter~\ref{chap:qr-oscillator}, original
work~\cite{petersen:qr-oscillator} on a new 2-parameter deformation of
the oscillator algebra is presented and its representations at a root
of unity are described.

Part~III is concerned with infinite dimensional quantum groups. In
chapter~\ref{chap:kac-moody}, the structure of an (untwisted) quantum
affine Kac-Moody algebra is discussed. As in the classical case, it
has both a Chevalley and a loop algebra presentation, which can be
shown to be isomorphic using braid group and translation
automorphisms. A quantum affine algebra has also a Heisenberg
subalgebra: I describe its Fock modules and their unitarisability.
Finally in chapter~\ref{chap:qkm-root-of-1}, I present original
results~\cite{petersen:qkm-root-of-1} on the specialisation of a
quantum affine algebra at an odd root of unity. I prove that a quantum
affine algebra at a root of unity has an infinite dimensional centre
and construct the central elements corresponding to the real and
imaginary roots. At the odd root of unity, some new infinite
dimensional representations of the algebra are shown to exist.

\clearpage


\setcounter{tocdepth}{1}
\tableofcontents

\pagestyle{headings}

\chapter{Introduction}

This thesis describes work I have done in the area of quantum groups.
It is presented in three parts.  Part~I is largely concerned with
simple quantum enveloping algebras and quantum groups.  Part~II is
devoted to $q$-oscillators. Infinite dimensional quantum affine
enveloping algebras are described in part~III. The next chapter sets
my notation for the rest of thesis.

I have tried to include references to the literature: the many
omissions, that inevitably have occurred, should be put down to
ignorance rather than ill-will on my part. Also the thesis has been
written in a quite mathematical style, even though my background has
been in physics: I hope that in the process I have not made it
difficult to read for both mathematicians and theoretical physicists!

\section{Quantum groups}

\subsection{}
Quantum groups were first discovered by theoretical physicists to
occur as symmetries of integrable 1+1-dimensional systems,
particularly through the quantum inverse scattering
mechanism~\cite{sklyanin-takhajan-faddeev:qism}. The strong relation
of quantum groups to the Yang-Baxter
equation~\cite{jimbo-review:ybe-integrable-sys} and transfer matrices
in statistical mechanical models~\cite{baxter:exactly-solved} was also
crucial in their development. It was Drinfeld who realised that
mathematically quantum groups are Hopf algebras. These beginnings led
to two major formulations of quantum groups that will be discussed in
this thesis, although the first will only be described briefly.

Faddeev, Reshetikhin and Takhajan~\cite{reshetikhin-takhajan-faddeev}
developed the theory of matrix quantum groups.  Drinfeld and Jimbo
introduced 1-parameter deformations of universal enveloping algebras
of semi-simple Lie algebras. Actually Drinfeld's and Jimbo's
constructions, although similar in structure, have rather different
origins. The motivation behind Drinfeld's approach is the quantisation
of so-called Poisson-Lie groups. The approach of Jimbo, which will be
the one followed in this thesis, is more Lie algebraic and leads to a
much richer representation theory.

{}From a quite different direction, namely that of non-commutative
geometry, Woro\-no\-wicz~\cite{woronowicz:cmp1} independently
initiated the study of quantum groups from a $C^\star$-algebra point
of view.  Manin~\cite{manin:qgps-preprint} constructed quantum groups
as the covariance symmetries of certain quantum (vector) spaces.  The
work of Woronowicz~\cite{woronowicz:cmp2} and Manin led to the field
of non-commutative differential geometry of quantum
groups~\cite{wess-zumino:npbproc,schupp-watts-zumino:diff-geom-lin-qgp}
and non-commutative spaces. Majid~\cite{majid:braided-group-intro}
introduced braided groups and algebras.

In chapter~\ref{chap:matrix-quantum-groups}, I discuss briefly some
simple examples of matrix quantum groups, which co-act on the quantum
plane. At a root of unity they have a large centre. At the end of the
chapter, a 2-parameter deformed matrix quantum
group~\cite{schirrmacher-wess-zumino:2-parameter-gl2} is mentioned.

\subsection{}
More recently it has been realised that quantum groups occur as
symmetries of a large number of systems in mathematical physics (see
for instance~\cite{infinite-analysis:rims,%
  curtright-fairlie-zachos:ANL-workshop}).  Notably quantum enveloping
algebras occur as symmetries in quantum spin chains and solvable
lattice models~\cite{jimbo-miwa:book}, in two dimensional conformal
field theory (see~\cite{fuchs:book}) and in massive integrable
systems~\cite{smirnov:book-form-factors}. Quantum groups also play an
important role in the theory of link invariants and
knots~\cite{yang-ge:braid-gp-knots-SM}. One can say that by now the
use of quantum groups in theoretical and mathematical physics has
become so wide-spread that almost every day new preprints appear
describing their structure and applications.

After Jimbo and Drinfeld introduced the quantum enveloping algebras of
simple and affine Lie algebras, pure mathematicians (particularly Lie
algebraists) gradually started to become very interested in these
structures as a new development in Lie theory. Around 1990 many deep
results on the representation theory of quantum enveloping algebras,
particularly at a root of unity, appeared. Particularly notable is the
work of Lusztig~\cite{lusztig:book} and of De~Concini, Kac and
Procesi. In their point of view, the specialisation of a quantum
enveloping algebra at a $p$-th root of unity is a quantum version of
the corresponding Lie algebra over a field of characteristic $p$.
Lusztig's approach to quantum groups at a root of unity involves the
specialisation of a subalgebra generated by divided powers with
$q$-factorial numbers (a $q$-analogue of the $\Z$-form of a classical
enveloping algebra of Chevalley, Kostant and Tits) --- this does not
give an enlarged centre at a root of unity. I will not have time to
describe his interesting approach. Kashiwara and Lusztig
independently introduced some important bases of quantum groups
in~\cite{kashiwara:crystal-bases,lusztig:canonical-bases}

In chapter~\ref{chap:quantum-enveloping-algebras}, I describe in some
detail the structure of the quantum enveloping algebra of a simple Lie
algebra. I introduce Lusztig's braid group automorphisms, that allow a
construction of a basis of root vectors of the quantum enveloping
algebra. Following De~Concini and Kac, I apply the braid group
automorphisms to construct the large centre of a quantum enveloping
algebra at an odd root of unity. The Verma modules at a root of unity
are reducible, they are `replaced' by finite dimensional diagonal and
(semicyclic) triangular modules. Cyclic modules also exist. The
example of quantum enveloping algebra of $\sltwo$, which first
appeared in~\cite{kulish-reshetikhin:sine-gordon}, is given and
studied at a root of unity.  Finally a 2-parameter deformation of
$\gl_2$ is mentioned.

\subsection{Representations at a root of unity}

When the deformation parameter is not a root of unity, the
representation theory of a quantum enveloping algebra is essentially
the same~\cite{lusztig:reps,rosso:cmp} as that of the corresponding
universal enveloping algebra of the simple Lie algebra.

One of the exciting aspects of quantum enveloping algebras at a root
of unity is the possibility of new types of representations that have
no classical analogue. The irreducible representations, in the case of
an odd $\l$-th root of unity, are parametrised by $m$ continuous
parameters (with $m$ equal to the dimension of the underlying Lie
algebra) and have maximal dimension $\l^N$ (where $N$ is the number of
positive roots of the Lie algebra).

There are three distinct types of irreducible representations:
nilpotent representations, semicyclic (semiperiodic) and cyclic
(periodic) representations, but mixtures of these three types can also
occur.

The nilpotent representations are deformations of classical
representations with a highest and lowest weight, in which the
positive and negative root vectors act nilpotently. They occur in 2d
conformal field theories~\cite{alvarez-gaume-gomez-sierra:knizhnik-book,%
  gomez-sierra:qgp-coulomb-gas,fuchs:book}

In a semicyclic representation, for each positive root, the associated
positive or negative root vector (but not both) acts injectively. They
are used in connection with lattice spin models and also relativistic
solitons~\cite{gomez-sierra:new-ybe-solns}. It is possible to modify
certain nilpotent representations (of appropriate dimension) so that
they become a semicyclic.

In a cyclic representation every root vector, corresponding to a
positive or negative root, acts injectively. Cyclic modules have
neither highest nor lowest weights, and their construction appears to
be much more difficult, though considerable progress on this has been
made (particularly for the case of cyclic modules of minimal
dimension)
\cite{date-jimbo-miki-miwa:cyclic-sln,arnaudon-chakrabarti:flat-periodic,%
  arnaudon-chakrabarti:su_n,chari-pressley:min-cyclic,%
  chari-pressley:fundamental-rep,chari-pressley:so(8),schnizer:jmp,%
  schnizer}.  Cyclic modules first occurred in the study of the
8-vertex model with the Sklyanin
algebra~\cite{sklyanin:alg-structures-reps}. Cyclic representations
occur~\cite{bazhanov-kashaev-mangazeev-stroganov,date-jimbo-miki-miwa:potts}
in the context of the generalised Potts models. This work gave rise to
new $\cs{R}$-matrices (solutions of the Yang-Baxter equation) which
intertwine certain tensor products of minimal cyclic representations,
which are the local state-spaces of the model, \ie these
$\cs{R}$-matrices are the Boltzmann weights of the model.

The difficult task of classifying the irreducible (cyclic) and
indecomposable~\cite{pasquier-saleur:common-structures,%
  alekseev-gluschenkov-lyakovskaya:reg-reps} representations of
a quantum enveloping algebra at a root of unity is still an open
problem.

\section{$q$-oscillators}

Since the invention of quantum mechanics, the quantum mechanical
oscillator algebra has remained a key model in the theory of quantum
physics and it is the basis of our understanding of quantum field
theory and canonical quantisation. The oscillator algebra is also
important in Lie algebra theory, since it allows the construction of
realisations of Lie algebras. In retrospect then, it is perhaps not
too surprising that $q$-analogues of the oscillator algebra were found
in the context of quantum groups and from them $q$-oscillator
realisations of quantum enveloping algebras could be constructed.

Chapter~\ref{chap:q-oscillators} has a detailed discussion of a
typical $q$-oscillator algebra~\cite{hayashi} and its representation
theory, and some comparisons with other $q$-oscillators. The algebra
has a large centre at a root of unity. Irreducible finite dimensional
quotients of the Fock modules exist. Semicyclic~\cite{sun-ge} and new
cyclic modules are also constructed. Remarkably this algebra enjoys a
unitary representation at even roots of unity. The quantum enveloping
algebra of a classical simple Lie algebra can be bosonised with this
algebra. I end this chapter with a new 2-parameter deformation of the
oscillator algebra (compare~\cite{wehrhahn-vraceanu:pq-oscillator}).


Chapter~\ref{chap:qr-oscillator} is a `mathematised' version of my
work, published in~\cite{petersen:qr-oscillator}, describing the
representations of a new quadratic (2-parameter) $qr$-deformation of
the oscillator algebra, inspired by a 2-parameter
deformation~\cite{fairlie:su2} of $\sltwo$.  There are a number of
improvements over the original work~\cite{petersen:qr-oscillator}: for
example the $qr$-oscillator algebra is considered at both odd and even
roots of unity (rather than just even roots of unity). In both cases
there exist irreducible finite dimensional quotients of its Fock
modules.  For real positive deformation parameters the Fock modules
are unitarisable.  Cyclic and semicyclic modules exist.  Bosonisation
maps for a number of well-known quadratic quantum algebras are
constructed into the Heisenberg-Weyl subalgebra. The Heisenberg-Weyl
subalgebra has a quantum group symmetry.

\section{Quantum affine algebras}

\subsection{} \label{intro:qkm-generic}
Affine Kac-Moody algebras~\cite{kac:book} are central to our present
understanding of conformal field theory (Wess-Zumino-Witten models),
solvable lattice models, and low dimensional integrable systems. From
the point view of Lie theory, they are also a generalisation of the
theory of simple (finite) Lie algebras. They are characterised by
Cartan matrices of affine type. To each affine Cartan matrix there is
a quantum enveloping algebra (due to Jimbo and Drinfeld), called a
quantum affine algebra.

Quantum affine algebras arise as symmetries of quantum spin chains,
solvable vertex and lattice models.
In~\cite{davies-foda-jimbo-miwa-nakayashiki:XXZ-diag} the XXZ-spin
chain model was studied in the thermodynamic limit. Using the
representation theory of quantum affine algebras, $q$-vertex
operators~\cite{frenkel-reshitikhin} and corner transfer matrix
methods, the authors were able to diagonalise the XXZ-Hamiltonian and
find its vacuum vectors. After they calculated correlation functions
for the model~\cite{jimbo-miki-miwa-nakayashiki:XXZ-correl} as traces
of bosonised~\cite{frenkel-jing} level 1 quantum affine algebra vertex
operators, a large number of authors studied the bosonisations
\cite{bernard:lmp,matsuo:boson,bourgourzi:unique,kimura,%
  awata-odake-shiraishi:sln,bourgourzi-weston:vertex} of quantum
affine algebras further and $q$-analogues of the Wakimoto construction
were obtained.  Quantum affine algebras are also symmetries of some
2~dimensional integrable quantum field theories. For example quantum
affine algebras are important in Toda
theory~\cite{bernard-leclair:qgp-symm,feigen-frenkel:toda}.  A further
interest in quantum affine algebras is in the programme to try to
develop a more geometrical understanding (``$q$-conformal field
theory'') \cite{frenkel-reshitikhin,matsuo:primary} and to better
understand its relation to 2d integrable field
theories~\cite{lukyanov}.



The representation theory of quantum affine algebras (for generic $q$)
is quite well understood. The results of Lusztig~\cite{lusztig:reps}
and Rosso~\cite{rosso:cmp}, describing the equivalence of the
irreducible representations of a finite quantum group (at generic $q$)
to the irreducible representations of the corresponding simple Lie
algebra, extend~\cite[33]{lusztig:book} to the case of a quantum
affine Kac-Moody algebra.  In analogy with the classical results
(see~\cite{kac:book}), it has been shown~\cite{zhang-gould:unitary}
that (a certain category of) integrable modules over a quantum affine
algebra (which includes all highest weight representations) are
completely reducible and that every irreducible integrable highest
weight module over a non-twisted quantum affine algebra with
$q\in\Rplus$ is unitary.  Chari and
Pressley~\cite{chari-pressley:eval-reps} have classified the
irreducible finite dimensional evaluation representations (IFDRs) of
the quantum affine algebra of $\sltwo$ and their tensor products.



In chapter~\ref{chap:kac-moody}, I describe the quantum affine algebra
of a (untwisted) affine Kac-Moody algebra and a recent extension of
the braid group automorphisms~\cite{beck:qkm-braid-group}, which
allowed an explicit proof that Drinfeld's loop algebra presentation is
isomorphic to the usual one.  Fock modules of the Heisenberg
subalgebra can be constructed. They are unitarisable when the
deformation parameter is positive real.

\subsection{Quantum affine algebras at a root of unity}
The structure of a quantum affine algebra at a root of unity is not
well understood.  Infinite dimensional representations of a quantum
affine algebra at a root of unity at non-zero level, \ie
representations which are not evaluation representations or loop
modules, have not been studied in the literature.

Finite dimensional minimal cyclic representations of quantum affine
algebras occur~\cite{bazhanov-kashaev-mangazeev-stroganov,%
  date-jimbo-miki-miwa:cyclic-kmsl3,%
  date-jimbo-miki-miwa:cyclic-twisted-qkmsl3} in the context of the
generalised Potts models. These representations are essentially
evaluation map representations of the quantum affine algebra at a root
of unity onto the cyclic (periodic) representation of the
corresponding finite quantum group.

In chapter~\ref{chap:qkm-root-of-1}, I present my
results~\cite{petersen:qkm-root-of-1} on a quantum affine algebra at a
root of unity.  I show that a quantum affine algebra has an infinite
dimensional centre at a root of unity. In particular I prove that at
an odd $\l$-th root of unity every real root vector in the algebra
raised to the $\l$-th power lies in the centre of the algebra.  Also
in the Heisenberg subalgebra, every imaginary root vector, whose mode
number is a multiple of $\l$ is also central.  Nevertheless a quantum
affine algebra at a root of unity is infinite dimensional over its
centre, and its modules with nonzero level (mod~$\l$) are necessarily
infinite dimensional.

An interesting problem would be to try to construct level~1 (and
higher) vertex operators at a $\l$-th root of unity.  I
hope that my results on a quantum affine algebra at a root of unity
may help with this problem and find applications in solvable lattice
models and integrable systems.  They may for example have some
relevance to the massless phase of the XXZ spin chain model (for which
$\modulus{q}=1$).

\section{Summary of original results}

Part~I of the thesis is essentially all review. Most of the results in
chapter~\ref{chap:q-oscillators} are well-known, though I believe
proposition~\ref{h4:classical-specialisation},
lemma~\ref{h4:CGST-oscill-trivial},
lemma~\ref{h4:odd-lem}, lemma~\ref{h4:Ue(h4')-even-semicyclic} and
section~\ref{h4:sec-cyclic-reps} are new.
Chapter~\ref{chap:qr-oscillator} is original work. Most of the results
of chapter~\ref{chap:kac-moody} have appeared already in the
literature or are natural generalisations of classical results, but I
have not seen lemma~\ref{km:centre}, lemma~\ref{km:quad-Serre-reln}
and most of section~\ref{km:heisenberg-section} before.
Chapter~\ref{chap:qkm-root-of-1} is original work, the most important
results being proposition~\ref{qkm-1:H-centre},
lemma~\ref{qkm-1:H-new-fock},
proposition~\ref{qkm-1:real-root-centre},
lemma~\ref{qkm-1:kmg-inf-over-Ze} and
proposition~\ref{qkm-1:V-infinite-dim}.

\section{Acknowledgements and thanks}

\subsection{Acknowledgements}
It is a pleasure to thank my supervisor, Chris Hull, for all his
encouragement, good advice, kindness and great patience with me, and
for directing me into the fascinating area of quantum groups.

I would like to thank many people for useful communications, in
particular conversations I had with following people were very
helpful: Daniel Arnaudon, David Fairlie, Tetsuji Miwa, Andrew
Pressley, Paul Wai and Cosmas Zachos.

I thank my examiners David Fairlie and Andrew Pressley for many
constructive comments and for pointing out a number of mistakes in the
original submitted version of the thesis, which resulted in this
improved final version.

I am greatly indebted to the authors of the following papers, without
which much of my work described in this thesis would not have been
possible: \cite{lusztig:geom-ded,deconcini-kac:unity,hayashi,jimbo:review,%
  kac:book,drinfeld:presentation,beck:qkm-braid-group}.

I am grateful to Professors Chris Isham and David Olive for the
inspiring lecture-courses, they gave at Imperial College,
London, on group theory, functional analysis and differential
geometry, and quantum field theory and Virasoro and Kac-Moody algebras
respectively, which I attended before and during my first year at
QMW. 

This work was supported by an U.K. \serc Research Studentship.

\subsection{Thanks}

This thesis would likely never have been finished without the constant
encouragement, support and partnership from my wife, Noriko Matsuo.

To Leonardo Palacios-Mor\'on, Zohora Khatun, Chinorat `Mo' Kobdaj,
Nelson Vanegas, Jos\'e-Luis Vazquez-Bello, Eduardo Ramos, Jos\'e
Figueroa-O'Farill and Bill Spence thank you for friendship, fun and
sometimes keeping me on the right path. I would like to thank
Professors Michael B. Green and John M. Charap, and Kate Shine, Christine
Ado and Cathy Boydon for their help during my time at QMW.

To my brother, mother, father and Noriko's and my family, I say thank
you for your support and love over the years.

Also hearty thanks to my close friends James Connolly, Kenji Kaneko
and Haitham Hindi.

Finally, on a more mundane note, I can say with gratitude to the GNU
software community that the preparation of this thesis was aided
enormously by the following excellent copylefted software systems:
\TeX, \LaTeX, BIB\TeX\ and \AmS-\LaTeX\ for typesetting, the powerful
GNU Emacs text-handling environment with the AUC-\TeX\ package for
editing \LaTeX, $X$-windows, and Linux: the free Unix-clone operating
system for personal computers.

\chapter{Preliminaries}

Let $\Z$ denote the set of integers, $\R$ the real numbers and $\C$
the field of complex numbers.  Let $\Zcross$, $\Rcross$ and $\Ccross$
denote the the corresponding sets with $\set{0}$ removed. For
$m,n\in\Z$, let $\range{m,n}$ denote the set of integers in the range
from $m$ to $n$ (empty if $m>n$). Let $\N$ denote the natural
numbers (non-negative integers). Let $\Zplus$ denote the positive
integers and let $\Zminus$ denote the negative integers. Let $\Rplus$
be the positive real numbers.

I start by recalling the definitions of some elementary algebraic
structures, to set up my notation.

\section{Groups, Rings and Fields}


\begin{definition}[group]
  A group $\structure{G,\cdot}$ is a set $G$ with an operation
  $\cdot$, such that

  (i)~there is an identity element $e$ in $G$
  \begin{displaymath}
    e\cdot g=g \quad \text{and} \quad g\cdot e= g \quad (\forall g\in G);
  \end{displaymath}

  (ii)~the operation $\cdot$ is closed
  \begin{displaymath}
    g\cdot g'\in G \quad (\forall g,g'\in G);
  \end{displaymath}

  (iii)~the operation is associative
  \begin{displaymath}
    g_1\cdot (g_2\cdot g_3) = (g_1\cdot g_2)\cdot g_3 \quad (\forall
    g_1,g_2,g_3\in G);
  \end{displaymath}

  (iv)~for every element of $g\in G$ there exists an inverse element
  $g\inv$
  \begin{displaymath}
    g\cdot g\inv = e \quad \text{and}\quad g\inv\cdot g = e \quad
    (\forall g\in G).
  \end{displaymath}
\end{definition}

\begin{definition}[ring]
  A ring $\structure{R,+,\cdot}$ is a set $R$ with an addition
  operation $+$ and a multiplication operation $\cdot$, such that (i)
  $\structure{R,+}$ is a commutative group, (ii) the multiplication is
  associative and (iii) the multiplication and addition are
  distributive:
  \begin{displaymath}
    (a+b)\cdot c = a \cdot c + b\cdot c
    \quad\text{and}\quad
    a\cdot(b+c)=a\cdot b + a\cdot c 
    \qquad (\forall a,b,c\in R).
  \end{displaymath}
  The identity element of $\structure{R,+}$ is usually denoted $0$.
  If a ring $R$ has a unity element~$1$ (such that $1\cdot x =x=x\cdot
  1$ ($x\in R$)) then $R$ is called a ring with unity
  $\structure{R,+,\cdot,1}$.

  A ring is said to be {\em commutative\/} if $ab=ba \quad \forall
  a,b\in R$.
\end{definition}

Let $\structure{R,+,\cdot,1}$ be a ring with unity. An element in $R$
is called a unit if it has both a left and a right multiplicative
inverse. A ring in which every element, except 0, is a unit is called
a division ring. A commutative division ring is called a {\em
  field\/}.

\begin{notation}
  From here on, unless otherwise stated, a field will always assumed
  to be of characteristic~$0$.
\end{notation}

\begin{example}
  Let $R$ be a ring. The ring $R\brackets{x}$ of polynomials in an
  indeterminant $x$ with coefficients valued in $R$ is an example of a
  commutative infinite dimensional ring. This can be extended to the
  commutative ring $R\brackets{x_1,x_2,\dots,x_n}$ of polynomials in $n$
  indeterminants with coefficients in $R$. Often it is also desirable
  to consider the ring $R\brackets{x,x\inv}$ of polynomials in an
  indeterminant $x$ and its inverse $x\inv$.
\end{example}

{}From a commutative ring $R$ with no zero divisors (equivalently an
entire ring $R$ or integral domain $R$), a quotient field $Q(R)$ can be
constructed.

\begin{proposition}[quotient field]
  Let $R$ be a commutative ring with no zero divisors and let
  $R':=R\backslash\set{0}$.  Consider all pairs $(a,b)\in R\times R'$.
  Then $(a,b)\sim(c,d)$, if $a\cdot d = b \cdot c$, defines an
  equivalence relation on $R\times R'$. Define $Q(R):=R\times R'/\sim$
  and denote the equivalence class containing $(a,b)$ by
  $\brackets{(a,b)}$. The addition and product in $Q(R)$ defined as
  \begin{displaymath}
    \brackets{(a,b)} \cdot \brackets{(c,d)}:= \brackets{(ac,bd)}
    \quad\text{and}\quad
    \brackets{(a,b)} + \brackets{(c,d)}:= \brackets{(ad+bc,bd)},
  \end{displaymath}
  give $Q(R)$ the structure of a field, the quotient field of $R$.
\end{proposition}

\begin{proof}
  See for instance~\cite[chapter~26]{fraleigh:book}.
\end{proof}

\begin{example}
  The quotient field of a ring of polynomials $R\brackets{x}$: eg
  $\C(q):= Q(\Cfree{q,q\inv})$ is the quotient field of the ring of
  polynomials $\Cfree{q,q\inv}$.
\end{example}

\section{Modules and Vector spaces}


Later when I come to consider the representations of quantum groups
and quantum algebras, the notation of modules and vector spaces will
be important.

\begin{definition}[module]
  Let $R$ be a ring. A {\em left-module\/} $\structure{M,+,R,\cdot}$
  over $R$ (or left $R$-module) is an abelian group $\structure{M,+}$
  and an operation of $\structure{R,\cdot}$ on $M$ ($R\times M \mapsto
  M$) such that
  \begin{displaymath}
    (a+b)\cdot u = a \cdot u + b\cdot u \quad\text{and}\quad
    a\cdot(u+v)=a\cdot u + a\cdot v \qquad (a,b\in R, u,v\in M).
  \end{displaymath}
  A module over a field $k$ is called a {\em vector space\/}
  ($k$-vector space).
\end{definition}

\section{Associative Algebras}

\begin{notation}
  Let $U$ be a space (set). Denote by $\id$ the trivial map
  $\id:U\mapto U$, which maps $u\mapsto u$ ($u\in U$). When necessary,
  I use a subscript $\id_U$ to emphasise the space on which the map is
  acting.
\end{notation}


\begin{definition}[algebra]
  Let $R$ be a commutative ring. An {\em algebra\/}
  $\structure{\cs{A},+,R,\cdot}$ over $R$ (or $R$-algebra) is an
  $R$-module $\cs{A}$ with a bilinear product map $m:\cs{A}\times
  \cs{A} \mapsto \cs{A}$.  An algebra is said to be associative if
  \begin{displaymath}
    x\cdot (y \cdot z) = (x\cdot y)\cdot z \quad  (x,y,z\in \cs{A}).
  \end{displaymath}
  If $\cs{A}$ has an element $\unity$ such that $x\cdot \unity=
  x=\unity\cdot x$ ($x\in \cs{A}$), then the algebra is called an
  $R$-algebra with unity, and there is an natural map $\mu:R\mapto
  \cs{A}$, which maps $\mu:a\mapsto a\cdot \boldsymbol{1}$ ($a\in R$) called the
  unity map, by definition it satisfies
  \begin{displaymath}
    m \compose (\mu\times\id) (a, x) = a\cdot x =
    m\compose (\id\times\mu) (x, a) \qquad  (a\in R,x\in \cs{A}).
  \end{displaymath}
\end{definition}

\begin{notation}
  From here on, unless otherwise stated, an algebra will always be
  assumed to be associative.
\end{notation}

\begin{remark}
  Let $\structure{\cs{A},+,R,\cdot}$ be an associative algebra. The
  bilinear product of the algebra maps $m:(x,y)\mapsto
  x\cdot y$. Then the associativity of algebra can be expressed as
  \begin{displaymath}
    m\compose (m \times \id) (x, y, z) = m\compose (\id \times m) (x,
    y, z)\quad (x,y,z\in \cs{A}).
  \end{displaymath}
\end{remark}

\begin{example}
  Let $V$ be a vector space. Denote by $\End(V)$ the set of all linear
  maps $V\mapto V$, the algebra of (linear) endomorphisms of $V$.
  Given a basis of $V$, the elements of $\End(V)$ can be realised as
  matrices.
\end{example}

\begin{definition}[free algebra]
  Let $X:=\set{x_1,x_2,\dots,x_n}$ be a set of $n$ distinct letters.
  Consider the set of finite (noncommutative) monomials in the
  elements of $X\union 1$ (``strings of letters in the alphabet
  $X$''). The set of linear combinations of these monomials with
  coefficients in a ring $R$ (noncommutative polynomials)
  $R\brackets{\brackets{x_1,x_2,\dots,x_n}}$ is naturally an abelian
  group $\structure{R\brackets{\brackets{x_1,x_2,\dots,x_n}},1,+}$. It
  can be given the structure of an associative unital algebra
  $\structure{R\brackets{\brackets{X}},+,1,\cdot}$ by defining the
  multiplication of any pair of monomials $x_{i_1}\cdots x_{i_r}$ and
  $x_{j_1}\cdots x_{j_s}$ ($i_1,\dots,i_r\in X$, $j_1,\dots, j_s\in X$
  and $r,s\in\Zplus$) in an obvious way to be
  \begin{align*}
    (x_{i_1}\cdots x_{i_r})\cdot (x_{j_1}\cdots x_{j_s}) &:=
    x_{i_1}\cdots x_{i_r} \cdot x_{j_1}\cdots x_{j_s},\\
    (x_{i_1}\cdots x_{i_r})\cdot 1 &:= (x_{i_1}\cdots x_{i_r}),\\
    1\cdot (x_{i_1}\cdots x_{i_r}) &:= (x_{i_1}\cdots x_{i_r}).
  \end{align*}
  This $R$-algebra is called the {\em free $R$-algebra\/} on the generators
  in the set $X$. 
\end{definition}

\begin{definition}[subalgebra]
  Let $\alg{A}$ be an $R$-algebra, a subset $\alg{B}\subset \alg{A}$
  is called a $R$-subalgebra of $\alg{A}$, if $\alg{B}$ forms a (closed) algebra in
  $\alg{A}$.
\end{definition}

\begin{remark}
  Let $S$ be a subring of $R$ and let $\alg{A}$ be a $R$-algebra, then
  there is another type of subalgebra possible: an $S$-subalgebra of
  $\alg{A}$.
\end{remark}

\begin{definition}
  Let $\alg{A}$ and $\alg{B}$ be $R$-algebras. A $R$-linear map
  $\phi:\alg{A} \mapto \alg{B}$, is called an $R$-algebra homomorphism
  if:
  \begin{displaymath}
    \phi(x\cdot y) = \phi(x) \cdot \phi(y)
    \quad\text{and}\quad
    \phi(x+y)=\phi(x) + \phi(y) \quad
    (x,y\in\alg{A}).
  \end{displaymath}
  If the homomorphism is injective (`one-to-one') and surjective
  (`onto'), the map is called an isomorphism and the algebras are said
  to be isomorphic.
\end{definition}

\begin{definition}[ideal]
  Let $\alg{A}$ be an algebra, a subalgebra $\alg{I}$ of $\alg{A}$ is
  called a left (respectively right) ideal, if $\forall x\in \alg{I}$,
  $\alg{A}\cdot x \subseteq \alg{I}$ (respectively $x\cdot
  \alg{A}\subseteq \alg{I}$). A subalgebra which is both a left and a
  right ideal, is called a two sided ideal.  If $\alg{I}\neq\alg{A}$,
  then $\alg{I}$ is a proper ideal of $\alg{A}$. If
  $\alg{I}\neq\set{0}$, then $\alg{I}$ is a nontrivial ideal of
  $\alg{A}$.
\end{definition}

\begin{notation}
  From here on, unless otherwise stated, an ideal of an algebra will
  always be a proper, non-trivial, two sided ideal.
\end{notation}

\begin{lemma}
  Let $\alg{A}$ be an $R$-algebra, with a (two sided) ideal $\alg{I}$.
  There is a canonical homomorphism
  $\phi:\alg{A}\mapto\alg{A}/\alg{I}$, called the projection map. It
  follows that the quotient $\alg{A}/\alg{I}$ is an algebra (the
  quotient algebra of $\cs{A}$ by $\cs{I}$).
\end{lemma}

\begin{proof}
  Consider elements $x,x',y,y'\in\alg{A}$, such that $x\neq x'$ and
  $y\neq y'$, but $\phi(x)=\phi(x')$ and $\phi(y)=\phi(y')$. Then I
  wish to check that the addition and multiplication are well defined
  in $\alg{A}/\alg{I}$, \ie
  \begin{displaymath}
    \phi(x+y)=\phi(x'+y') \quad \text{and}\quad
    \phi(x\cdot y)=\phi(x'\cdot y').
  \end{displaymath}
  Let $a=x-x'$ and $b=y-y'$. Clearly $a,b\in\alg{I}$. Then
  $x+y=x'+y'+a+b$ and $x\cdot y=x'\cdot y'+x'\cdot b+a\cdot y'$.
  But $a+b,x'\cdot b,a\cdot y'\in\alg{I}$ and the lemma is proven.
\end{proof}

\begin{lemma} \label{prel:lem:free-quotient}
  Let $\alg{A}:=k\brackets{\brackets{x_1,\dots,x_n}}$ be a free
  $k$-algebra (on $n$ generators). Let $\set{r_1,\dots,r_m}$ be $m$
  distinct elements in $\alg{A}$ and let $\alg{J}:=
  \text{Span}_k(r_1,\dots r_m)$. Then $\alg{I}:=\alg{A}\cdot
  \alg{J}\cdot \alg{A}$ is a two-sided ideal in $\alg{A}$. Therefore
  the quotient $\alg{A}/\alg{I}$ is an algebra.
\end{lemma}

\begin{definition}[centre]
  Let $\alg{A}$ be an $R$-algebra. The centre $\Zcentre(\alg{A})$ of
  $\alg{A}$ is defined to be the set $\set{x\in\alg{A}\mid x\cdot y =
    y\cdot x, \;\;\forall y\in \alg{A}}$.
\end{definition}

\section{Combining Vector spaces and Tensor algebras}

\begin{definition}[Cartesian product]
  Let $V$ and $W$ be two spaces. Then their Cartesian product
  is defined to be $V\times W:=\set{(v,w)\mid  v\in V,w\in W}$.
\end{definition}

Let $V$ and $W$ be $R$-modules. Two types of $R$-module structure can
be naturally defined on their Cartesian product $V\times W$.

\subsection{Direct sum}
Let $V$ and $W$ be $R$-modules and $V\times W$ be their
Cartesian product. $V\times W$ can be endowed with the structure of a
$R$-module $V\directsum W$, the {\em direct sum\/} of $V$ and $W$,
by defining
\begin{displaymath}
  \begin{aligned}
    a\cdot (v,w) &:= (a\cdot v,a\cdot w)\\
    (v,w)+(v',w') &:= (v+v',w+w')
  \end{aligned}
  \qquad  a\in R , v,v'\in V , w,w'\in W.
\end{displaymath}

\subsection{Tensor product}
Let $R$ be a ring. Let $V$ and $W$ be $R$-modules. Let $M$ be the free
module generated over $R$ by the set $V\times W$. Let $N$ be the
submodule of $M$ defined by
\begin{multline*}
  N:=\set{(v+v',w)- (v,w)- (v',w),\; (v,w+w')-
    (v,w)- (v,w'),\;\\
    (a\cdot v,w)- a(v,w),\; (v,a\cdot w)-
    a(v,w)\mid \forall v,v'\in V, w,w'\in W, a\in R}.
\end{multline*}
Then the canonical bilinear map $\phi:V\times W\mapto M/N:= V\tensor
W$ defines the $R$-tensor product of $V$ and $W$.  Let $v\in V$ and
$w\in W$. Denote $\phi(v,w)$ by $v\tensor w$.  By construction the
following identities hold in $V\tensor W$
\begin{displaymath}
  \begin{aligned}
    (a\cdot v)\tensor w &\equiv a\cdot (v\tensor w) \equiv
    v\tensor(a\cdot w),\\ 
    (v+v')\tensor w &\equiv v\tensor w + v'\tensor w,\\
    v\tensor(w+w') &\equiv v\tensor w + v\tensor w'.
  \end{aligned}
  \qquad  a\in R , v,v'\in V , w,w'\in W
\end{displaymath}
The tensor product of two modules is a module and the tensor product
over a field $k$ of two $k$-vector spaces is a vector space.

Having taken the tensor product $U\tensor V$ of two $R$-modules $U$
and $V$, the process can be repeated: I can take the tensor product of
$U\tensor V$ with another $R$-module $W$ to create $(U\tensor
V)\tensor W$ (or $W\tensor (U\tensor V)$).

\begin{proposition}[associativity] \label{prel:prop:tensor-assoc}
  Let $U$, $V$ and $W$ be $k$-vector spaces. There is a unique
  isomorphism $(U\tensor V)\tensor W \mapto U\tensor (V\tensor W)$,
  which maps
  \begin{displaymath}
    (u\tensor v)\tensor w \mapsto u\tensor (v\tensor w)
  \end{displaymath}
\end{proposition}
\begin{proof}
  See for example~\cite[chapter~XVI]{lang:algebra}.
\end{proof}

\begin{lemma}[twist]
  Let $V$ and $W$ be $k$-vector spaces. There is a unique isomorphism
  $V\tensor W \mapto W\tensor V$, $v\tensor w \mapsto w\tensor v$.
\end{lemma}

\begin{lemma}[tensor product of algebras] Let $\alg{A}$ and $\alg{B}$
  be $k$-algebras. Consider then their tensor product
  $\alg{A}\tensor\alg{B}$ as vector spaces. This vector space can be
  given the structure of a $k$-algebra by defining the product in
  $\alg{A}\tensor\alg{B}$ to be
  \begin{displaymath}
    (v_1\tensor w_1)\cdot(v_2\tensor w_2):=(v_1\cdot
    v_2)\tensor(w_1\cdot w_2).
  \end{displaymath}
\end{lemma}

A very important construction among associative algebras is that of
the tensor algebra of a vector space. The universality of tensor
algebras is essentially contained in the following fact: every
(finitely generated) associative algebra is the quotient of the tensor
algebra of some (finite dimensional) vector space.

\begin{definition}[tensor algebra]
  Let $V$ be a $k$-vector space. For each integer $n\in\Zplus$ define
  $T^n(V):=\Tensor_{i=1}^n V$. Define $T^0(V):=k$. Because the tensor
  product is associative (proposition~\ref{prel:prop:tensor-assoc}),
  there is a bilinear map $T^r(V)\times T^s(V)\mapto T^{r+s}(V)$.
  Using this I can give the space
  \begin{displaymath}
    T(V):=\Directsum_{n=0}^\infty V^{\tensor n}\equiv k\directsum
    (V) \directsum (V\tensor V) \directsum (V\tensor V\tensor V)
    \directsum \cdots
  \end{displaymath}
  the structure of a $k$-algebra the {\em tensor algebra\/} $T(V)$ of
  $V$. The product in $T(V)$ is again denoted $\tensor$.
\end{definition}

\begin{lemma}
  (i) Let $T(V)$ be the tensor algebra of a $k$-vector space $V$.
  $T(V)$ is an associative, infinite dimensional $k$-algebra.
  
  (ii) Let $W$ be a $k$-vector space such that $\dim(W)=\dim(V)$. Then
  $T(V)$ and $T(W)$ are isomorphic as $k$-algebras.
\end{lemma}


\begin{lemma}
  Let $k\brackets{\brackets{x_1,x_2,\dots,x_n}}$ be a free $k$-algebra
  with unity and let $V$ be an $n$-dimensional $k$-vector space.  Then
  $k\brackets{\brackets{x_1,x_2,\dots,x_n}}$ is isomorphic as a
  $k$-algebra to $T(V)$.
\end{lemma}

  Let $\set{v_1,v_2,\dots,v_n}$ be a linear basis of $V$. Then the map
  \begin{align*}
    k\brackets{\brackets{x_1,x_2,\dots,x_n}} &\mapto T(V),\\
    x_i &\mapsto v_i,
  \end{align*}
  is an $k$-algebra isomorphism.

\begin{notation}
  Let $\alg{A}$ be a $k$-algebra. I will denote the two-sided ideal
  $\alg{I}$ in $\alg{A}$ (freely) {\em generated\/} over $k$ by
  elements $\set{r_1,\dots,r_n}\subset \alg{A}$
  (cf.~\ref{prel:lem:free-quotient}) by the notation
  \begin{displaymath}
    \ideal{r_1,\dots,r_n}:= \alg{I}\equiv
    \alg{A}\cdot \text{Span}(\set{r_1,\dots,r_n}) \cdot \alg{A}.
  \end{displaymath}
\end{notation}

\begin{remark}
  A important problem when considering the quotient of a tensor or
  free algebra by a two-sided ideal is whether the quotient algebra is
  associative or not.
\end{remark}

I briefly recall the construction of the symmetric algebra from the
tensor algebra of a vector space.

\begin{proposition}[symmetric algebra]
  Let $V$ be a $k$-vector space and $T(V)$ its tensor algebra. The the
  elements $\set{x\tensor y - y\tensor x\mid x,y\in V}$ generate an
  ideal in $T(V)$, $I_{\text{sym}}:= \ideal{x\tensor y - y\tensor
    x\mid x,y\in V}$. The quotient algebra $S(V):=T(V)/I_{\text{sym}}$
  is a commutative infinite dimensional algebra, the {\em symmetric
    algebra\/} of $V$.
\end{proposition}

\begin{proof}
  See for instance~\cite[chapter V]{humphreys}.
\end{proof}

\begin{remark}
  Let $V$ be a $n$-dimensional $k$-vector space. Quotienting $T(V)$ by
  an anti-symmetrising ideal, the exterior (grassmann) algebra
  \begin{displaymath}
    A(V):=T(V)/\ideal{a\tensor b + b\tensor a\mid a,b\in V},
  \end{displaymath}
   on $V$ is constructed. $A(V)$ is a
  $2^n$ dimensional associative $k$-algebra.
\end{remark}

\section{Lie algebras and Universal enveloping algebras}

\begin{definition}[Lie algebra] 
  A {\em Lie algebra\/} $\structure{\g,\brackets{\cdot,\cdot}}$ is a
  $k$-algebra with a $k$-bilinear product map
  $\brackets{\cdot,\cdot}:\g\times \g\mapto \g$ satisfying
  \begin{gather}
    \brackets{x,x} =0 \qquad (x\in \g), \tag{antisymmetry} \\
    \brackets{x,\brackets{y,z}}+ \brackets{y,\brackets{z,x}}+
    \brackets{z,\brackets{x,y}}=0\quad (x,y,z\in \g). \tag{Jacobi
      identity}
  \end{gather}
\end{definition}

\begin{remark}
  The antisymmetry property $\brackets{x,x}=0$ ($x\in \g$) implies
  that $\brackets{x,y}=-\brackets{y,x}$. This is easily seen by
  considering $\brackets{x+y,x+y}$.
\end{remark}

\begin{lemma}
  Let $\structure{\alg{A},\cdot,k}$ be an associative $k$-algebra. 
  Defining the commutator $\brackets{\cdot,\cdot}:\alg{A}\times
  \alg{A} \mapto \alg{A}$
  \begin{displaymath}
    \brackets{x,y}:= x\cdot y - y\cdot x \qquad  (x,y\in \alg{A}),
  \end{displaymath}
  then $\structure{\alg{A},\brackets{\cdot,\cdot}}$ is a Lie algebra.
\end{lemma}

\begin{proof}
  Antisymmetry follows immediately from the definition of the
  commutator and the Jacobi identity from the associativity of $\alg{A}$.
\end{proof}

\begin{definition}[Lie subalgebra]
  Let $\g$ be a Lie algebra. Let $\h$ be a linear subspace of $\g$.
  If $\h$ itself forms a closed Lie algebra, then $\h$ is called a Lie
  subalgebra of $\g$.
\end{definition}

\begin{example} Let $n\in\Zplus$. Consider the $k$-algebra of $n\times
  n$-matrices $\alg{M}_n$ (with entries valued in $k$). Then
  $\structure{\alg{M}_n,\brackets{\cdot,\cdot}}$ is a Lie algebra,
  usually denoted $\gl_n(k)$. The subset of $\alg{M}_n$ of traceless
  matrices, forms a Lie subalgebra, denoted $\sln(k)$.
\end{example}

The standard way of analysing the structure of Lie algebras is by
consideration of their ideals.

\begin{definition}[ideal]
  Let $\g$ be a Lie algebra and $\h$ an Lie subalgebra of $\g$. If
  $\brackets{u,x}\in \h$ ($u\in \h$ and $x\in \g$), then $\h$ is
  called an ideal of $\g$.
\end{definition}

\begin{definition}[simple]
  Let $\g$ be a Lie algebra, which as a vector space has dimension
  greater than 1. The Lie algebra $\g$ is called {\em simple\/}, if it
  has no (proper, non-trivial) ideals.
\end{definition}

\begin{lemma}[quotient algebra]
  Let $\g$ be a Lie algebra and $\h$ an ideal in $\g$. Then the quotient
  algebra $\g/\h$ is a Lie algebra.
\end{lemma}

\begin{lemma}[derived Lie algebra] \label{prel:derived-Lie-algebra}
  Let $\g$ be a Lie algebra. The subset $\g':=\set{\brackets{x,y}\mid
    x,y\in \g} =:\brackets{\g,\g}$ forms an ideal in $\g$.
\end{lemma}

\begin{proof} Clearly $\g'$ is a Lie subalgebra of $\g$, since
  \begin{displaymath}
    \brackets{\brackets{x,y}, \brackets{x',y'}}\in \g' \qquad
    x,x',y,y'\in \g.
  \end{displaymath}
  That $\g'$ is an ideal of $\g$, follows since
  $\brackets{\g,\g'}\subseteq \g'$
\end{proof}

It is well-known that universal enveloping algebras play an important
role in Lie algebra and Lie group theory, specially in their
representation theory (see for example~\cite{humphreys}).


\begin{proposition}[universal enveloping algebra]
  \label{prel:U(g)-def}
  Let $\structure{\g,\brackets{\cdot,\cdot},+}$ be a Lie algebra. Let
  $T(\g)$ be the tensor algebra of the $k$-vector space
  $\structure{\g,+}$. Consider then the following ideal in $T(\g)$
  \begin{displaymath}
    I_{\g}:=\ideal{x\tensor y - y\tensor x - \brackets{x,y} \mid x,y\in \g} 
  \end{displaymath}
  The quotient algebra $\U(\g):=T(\g)/I_{\g}$ is called the {\em universal
    enveloping algebra\/}. $\U(\g)$ is an associative infinite
  dimensional algebra with unity.
\end{proposition}

\section[Representations]{Representations and Reducibility}

\begin{definition}[representation] 
  Let $V$ be a $k$-vector space and let $\alg{A}$ be a $k$-algebra. If
  $\pi:\alg{A}\mapto\End(V)$ is a $k$-algebra homomorphism:
  \begin{equation*}
    \pi(x\cdot y) \cdot v = \pi(x)\cdot\pi(y)\cdot v ,\qquad  x,y\in
    \alg{A}, v\in V,
  \end{equation*}
  then the pair $\structure{V,\pi}$ is called a (linear)
  representation of $\alg{A}$ on $V$.  The vector space $V$ carrying
  the representation $\pi$ of $\alg{A}$ is also called an {\em
    $\alg{A}$-module}.
\end{definition}

\begin{definition}[intertwiner]
  Let $V$ and $W$ be $\alg{A}$-modules.  A vector space homomorphism
  $\phi:V \to W$ is called an $\alg{A}$-module homomorphism, if $\phi$
  is such that:
\begin{equation*}
  \phi(\pi_V (a)\cdot v) = \pi_W(a) \cdot \phi(v) ,  a\in\alg{A} ,
  v\in V.
\end{equation*}
Such a map is also called an intertwining map (intertwiner) from $V$
to $W$.  If further $\phi$ is an isomorphism of vector spaces, then
$\phi$ is called an $\alg{A}$-module isomorphism and the two modules
$V$ and $W$ are {\em equivalent\/} as $\alg{A}$-modules.
\end{definition}

\begin{definition}[submodule]
  Let $V$ be an $\alg{A}$-module. If $V$ contains a proper linear
  subspace $W$, such that $W$ is closed under the action of $\alg{A}$
  (\ie $\pi(\alg{A})\cdot W \subseteq W$) then $W$ is said to carry a
  {\em subrepresentation} of $\alg{A}$ and $W$ is a {\em
    $\alg{A}$-submodule} of $V$.
\end{definition}

\begin{definition}[irreducible]
  An $\alg{A}$-module is called {\em irreducible}, if it has no
  (proper, non-trivial) submodules, otherwise it is called {\em
    reducible}.
\end{definition}

\begin{definition}[completely reducible]
  An $\alg{A}$-module is called {\em completely reducible\/}, if
  it is equivalent to a direct sum of irreducible $\alg{A}$-modules.
\end{definition}

\begin{definition}[indecomposable]
  An $\alg{A}$-module is called {\em indecomposable}, if it is not
  equivalent to a direct sum of (two or more) $\alg{A}$-modules,
  otherwise the module is called {\em decomposable}.
\end{definition}


An irreducible $\alg{A}$-module is always indecomposable, but in
general there will exist indecomposable $\alg{A}$-modules which are
not irreducible. Of course the full reduction of an indecomposable
representation gives an irreducible one.

\section{Hopf Algebras}
\label{prel:hopf-algebras}

I come now to the definition of some of the key objects, examples of
which will be discussed in detail throughout this thesis. Earlier I
wrote down the definition of an algebra. I define next a dual
structure. It is dual in the sense of category theory and because it
basically defines a structure with an ``unmultiplication'' map.

\begin{notation}
  Let $V$ be an $R$-module. The map $\sigma:V\tensor V\mapto V\tensor
  V$, which maps $x\tensor y\mapsto y\tensor x$ ($x, y\in V$), is
  called the twist map.
\end{notation}

\begin{definition}[coalgebra]
  Let $R$ be a ring.  A {\em coalgebra\/}
  $\structure{\alg{C},+,R,\Delta}$ over $R$ ($R$-coalgebra) is a
  $R$-module $\alg{C}$ with a linear map $\Delta:\alg{C}\mapto
  \alg{C}\tensor \alg{C}$ called the coproduct (or comultiplication
  map). If
  \begin{displaymath}
    (\id \times \Delta)\compose \Delta (x)=
    (\Delta \times \id)\compose \Delta (x) \quad  (\forall x\in \alg{C})
  \end{displaymath}
  then the coproduct $\Delta$ and the coalgebra are called
  coassociative. If there exists a map $\epsilon:\alg{C}\mapto R$ such that
  \begin{displaymath}
    (\epsilon\times \id) \compose \Delta (x)= x =
    (\id\times\epsilon) \compose \Delta (x) \qquad (\forall x\in \alg{C}),
  \end{displaymath}
  then the map $\epsilon$ is called a counit map and the coalgebra is
  called a coalgebra with counit $\structure{\alg{C},+,R,\Delta,\epsilon}$.
  A coalgebra $\structure{\alg{C},\Delta}$ is called {\em cocommutative\/} if
  $\sigma\compose \Delta(x) = \Delta(x)$ ($\forall x\in \alg{C}$).
\end{definition}

\begin{lemma}
  Let $\structure{\alg{A},+,m;k}$ be a (finite) $k$-algebra and let
  $\structure{\alg{A}^*,+;k}$ be the $k$-vector space dual to
  $\structure{\alg{A},+;k}$, with the dual pairing
  $\pairing{\cdot,\cdot}:\alg{A}^*\tensor \alg{A}\mapto k$. Then the
  dual pairing induces a coalgebra structure
  $\structure{\alg{A}^*,+,\Delta;k}$ on $\alg{A}^*$
  \begin{displaymath}
    \pairing{\Delta^*(x^*),x\tensor y}:= \pairing{x^*,m(x\tensor y)}
    \qquad (x^*\in \alg{A}^*; x,y\in \alg{A}).
  \end{displaymath}
\end{lemma}

Many examples of coalgebras are also at the same time algebras. This
motivates the following definition.

\begin{definition}[bialgebra]
  Let $R$ be a ring and $\alg{B}$ an $R$-algebra. $\alg{B}$ has the
  structure of a {\em bialgebra\/} $\structure{\alg{B},+,m,\Delta}$,
  if there exists a coproduct map $\Delta$, that is an algebra
  homomorphism.  Then $\structure{\alg{B},\Delta;R}$ is a
  $R$-coalgebra.
\end{definition}

\begin{remark}[uniqueness]
  Usually a bialgebra is constructed by giving an algebra a coalgebra
  structure. However a coalgebra structure on an algebra is seldom
  unique. For any noncocommutative coproduct $\Delta$ on an algebra, the
  twisted coproduct $\Delta':=\sigma\compose\Delta$ affords another
  coproduct on the algebra. These two coproducts lead to two different
  bialgebras.
\end{remark}

Most bialgebras, that I shall consider, will have a unity $\mu$ and
a counit $\epsilon$ homomorphism and an inverse map $S$
(anti-automorphism), which should be likened to the inverse map of a
group $G$, that maps $g\mapsto g\inv$ ($g\in G$).

\begin{definition}[antipode]
  Let $\structure{B,+,R,m,\Delta,\mu,\epsilon}$ be a bialgebra with
  unity $\mu$ and counit $\epsilon$ homomorphisms. An algebra
  anti-automorphism $S:B\mapto B$ which satisfies
  \begin{gather*}
    S\compose m (x\tensor y) = m\compose(S(y)\tensor S(x))\quad
     x,y\in B\\
    m\compose (S\times \id)\compose \Delta (a) =
    \mu \compose \epsilon (a) =
    m\compose (\id\times S)\compose \Delta (a)
  \end{gather*}
  is called an {\em antipode\/} map.
\end{definition}

\begin{remark}
  In contrast with the bialgebra structure of an algebra, which in
  general is not unique, the antipode map of a bialgebra is unique.
\end{remark}


Finally I am in a position to be able to make the following

\begin{definition}[Hopf algebra]
  A {\em Hopf algebra\/}
  $\structure{\alg{H},+,R,m,\Delta,\mu,\epsilon,S}$ is a bialgebra
  with a unity ($\mu:R\mapto \alg{H}$), counit
  ($\epsilon:\alg{H}\mapto R$) and an antipode anti-automorphism $S$.
\end{definition}

  

\begin{lemma} \label{prel:Hopf-U(g)}
  Let $\g$ be a $k$-Lie algebra and $\U(\g)$ its universal enveloping
  algebra. Then $\U(\g)$ has the structure of a Hopf algebra by
  extending the maps
  \begin{displaymath}
    \begin{alignedat}{2}
      \Delta(x) &:= x\tensor\unity + \unity\tensor x, \quad&
      \Delta(a\cdot\unity)&:= a\cdot \unity\tensor\unity,\\
      \mu(a)&:= a\unity, &&\\ 
      \epsilon(x)&:= 0, &
      \epsilon(a\cdot\unity)&:= a,\\
      S(x)&:= -x, &
      S(\unity)&:= 1,
    \end{alignedat}
     \qquad  (\forall x\in \g, a\in k)
  \end{displaymath}
  homomorphically to all of $\U(\g)$.
\end{lemma}

\begin{proof}
  It is necessary to show that all the Hopf algebra maps are
  algebra homomorphisms. For the coproduct
  \begin{align*}
    \Delta(x\cdot y - y \cdot x)&=
    \Delta(x)\cdot \Delta(y) - \Delta(y)\cdot \Delta(x)\\
    &=(x\cdot y)\tensor\unity + x\tensor y + y\tensor x + \unity\tensor
    (x\cdot y)\\
    &\qquad -(y\cdot x)\tensor\unity- y\tensor x- x\tensor y- \unity
    \tensor(y\cdot x)\\ 
    &=(x\cdot y - y\cdot x)\tensor \unity + \unity\tensor(x\cdot y -
    y\cdot x)\\
    &=\Delta(\brackets{x,y})
  \end{align*}
\end{proof}
\section{Representations of algebras and Hopf algebras}

One reason for the importance of Hopf algebras is that the Hopf
structure of an algebra facilitates the construction of
representations of the algebra, as I now explain.

A trivial representation of an algebra is one dimensional.

\begin{lemma}[trivial representation]
  Let $\alg{A}$ be a $k$-algebra (bialgebra) with a counit
  homomorphism map $\epsilon:\alg{A}\mapto k$. Then the counit map
  $\epsilon$ gives a trivial one dimensional representation of
  $\alg{A}$ on $k$.
\end{lemma}

\begin{proposition}[tensor product]
  Let $\alg{A}$ be a $k$-algebra with a bialgebra structure
  $\structure{\alg{A},+,\cdot,\Delta;k}$. Let
  $\structure{\pi_1,V_1}$ and $\structure{\pi_2,V_2}$ be
  representations of $\alg{A}$. Then
  $\structure{\pi_{12},V_1\tensor V_2}$, with
  \begin{displaymath}
    \pi_{12}:=(\pi_1\tensor \pi_2)\compose \Delta,
  \end{displaymath}
  is an tensor product representation of the two representations.
\end{proposition}

The antipode map allows the construction of dual representations.

\begin{lemma}[dual representations]
  Let $\structure{\alg{A},\cdot;k}$ be a $k$-algebra, with a Hopf
  algebra structure $\structure{\alg{A},+,\cdot,\Delta,S;k}$. Let $\w$
  be an algebra anti-automorphism of the algebra $\alg{A}$. Let
  $\structure{\pi,V}$ be a representation of the algebra $\alg{A}$ and
  let $V^*$ be the dual vector space of $V$, with the dual pairing
  $\pairing{\cdot,\cdot}$. Defining
  \begin{displaymath}
    \pairing{\pi^\w(x)\cdot v^*,v}:= \pairing{v^*,\pi(\w(x))\cdot v},
  \end{displaymath}
  then $\structure{\pi^\w,V^*}$ is the dual representation (dual
  $\alg{A}$-module) to $\structure{\pi,V}$ with respect to $\w$. Note
  that the dual representation $\alg{A}\mapto\End(V^*)$ is also a left
  representation: $\pi^*(x\cdot y)= \pi^*(x)\cdot \pi^*(y)$.
\end{lemma}

In particular choosing $\w=S$, the dual representation $(\pi^S,V^*)$
is obtained. If the inverse of the antipode $S\inv$ exists, $\w=S\inv$
leads to another dual representation $(\pi^{S\inv},V^*)$.

\begin{definition}[corepresentations] \label{prel:corepresentation}
  Let $\structure{\alg{C},\Delta}$ be a $k$-coalgebra and let $V$ be a
  $k$-vector space. A {\em right corepresentation\/} of
  $\structure{\alg{C},\Delta}$ on $V$ is a map $\Delta_R:V\mapto
  V\tensor\alg{C}$ such that
  \begin{displaymath}
    (\Delta_R\times \id)\compose \Delta_R(v) =
    (\id\times\Delta)\compose \Delta_R(v) \qquad (v\in V).
  \end{displaymath}
  If further $\alg{C}$ is a coalgebra with counit, then a
  corepresentation $\Delta_R$ of $\structure{\alg{C},\Delta,\epsilon}$ is
  required to satisfy
  \begin{displaymath}
    (\id\times\epsilon)\compose \Delta_R(x) = \id (x)\qquad
    (x\in\alg{C}).
  \end{displaymath}
\end{definition}

\begin{example}
  Let $\structure{\alg{A},m}$ be an $k$-algebra and
  let $\structure{\alg{A}^*,\Delta}$ be the dual coalgebra. Let
  $\structure{\pi,V}$ be a representation of $\alg{A}$ and $V^*$ the
  linear vector space dual to $V$, then the map $\Delta_L$
  \begin{displaymath}
    \pairing{\Delta_L(v^*),x\tensor v}_{\alg{A}\tensor V}:=
    \pairing{v^*,\pi(x)\cdot v}
  \end{displaymath}
  is a left corepresentation of $\alg{A}^*$.
\end{example}

\section{Yang-Baxter equations and $\cs{R}$-matrices}


When considering the tensor product representations of a (bi)algebra,
an interesting problem is determining under what conditions
$\structure{\pi_{12},V_1\tensor V_2}$ and
$\structure{\pi_{21},V_2\tensor V_1}$ are equivalent. It is perhaps
surprising that for a general Hopf algebra it is a nontrivial
problem to find an intertwiner between these two tensor product
representations.

\begin{definition}[$\cs{R}$-matrix]
  Let $\structure{\alg{A},\cdot,\Delta;k}$ be a $k$-bialgebra. Let
  $\structure{\pi_1,V_1}$ and $\structure{\pi_2,V_2}$ be
  representations of the $k$-algebra $\structure{\alg{A},\cdot}$.  If
  the tensor product representations
  $\pi_{12}:=(\pi_1\times\pi_2)\compose\Delta$ and
  $\pi_{21}:=(\pi_2\times\pi_1)\compose \Delta$ are equivalent, then
  there exists an intertwiner $\cs{R}_{12}:V_1\tensor V_2\mapto
  V_2\tensor V_1$:
  \begin{displaymath}
    \cs{R}_{12}\compose(\pi_1\tensor\pi_2)\compose\Delta(x)=
    (\pi_2\tensor\pi_1)\compose \cs{R}_{12}\compose\Delta(x)\quad 
    x\in \alg{A}.
  \end{displaymath}
\end{definition}

\begin{lemma}
  Let $\structure{\alg{A},\cdot,\Delta;k}$ be a $k$-bialgebra and
  $\structure{\pi_i,V_i}$ $(i\in\set{1,2,3})$ be representations of
  the $k$-algebra $\structure{\alg{A},\cdot}$.

  If the following conditions are satisfied
  \begin{enumerate}
  \item there exist intertwiners $\cs{R}_{ij}:V_i\tensor V_j\mapto
    V_j\tensor V_i$ (for all pairs $(i,j)$ such that $i\neq j$) from the
    tensor product representation $\structure{\pi_{ij},V_i\tensor V_j}$
    to $\structure{\pi_{ji},V_j\tensor V_i}$ and
  \item the representations $\structure{\pi_{ijk},V_i\tensor V_j\tensor
      V_k}$ ($\pi_{ijk}:= (\pi_i\times\pi_j\times\pi_k)\compose
    (\Delta\times\id)\compose \Delta:\alg{A}\mapto \End(V_i)\tensor
    \End(V_j)\tensor \End(V_k)$) are irreducible,
  \end{enumerate}
  then two intertwiners from the representation
  $\structure{\pi_{123},V_1\tensor V_2 \tensor V_3}$ to
  $\structure{\pi_{321},V_3\tensor V_2\tensor V_1}$ can be constructed
  from the R-matrices $\cs{R}_{ij}$ and their equivalence (up to a
  scalar $\rho\in k$) is expressed as
  \begin{displaymath}
    (\cs{R}_{23}\times\id)\compose (\id\times\cs{R}_{13})\compose
    (\cs{R}_{12}\times\id) =
    \rho(\id\times\cs{R}_{12})\compose (\cs{R}_{13}\times\id)\compose
    (\id\times\cs{R}_{23}).
  \end{displaymath}
\end{lemma}

This is called the {\em quantum Yang-Baxter Equation without spectral
  parameter\/}. The lemma is proved for example
in~\cite[2.1]{jimbo:review}.

\part{Finite Quantum Groups}

\chapter{Matrix Quantum Groups and Quantum Spaces}
\label{chap:matrix-quantum-groups}

In classical differential geometry it is the commutative $k$-algebra
$C^\infty (M,k)$ of smooth $k$-valued functions on a smooth manifold
$M$ that is of central importance.  $C^\infty (M,k)$ contains in
particular the functions, whose restriction to local open sets in $M$,
gives $M$ a local coordinate structure.  In noncommutative
differential geometry, noncommutative algebras take the role
that $C^\infty (M,k)$ had classically.

\section{The quantum groups of $\GL_2$ and $\SL_2$}

I start with two simple examples of matrix quantum groups: the quantum
groups corresponding to the Lie groups $\GL_2$ and $\SL_2$.

\subsection{}
Recall that the Lie group $\GL_2(\C)$ (from here on denoted $\GL_2$)
of complex $2\times2$ matrices with nonzero determinant can also be defined
in terms of its natural coordinates $\set{\bar{a}, \bar{b}, \bar{c},
  \bar{d}}$, whose values on a matrix $g\in\GL_2$ are the entries in
the matrix:
\begin{displaymath}
  g=\twomatrix{\bar{a}(g)}{\bar{b}(g)}{\bar{c}(g)}{\bar{d}(g)}.
\end{displaymath}
Of course the algebra $\fun(\GL_2,\C)$ of polynomial functions (in
$\set{\bar{a}, \bar{b}, \bar{c}, \bar{d}}$) is commutative: the
coordinate functions commute. They also satisfy the relation
$\bar{a}(g)\bar{d}(g)- \bar{b}(g)\bar{d}(g)\neq 0$ for all $g\in
\GL_2$.

The Lie group $\SL_2$ is defined to be the group formed by the subset
$\set{g\in \GL_2\mid \det(g)=1}$ of $\GL_2$, the elements of $\GL_2$
with unit determinant. Since $\SL_2$ is a subgroup of $\GL_2$, I
define $\fun(\SL_2,\C)$ to be the restriction of $\fun(\GL_2,\C)$ to
$\SL_2$: $\fun(\SL_2,\C):= \fun(\GL_2,\C)_{|\SL_2}$.  I use the same
symbols $\set{\bar{a}, \bar{b}, \bar{c}, \bar{d}}$ to denote the
commutative coordinate functions of $\GL_2$ when restricted to
$\SL_2$. They satisfy the relation $\bar{a}(g)\bar{d}(g)-
\bar{b}(g)\bar{d}(g)= 1$ for all $g\in \SL_2$. From here on, write
$\fun(\GL_2)$ and $\fun(\SL_2)$ respectively for $\fun(\GL_2,\C)$ and
$\fun(\SL_2,\C)$.

\begin{notation}
  Let $\cs{A}$ and $\cs{B}$ be associative $k$-algebra. Consider then
  two $n\times n$ matrices $U=(u_{ij})_{i,j\in\range{1,n}}$ and
  $V=(v_{ij})_{i,j\in\range{1,n}}$, with elements $u_{ij}\in \cs{A},
  v_{ij}\in \cs{B}$. Then I define the {\em matrix tensor product\/}
  $U\dottensor V$ of $U$ and $V$ to be the $n\times n$ matrix with
  entries
  \begin{displaymath}
    (U \dottensor V)_{ik}:= \sum_{j=1}^n u_{ij}\tensor v_{jk}
    \quad \in \cs{A}\tensor \cs{B}.
  \end{displaymath}
\end{notation}

\begin{proposition}
  The polynomial function algebra $\fun(\SL_2)$ is a commutative,
  noncocommutative Hopf algebra.

  The coproduct $\Delta:\fun(\SL_2)\mapto \fun(\SL_2)\tensor
  \fun(\SL_2)$ is given by
  \begin{displaymath}
    \Delta:\twomatrix{\bar{a}}{\bar{b}}{\bar{c}}{\bar{d}} \mapsto
    \twomatrix{\bar{a}}{\bar{b}}{\bar{c}}{\bar{d}}\dottensor
    \twomatrix{\bar{a}}{\bar{b}}{\bar{c}}{\bar{d}},\qquad
    \Delta:\unity\mapsto \unity\tensor \unity.
  \end{displaymath}

  The antipode map $S:\fun(\SL_2)\mapto \fun(\SL_2)$ is
  \begin{displaymath}
    S:\twomatrix{\bar{a}}{\bar{b}}{\bar{c}}{\bar{d}}\mapsto
    \twomatrix{\bar{d}}{-\bar{b}}{-\bar{c}}{\bar{a}},\qquad
    S:\unity\mapsto \unity \in \fun(\SL_2).
  \end{displaymath}

  The counit $\epsilon:\fun(\SL_2)\mapto \Cfree{q,q\inv}$ is
  \begin{displaymath}
    \epsilon:\twomatrix{\bar{a}}{\bar{b}}{\bar{c}}{\bar{d}} \mapsto
    \twomatrix{1}{0}{0}{1},\qquad
    \epsilon:\unity\mapsto 1\in \C.
  \end{displaymath}
  The coproduct and counit extend homomorphically and the
  antipode antihomomorphically to the whole of $\fun(\SL_2)$.
\end{proposition}

\begin{remark}
  In order to extend this Hopf algebra structure to $\fun(\GL_2)$ it
  is necessary to `complete' it by adding an element $\bar{D}\inv$ to
  it. $\bar{D}\inv$ is the essentially the reciprocal of the
  determinant function of $\fun(\GL_2)$, which cannot be expressed as
  a polynomial of the coordinate functions $\set{\bar{a}, \bar{b},
    \bar{c}, \bar{d}}$.
\end{remark}

\begin{definition}[$\funq(\GL_2)$] \label{qgps:funq-GLtwo}
  Let $q$ be an indeterminant with inverse $q\inv$.  The quantum
  matrix group (or quantised polynomial function algebra)
  $\funq(\GL_2)$ is defined to be the associative unital
  $\Cfree{q,q\inv}$-algebra with generators $\set{a,b,c,d,D\inv}$
  which satisfy the relations
  \begin{alignat*}{2}
    \text{$D\inv$ is central},\\
    a\cdot b &= q\; b\cdot a, \qquad\qquad
    & a\cdot c &= q\; c\cdot a,\\ 
    b\cdot d &= q\; d\cdot b,
    & c\cdot d &= q\; d\cdot c,\\ 
    b\cdot c &= c\cdot b,
    & a\cdot d- d\cdot a &= (q- q\inv)\;b\cdot c,\\
    D\inv \cdot (ad-qbc) &=1,
    & (ad-qbc)\cdot D\inv &=1.
  \end{alignat*}

  So $\funq(\GL_2):=\Cfree{q,q\inv}\freeNC{a,b,c,d,D\inv}/ \alg{I}_q$,
  where $\alg{I}_q$ is the two-sided ideal generated by the above
  relations.
\end{definition}

\begin{remark}
  $\funq(\GL_2)=\sum_{k,l,m,n,p\in\N} \Cfree{q,q\inv}a^k b^l c^m d^n
  D^{-n}$ as a $\Cfree{q,q\inv}$-module (though this basis is
  overcomplete).
\end{remark}

\subsection{}
It is often useful to do calculations in $\funq(\GL_2)$ using the
following matrix of the generators $\set{a,b,c,d}$
\begin{displaymath}
  T:=\twomatrix{a}{b}{c}{d}.
\end{displaymath}

\begin{lemma}
  The element $\det_q(T):=ad- qbc\equiv da- q\inv bc$ is central in
  $\funq(\GL_2)$. It is called the quantum determinant of
  $\funq(\GL_2)$.
\end{lemma}

The generator $D\inv$ can be thought of as the inverse of the quantum
determinant.

\begin{proposition}
  The quantum group function algebra $\funq(\GL_2)$ has the structure
  of a noncommutative, noncocommutative Hopf algebra. 

  The coproduct $\Delta:\funq(\GL_2)\mapto
  \funq(\GL_2)\tensor \funq(\GL_2)$ is given by
  \begin{displaymath}
    \Delta:\twomatrix{a}{b}{c}{d} \mapsto
    \twomatrix{a}{b}{c}{d}\dottensor \twomatrix{a}{b}{c}{d}, \qquad
    \Delta:D\inv \mapsto D\inv \tensor D\inv.
  \end{displaymath}

  The antipode map $S:\funq(\GL_2)\mapto \funq(\GL_2)$ is
  \begin{displaymath}
    S:\twomatrix{a}{b}{c}{d}\mapsto D\inv \twomatrix{d}{-q\inv b}{-q
      c}{a}, \qquad
    S: D\inv \mapsto ad - qbc.
  \end{displaymath}

  The counit $\epsilon:\funq(\GL_2)\mapto \Cfree{q,q\inv}$ is
  \begin{displaymath}
    \epsilon:\twomatrix{a}{b}{c}{d} \mapsto \twomatrix{1}{0}{0}{1}, \qquad
    \epsilon: D\inv \mapsto 1.
  \end{displaymath}
  The coproduct and counit extend homomorphically and the antipode
  antihomomorphically to the whole of $\funq(\GL_2)$.  The Hopf
  algebra maps act on the unity as $\Delta(1)= 1\tensor 1$,
  $\epsilon(1)=1\in \Cfree{q,q\inv}$ and $S(1)=1\in \funq(\GL_2)$.

  Note also the matrix products: $T\cdot S(T)=1$ and $S(T)\cdot T=1$.
\end{proposition}

\begin{proof}
  The proposition is easily proved by checking that all the maps are
  consistent and satisfy the relations required of a Hopf algebra.
\end{proof}

\subsection{The quantum group of $\SL_2$} \label{qgps:funq(SL2)-def}
\cite{faddeev-takhtajan:lattice-liouville} The quantum group
$\funq(\SL_2)$ is defined to be the quotient algebra of $\funq(\GL_2)$
generated by only $\set{a,b,c,d}$ with unit quantum determinant
($\det_q(T)=1$) satisfying the relations of $\funq(\GL_2)$
(see~\ref{qgps:funq-GLtwo}). So
\begin{displaymath}
  \funq(\SL_2):= \funq(\GL_2)/\ideal{D\inv- 1,\; ad-qbc- 1}.
\end{displaymath}
So the quantum group $\funq(\SL_2)$ is a quotient algebra of the
quantum group $\funq(\GL_2)$.

\begin{proposition} \label{qgps:left-&-right-actions}
  Let $\set{a,b,c,d}$ and $\set{a',b',c',d'}$ be two sets of
  noncommutative coordinates of $\funq(\SL_2)$, both satisfying the
  relations of~\ref{qgps:funq-GLtwo}.  Let
  \begin{displaymath}
    T:=\twomatrix{a}{b}{c}{d} \quad \text{and}\quad
    T':=\twomatrix{a'}{b'}{c'}{d'}.
  \end{displaymath}
  Consider the maps $L,R:\funq(\SL_2)\mapto \funq(\SL_2)\tensor
  \funq(\SL_2)$ given by the matrix tensor products $L:T\mapsto T
  \dottensor T'$ and $R:T\mapsto T'\dottensor T$
  \begin{align*}
    L: \twomatrix{a}{b}{c}{d} &\mapsto \twomatrix{a}{b}{c}{d}
    \dottensor \twomatrix{a'}{b'}{c'}{d'},\\ 
    R: \twomatrix{a}{b}{c}{d} &\mapsto
    \twomatrix{a'}{b'}{c'}{d'}\dottensor \twomatrix{a}{b}{c}{d}.
  \end{align*}
  The maps $L$ and $R$ are $\Cfree{q,q\inv}$-algebra homomorphisms.
\end{proposition}

A special case of this result, is the coproduct map $\Delta:T \mapsto
T \dottensor T$, which is a homomorphism $\funq(\SL_2)\mapto
\funq(\SL_2)\tensor \funq(\SL_2)$.

\subsection{Classical limit}
Let $\e\in \Ccross$. I define the specialisation $\fune(\SL_2)$ of
$\funq(\SL_2)$ at $q=\e$, to be $\fune(\SL_2):=\funq(\SL_2)/
\ideal{q-\e}$.

A particularly important specialisation of $\funq(\SL_2)$ is
$\fun_1(\SL_2)$ at $q=1$.
\begin{lemma}
  The $\C$-algebra $\fun_1(\SL_2)$ is a commutative algebra. It is
  Hopf algebra isomorphic to $\fun(\SL_2)$.
\end{lemma}

\begin{proof}
  It is clear from the definition of $\fune(\SL_2)$ that at $\e=1$,
  $\fune(\SL_2)$ becomes a commutative algebra. Further $ad-bc=1$.
  This proves the algebra isomorphism $\fun_1(\SL_2)\isomorphic
  \fun(\SL_2)$. The Hopf algebra isomorphism follows similarly.
\end{proof}

\section{The Quantum Plane}

Lie groups occur often as the transformation or symmetry groups of a
vector space (perhaps with certain additional structures (for example
inner products or a symplectic structure), which may also be invariant
under the group action). It is interesting to consider what the
analogue of an action on a linear space is for a matrix quantum group.

\subsection{} Consider an $n$-dimensional $\C$-vector space $V$.
Let $V^*$ be the vector space dual to $V$.  A (dual) basis in $V^*$
forms a set of linear coordinates on $V$. Let $\set{\bar{x}_i\mid
  i=1,\dots,n}$ be a basis of $V^*$. The $\C$-span of all monomials in
these coordinates has a natural structure of a commutative algebra
$\fun(V,\C)$ ($\fun(V)$) (the polynomial function algebra on $V$) and
is isomorphic to the symmetric algebra $S(V^*)$ over $V^*$.

\begin{definition}~\cite{manin:cmp}
  Let $q$ be an indeterminant with inverse $q\inv$. The {\em quantum
    vector space\/} (or quantum function space) $\funq(V)$ of
  dimension $n$ is an associative $\Cfree{q,q\inv}$-algebra with unity
  and $n$ generators $\set{x_1,\dots,x_n}$ which satisfy the
  relations:
  \begin{displaymath}
    x_i \cdot x_j = q x_j \cdot x_i \qquad (1\leq i<j \leq n).
  \end{displaymath}
  Then $\funq(V):= \Cfree{q,q\inv}\freeNC{x_1,\dots,x_n}/ \cs{J}_q$,
  where $\cs{J}_q$ is the two-sided ideal generated by the above
  relations.
\end{definition}

\section{Coaction of $\funq(\SL_2)$ on the quantum plane
  $\funq(V_2)$}

Next I consider how a quantum group function algebra co-acts on a
quantum vector space. Since the objects are defined in terms of the
dual (noncommutative) function algebras, the arrows in the maps are
reversed relative to the usual maps of a (Lie) group acting on a
vector space.

\begin{definition}[coaction]
  Let $k$ be a field. A left coaction of a $k$-bialgebra
  $\cs{B}$ (with counit $\epsilon$) on a $k$-algebra $\cs{V}$, is a
  $k$-algebra homomorphism $\Delta_L:\cs{V}\mapto \cs{B}\tensor
  \cs{V}$, which satisfies
  \begin{alignat*}{2}
    (\id \times \Delta_L)\compose \Delta_L &= (\Delta \times
    \id)\compose \Delta_L \quad &
    (\cs{V} &\mapsto \cs{B}\tensor\cs{B}\tensor \cs{V})\\
    (\epsilon\times \id)\compose \Delta_L &= i_k \qquad &
    (\cs{V} &\mapto k \tensor \cs{V}).
  \end{alignat*}
  Here $i_k$ is the isomorphism
  $\cs{V}\overset{\isomorphic}{\longrightarrow} k\tensor \cs{V}$.
  When the bialgebra $\cs{B}$ co-acts on $\cs{V}$, then $\cs{V}$ is
  called a left $\cs{B}$-comodule (or a left corepresentation) of
  $\cs{B}$. A right coaction ($\cs{V}\mapto \cs{V}\tensor \cs{B}$) and
  right comodule are similarly defined (\cf~\ref{prel:corepresentation}).
\end{definition}

Consider now a 2 dimensional $\C$-vector space $V_2$ and the quantum
vector space $\funq(V_2)$. Call $\funq(V_2)$ the quantum plane.

\begin{lemma}
  Write $\set{x,y}$ $(xy=qyx)$ for the generators $\set{x_1,x_2}$ of
  the quantum plane $\funq(V_2)$.
  The quantum plane $\funq(V_2)$ is a $\funq(\SL_2)$-comodule. The
  left coaction $\Delta_L:\funq(V_2)\mapto \funq(\SL_2)\tensor
  \funq(V_2)$ is given by
  \begin{displaymath}
    \Delta_L:
    \begin{pmatrix}
      x \\ y
    \end{pmatrix}
    \mapsto \twomatrix{a}{b}{c}{d} \dottensor
    \begin{pmatrix}
      x \\ y
    \end{pmatrix}.
  \end{displaymath}
\end{lemma}

\begin{proof}
  I check that $\Delta_L(x\cdot y)= q\Delta_L(y\cdot x)$.
  \begin{align*}
    \Delta_L(x y) &= \Delta_L(x)\cdot \Delta_L(y)\\
    &= (a\tensor x+ b\tensor y) \cdot (c\tensor x+ d\tensor y)\\
    &= ac\tensor xx+ ad\tensor xy+ bc\tensor yx + bd\tensor yy\\
    &= q(ca\tensor xx)+ q(da+ (q- q\inv)bc)\tensor yx+ bc\tensor yx+
    q(db\tensor yy)\\
    &= q(ca\tensor xx+ da\tensor yx+ bc\tensor xy+ db\tensor yy)\\
    &= q(c\tensor x+ d\tensor y)\cdot (a\tensor x+ b\tensor y)\\
    &= q \Delta_L(y) \cdot \Delta_L(x).
  \end{align*}
\end{proof}

\subsection{}
Let $V$ be a $\C$-vector space and let $\e\in \Ccross$. Define the
specialisation $\fune(V)$ of $\funq(V)$ at $q=\e$ to be the $\C$-algebra
$\fune(V):= \funq(V)/ \ideal{q-\e}$.

\begin{lemma}
  $\fun_1(V)\isomorphic \fun(V)$ as $\C$-algebras.
\end{lemma}

\section{Representations of Quantum groups}


\begin{notation}
  In a bialgebra, let the notation
  \begin{displaymath}
    \Delta(x)= \sum_i x_{(1,i)}\tensor x_{(2,i)},
  \end{displaymath}
  denote the components of the coproduct.
\end{notation}

Recall that in a Lie group $G$ there are some natural actions of the
group on itself. There is the left action $l_g:g'\mapsto gg'$ and the
right action $r_g:g'\mapsto g'g$ and the adjoint action
$\Ad_g:g'\mapsto gg'g\inv$. The adjoint map $\Ad_g$ acts as a
automorphism of $G$ for each $g\in G$. For a quantum group the
equivalent of the left and right actions is essentially given by the
coproduct (see~\ref{qgps:left-&-right-actions}).


\begin{definition}[adjoint action]
  The {\em quantum adjoint action\/} of a Hopf algebra $\cs{H}$ on
  itself is defined by
  \begin{displaymath}
    \begin{aligned}
      \Ad(x): \cs{H} &\mapto \cs{H},\\
      \Ad(x):y &\mapsto \sum_i x_{(1,i)} y S(x_{(2,i)}),
    \end{aligned}
    \qquad (x,y\in \cs{H}).
  \end{displaymath}
\end{definition}

The map $\Ad(x):\funq(\SL_2)\mapto \funq(\SL_2)$ ($x\in \funq(\SL_2)$)
gives the adjoint representation of $\funq(\SL_2)$ on itself.


\section{$\fune(\SLtwo)$ and $\fune(V_2)$ at a root of unity}

\begin{proposition}
  Let $\l$ be an integer such that $\l>2$. Let $\e=e^{\frac{2\pi
      i}{\l}}$ (a primitive $\l$-th root of unity).

  (a) In $\fune(V)$, the elements $\set{x_i^\l\mid i\in\range{1,n}}$
  are central.

  (b) In $\fune(\SL_2)$, the elements $\set{a^\l, b^\l, c^\l, d^\l}$ are
  central.
\end{proposition}

\begin{proof}
  (a) In $\fune(V_2)$, $x^\l y= \e^\l y x^\l= y x^\l$, so $x^\l$ is
  central. Similarly $y^\l$ is central.
  (b) That $b^\l$ and $c^\l$ are central, follows from (a). To show that
  $a^\l$ is central it is necessary to check only that $a^\l d= d a^\l$,
  since it clearly commutes with $b$ and $c$. Using the relation $ad-da=(q-
  q\inv)bc$, the following identity is proved by induction
  \begin{displaymath}
    a^\l\cdot d = d\cdot a^\l + (\e- \e\inv) \e^{\l-1}\brackets{\l}_\e
    b\cdot c\cdot a^{\l-1}.
  \end{displaymath}
  Since $\brackets{\l}_\e=0$, $a^\l$ is central in $\fune(\SL_2)$. The
  proof that $d^\l$ is central is similar.
\end{proof}

\begin{corollary}
  (a) Let $\Zcentre_V$ be the central subalgebra of $\fune(V)$
  generated by $\set{x_i^\l\mid i\in\range{1,n}}$.  $\fune(V)$ is
  finite dimensional over $\Zcentre_V$, \ie $\fune(V)$ is a free
  module over $\Zcentre_V$ with basis $x_1^{m_1} x_2^{m_2}\cdots
  x_n^{m_n}$ ($m_i\in \range{0,\l-1}$, $i\in\range{1,n}$).

  (b) Let $\Zcentre_{\SL_2}$ be the central subalgebra of
  $\fune(\SL_2)$ generated by $\set{a^\l, b^\l, c^\l, d^\l}$.  Then
  $\fune(\SL_2)$ is finite dimensional over $\Zcentre_{\SL_2}$:
  $\fune(\SL_2)$ is a free module over $\Zcentre_{\SL_2}$ with basis
  $a^{m_1} b^{m_2} c^{m_3} d^{m_4}$ ($m_1,m_2,m_3,m_4\in
  \range{0,\l-1}$).
\end{corollary}

\begin{remark}
  Representations of $\fune(\SL_2)$ at a root of unity have been
  studied in~\cite{kondratowicz-podles:SL2-root-of-1-reps}. The case
  of a general quantum matrix group at a root of unity is described
  in~\cite{deconcini-lyubashenko:quantum-function-alg-root-1}.
\end{remark}








\begin{lemma}
  The map $\omega:\funq(\SL_2)\mapto \funq(\SL_2)$ given by
  \begin{displaymath}
    \omega:T \mapsto S(T)^t, \qquad \omega:q\mapsto q,
  \end{displaymath}
  is an anti-automorphism of $\SL_2$.
\end{lemma}

\begin{proof}
  It is easily checked that $\omega$ is an anti-automorphism of
  $\funq(\SL_2)$. Here I check that
  \begin{align*}
    \omega(ab) - q\omega(ba)&= \omega(b) \omega(a) -q\omega(a)\omega(b)\\
    &= -q\inv( cd- q dc)\\
    &= 0,
  \end{align*}
  and
  \begin{align*}
    \omega(ad-da -(q- q\inv)bc) &= ad-da- (q- q\inv)bc\\
    &= 0.
  \end{align*}
  The rest of the relations of $\funq(\SL_2)$ are similarly
  shown to map into each other under $\omega$.
\end{proof}


\section{The 2-parameter Quantum Group $\funpq(\GL_2)$}

\subsection{}
Let $p,q$ be indeterminants with inverses $p\inv,q\inv$. Let
$\Cfree{p,p\inv,q,q\inv}$ be the ring of polynomials in $p$, $q$ and
their inverses. Consider the associative
$\Cfree{p,p\inv,q,q\inv}$-algebra $\funpq(\GL_2)$ generated by
$\set{a, b, c, d, D\inv}$ satisfying the relations
\begin{alignat*}{2}
  a\cdot b &= p\; b\cdot a, \qquad\qquad
  & a\cdot c &= q\; c\cdot a,\\ 
  b\cdot d &= q\; d\cdot b,
  & c\cdot d &= p\; d\cdot c,\\ 
  p\;b\cdot c &= q\; c\cdot b,
  & a\cdot d- d\cdot a &= (p- q\inv)\;b\cdot c,\\ 
  D\inv\cdot a &= a\cdot D\inv,
  & p\; D\inv\cdot b &= q\; b\cdot D\inv,\\
  q\; D\inv\cdot c &= p\; c\cdot D\inv,
  & D\inv\cdot d &= d\cdot D\inv,\\
  D\inv\cdot (a\cdot d- q\; c\cdot b) &= 1.
\end{alignat*}
This is a 2-parameter deformation of $\fun(\GL_2)$. Note that $ad- q
cb$ is not central in $\funpq(\GL_2)$, so there is no simple way of
defining a 2-parameter deformation of $\fun(\SL_2)$ as a quotient of
this algebra.

\subsection{}
$\funpq(\GL_2)$ is a Hopf algebra. The coproduct
$\Delta:\funpq(\GL_2)\mapto \funpq(\GL_2)\tensor \funpq(\GL_2)$ is
given by
\begin{displaymath}
  \Delta:\twomatrix{a}{b}{c}{d} \mapsto
  \twomatrix{a}{b}{c}{d}\dottensor \twomatrix{a}{b}{c}{d}, \qquad
  \Delta:D\inv \mapsto D\inv \tensor D\inv.
\end{displaymath}

The antipode map $S:\funpq(\GL_2)\mapto \funpq(\GL_2)$ is
\begin{displaymath}
  S:\twomatrix{a}{b}{c}{d}\mapsto D\inv \twomatrix{d}{-q\inv b}{-q
    c}{a}, \qquad
  S: D\inv \mapsto ad - pbc.
\end{displaymath}

The counit $\epsilon:\funpq(\GL_2)\mapto \Cfree{q,q\inv}$ is
\begin{displaymath}
  \epsilon:\twomatrix{a}{b}{c}{d} \mapsto \twomatrix{1}{0}{0}{1},
  \qquad
  \epsilon: D\inv \mapsto 1.
\end{displaymath}
The Hopf algebra maps act on the unity as $\Delta(1)= 1\tensor 1$,
$\epsilon(1)=1\in \Cfree{p,p\inv,q,q\inv}$ and $S(1)=1\in \funpq(\GL_2)$.

\begin{remark}
  There is a Hopf algebra isomorphism
  $\funpq(\GL_2)/\ideal{p-q}\isomorphic \funq(\GL_2)$.
\end{remark}

\section{Quantum phase space}
The algebra $Q_{\hbar}:=\CfreeNC{x,p}/\ideal{xp-px-\hbar i}$ can
heuristically be thought of as the general `quantum mechanical phase
space of a particle in one dimension' (with `noncommutative phase
space coordinates' $x$ and $p$) and
$Q_{q,\hbar}:=\CfreeNC{x,p}/\ideal{xp-qpx-\hbar i}$ as a `deformed
quantum phase space'. 1-parameter deformations of bosonic and
fermionic quantum mechanical phase space and their symmetries have
been studied by Zumino~\cite{zumino:mpla} in the $\cs{R}$-matrix
formalism.

\chapter{Quantum enveloping algebras}
\label{chap:quantum-enveloping-algebras}

\section{Introduction}

In 1985 Jimbo and Drinfeld independently
introduced~\cite{drinfeld:hopf-ybe,jimbo:lmp1} a $q$-analogue of the
universal enveloping algebra of every simple (and affine) Lie
algebra.
In this chapter I describe some results, due mainly to Lusztig and to
De~Concini and Kac, on the quantum enveloping algebra of a simple
(finite) Lie algebra and its `specialisation' at an odd primitive root
of unity. The techniques and results described here will be assumed in
chapters~\ref{chap:kac-moody} and~\ref{chap:qkm-root-of-1}, when I
come to consider quantum affine (Kac-Moody) algebras and their
specialisation at a root of unity.  Let me summarise the ideas and
contents of this chapter.

\subsection{}
In the spirit of Kac, to every symmetrisable, positive definite Cartan
matrix $(a_{ij})$, there exists an associated unique simple finite
dimensional Lie algebra $\g$ over $\C$ and its universal enveloping
algebra $\U(\g)$. Also associated to $(a_{ij})$, there is a set of
fundamental weights, simple roots, a weight lattice, a root lattice,
a braid group, a Weyl group and so on. The Weyl group acts naturally on
the root lattice and induces a corresponding automorphic action of the
braid group on $\g$ and $\U(\g)$. In particular the Weyl group action
on the simple roots, which generates the complete root system of $\g$,
lifts to an action of the braid group on the Chevalley generators.
This action generates all the root vectors of $\g$. To each reduced
expression of the longest element of the Weyl group, there is an
associated basis of the root vectors of $\g$. Finally, as I mentioned
in~\ref{prel:Hopf-U(g)}, $\U(\g)$ is a cocommutative Hopf algebra.

In the quantum case, Drinfeld and Jimbo have associated to each
$(a_{ij})$ a quantum enveloping algebra $\Uq(\g)$ with relations in
terms of a quantum Chevalley presentation. In this chapter I
concentrate on the case when the Cartan matrix is of finite type,
\ie~that of a simple finite dimensional Lie algebra $\g$. (The quantum
groups associated to affine Cartan matrices are considered in
part~III.)  The quantum enveloping algebra $\Uq(\g)$ is a
noncocommutative Hopf algebra. At the specialisation $q=1$, it can be
shown rigorously that a certain subalgebra of $\Uq(\g)$ becomes
isomorphic (modulo certain central elements) to the classical
universal enveloping algebra $\U(\g)$. In the quantum case the problem
of finding a basis of root vectors is a rather more difficult problem
than for a classical Lie algebra. This is basically because the Hopf
algebra is noncocommutative (or equivalently the quantum adjoint
action acts essentially like a $q$-deformed commutator). This problem
was solved by Lusztig, who proved that the Weyl group action on simple
roots can also be lifted to an automorphic action of the braid group
on $\Uq(\g)$. He further showed that, in analogy with the classical
case, to each reduced expression of the longest element of the Weyl
group there is an associated basis of root vectors of $\Uq(\g)$.

The representation theory of $\Uq(\g)$ has been developed to a quite
mature state (at least in the case of $q$ generic). The fundamental
highest weight modules are the Verma modules over $\Uq(\g)$. All other
highest weight representations generated by a single vector can be
constructed out of the Verma modules and their quotients.

\subsection{}
A remarkable facet of quantum group theory is the case when $q$ is a
primitive $\l$-th root of unity $\e$. There exists a specialisation
$\Ue(\g)$ of $\Uq(\g)$ at $q=\e$. In this case the centre of $\Ue(\g)$
is much enlarged. The large centre is most easily constructed as
follows. First one proves that the Chevalley generators to the $\l$-th
power lie in the centre of $\Ue(\g)$. Then by successively applying
the braid group generators to these central elements, all the new
central elements are found.  Lusztig, De Concini, Kac and Procesi
think of the root of unity case as a $q$-analogue of a simple finite
Lie algebra over a field of positive characteristic $\l$.

With this information it is possible to study the representations of
$\Ue(\g)$ at a root of unity. Because of the large centre it turns out
that every Verma module over $\Ue(\g)$ is reducible, and has a natural
reduction to a (finite dimensional) `diagonal module'.  More
interestingly at a root of unity there exist finite dimensional
modules over $\Ue(\g)$ that are not quotients of a Verma module: these
include `cyclic' (`periodic') modules and `semicyclic'
(`semiperiodic') modules, where all or some of the Chevalley
generators act injectively. Unfortunately I do not have time or
space to describe the parametrisation of the irreducible finite
dimensional representations of $\Ue(\g)$ at a root of unity, due to De
Concini, Kac and Procesi.


The case of $\Uq(\sltwo)$ is the simplest. In this case it is rather
straightforward to write down all the irreducible finite dimensional
representation both at $q$ generic and at $q$ a root of unity.

\section{The Lie algebra $\g$}

\subsection{}
Let $(a_{ij})_{i,j\in \range{1,r}}$ be an $r\times r$ matrix with
integer entries, such that the diagonal elements $a_{ii}=2$, the
off-diagonal elements $a_{ij}\in \set{0,-1,-2,-3}$ ($i\neq j$) and
there exist $r$ positive coprime integers $d_i$ ($i\in\range{1,r}$)
such that $(d_i a_{ij})$ is a symmetric positive definite matrix. Then
$(a_{ij})$ is the Cartan matrix of a simple finite Lie algebra $\g$ of
rank $r$.

\begin{notation}
  As usual I define in a Lie algebra the adjoint map to be
  $\ad(x)y:=\comm{x,y}$.
\end{notation}

\subsection{} \label{quea:g-assoc-to-aij}
The Lie algebra $\g$ associated to the Cartan matrix $(a_{ij})$ has the
following relations over $\C$ among its Chevalley presentation generators
$\set{\bar{h}_i, \bar{e}_i, \bar{f}_i\mid i\in\range{1,r}}$:
\begin{alignat*}{2}
  \comm{\bar{h}_i,\bar{h}_j}&=0, \qquad&
  \comm{\bar{e}_i,\bar{f}_j} &= \delta_{ij} \bar{h}_i,\\ 
  \comm{\bar{h}_i,\bar{e}_j} &= a_{ij} \bar{e}_j, &
  \comm{\bar{h}_i,\bar{f}_j} &= -a_{ij} \bar{f}_j,\\ 
  \ad(\bar{e}_i)^{1-a_{ij}} (\bar{e}_j) &= 0, &
  \ad(\bar{f}_i)^{1-a_{ij}} (\bar{f}_j) &= 0.
\end{alignat*}
The last two relations are called the Serre-Chevalley relations (or
Serre relations for short). Let $\U(\g)$ be the universal enveloping
algebra of $\g$, as defined in~\ref{prel:U(g)-def}.

\section{Lattices, Braid group and Weyl group}
\label{quea:lattices}

\subsection{Weight lattice}
Let $P$ be a free abelian group over $\Z$ (or $\Z$-lattice) with generators
$\set{\omega_i\mid i\in\range{1,r}}$
\begin{displaymath}
  P:=\sum_{i\in\range{1,r}} \Z \omega_i.
\end{displaymath}
$P$ is called the weight lattice of $\g$.

Let $Q\czek:=\Hom(P,\Z)$ be the dual lattice to $P$ with a dual basis
$\Pi\czek:=\set{\alpha\czek_i\mid i\in \range{1,r}}$, so that under the
dual pairing $\pairing{\cdot, \cdot}:P\times Q\czek\mapto \Z$
\begin{displaymath}
  \pairing{\omega_i, \alpha\czek_j}= \delta_{ij}.
\end{displaymath}
$Q\czek$ is called the coroot lattice of $\g$.

\subsection{Root lattice}
Define the elements 
\begin{displaymath}
  \alpha_i:=\sum_{j\in \range{1,r}} a_{ij} \omega_j \; \in P.
\end{displaymath}
They generate a free abelian subgroup (sub-lattice) of $P$:
\begin{displaymath}
  Q:=\sum_{i\in\range{1,r}} \Z \alpha_i \subset P,
\end{displaymath}
called the root lattice of $\g$. The pairing on $P\times Q\czek$
restricts to $Q\times Q\czek$ as
\begin{displaymath}
  \pairing{\alpha_i,\alpha\czek_j}=a_{ij}.
\end{displaymath}

Let $P_+:=\sum_{i\in \range{1,r}} \N \omega_i$ (the positive weights)
and $Q_+:=\sum_{i\in \range{1,r}} \N \alpha_i$. Define a partial
ordering $\geq$ in $P$ by $\lambda\geq \mu$, if $\lambda- \mu\in P_+$.


\subsection{Symmetric form}
Define a bilinear map $(\cdot,\cdot): P\times Q \mapto \Z$ by
\begin{displaymath}
  (\omega_i,\alpha_j):= d_i \delta_{ij}.
\end{displaymath}
Note that the restriction of this map to $Q\times Q$ defines
a symmetric $\Z$-valued bilinear form on $Q$, since
\begin{displaymath}
  (\alpha_i,\alpha_j)= d_i a_{ij}.
\end{displaymath}

\subsection{}
Define the following numbers from the Cartan matrix
\begin{displaymath}
  m_{ij}:= \frac{\pi}{\arccos(\frac{\surd(a_{ij}a_{ji})}{2})} \qquad
  (i\neq j).
\end{displaymath}
Then $m_{ij}=2,3,4,6$ ($i\neq j$), when $a_{ij} a_{ji}= 0,1,2,3$
respectively. Note that $m_{ij}=m_{ji}$.

\begin{definition}[braid group]
  The Braid group $B$ associated to the Cartan matrix $(a_{ij})$ is
  generated by $\set{T_i, {T_i}\inv, 1 \mid i\in\range{1,r}}$. They
  satisfy the relations
  \begin{align*}
    T_i {T_i}\inv &= 1= {T_i}\inv T_i,\\
    \underbrace{T_i T_j T_i\cdots}_{\text{$m_{ij}$ factors}} &=
    \underbrace{T_j T_i T_j\cdots}_{\text{$m_{ij}$ factors}} \qquad\quad
    (i\neq j).
  \end{align*}
\end{definition}
The Braid group $B$ is infinite.

\begin{definition}[Weyl group] \label{quea:g-Weyl-group}
  The Weyl group $W$ associated to the Cartan matrix $(a_{ij})$ is the
  quotient of the Braid group $B$ by the normal subgroup generated by
  $T_iT_i$ and ${T_i}\inv{T_i}\inv$ ($i\in\range{1,r}$), 
  corresponding to relations $T_i={T_i}\inv$ ($i\in \range{1,r}$) in $W$:
  \begin{displaymath}
    W:=B/\ideal{T_i^2, {T_i}^{-2}\mid i\in \range{1,r}}.
  \end{displaymath}
  I denote the image of $T_i$ under the canonical map $B\mapto W$ by
  $s_i$. For completeness I write down the relations that these
  generators of $W$ satisfy:
  \begin{align*}
    {s_i}^2 &= 1 \qquad\qquad (i\in\range{1,r}),\\
    \underbrace{s_i s_j s_i\cdots}_{\text{$m_{ij}$ factors}} &=
    \underbrace{s_j s_i s_j\cdots}_{\text{$m_{ij}$ factors}} \qquad
    (i\neq j).
  \end{align*}
\end{definition}
The Weyl group $W$ of the simple Lie algebra $\g$ is finite. $W$ is an
example of a Coxeter group.

\begin{proposition} \label{quea:W-action-on-P}
  Let $W$ be the Weyl group associated to $(a_{ij})$. The following
  action of $W$ on $P$
  \begin{displaymath}
    s_i:x\mapsto x- \pairing{x, \alpha\czek_i}\alpha_i \quad (x\in P),
  \end{displaymath}
  is a group homomorphism $W\mapto \Aut(P)=\GL(P)$, \ie this action
  defines a representation of $W$ on $P$.
\end{proposition}

\begin{proof}
  Since the map $s_i$ is linear in $x$, it is clear that the action is
  automorphic. To prove that this defines a representation of $W$, it
  is necessary to check that the relations of $W$ are satisfied on
  $P$. I show that ${s_i}^2(x)=x$ ($x\in P$).
  \begin{align*}
    s_i^2(x) &= s_i(x- \pairing{x, \alpha\czek_i}\alpha_i)\\
    &= s_i(x)- \pairing{x, \alpha\czek_i} s_i(\alpha_i)\\
    &= x,
  \end{align*}
  since $s_i(\alpha_i)=-\alpha_i$. Hence $s_i$ acts as a reflection in
  $P$. The braid relations of $W$ are similarly checked case by case.
\end{proof}

Note that the root sub-lattice $Q$ of the weight lattice $P$ is
invariant under the action of $W$, \ie $W:Q\mapto Q\subset P$. In
particular $W$ acts on the generators of $Q$ as
\begin{displaymath}
  s_j:\alpha_i\mapsto \alpha_i- \pairing{\alpha_i,\alpha\czek_j}\alpha_j.
\end{displaymath}
Hence the restriction of the action of $W$ to $Q$, gives a
subrepresentation of $W$ on $Q$.

\begin{lemma}
  The following action of $W$ on $Q\czek$ gives a representation of
  $W$ on the coroot lattice.
  \begin{displaymath}
    s_i:\alpha\czek_j \mapsto \alpha\czek_j-
    \pairing{\alpha_i,\alpha\czek_j} \alpha\czek_i,
  \end{displaymath}
\end{lemma}

\begin{proof}
  The lemma is proved analogously to~\ref{quea:W-action-on-P}.
\end{proof}

\subsection{Roots}
Let $\Pi:=\set{\alpha_i\mid i\in\range{1,r}}$, the simple roots of
$\g$.  Define the root system of $\g$ corresponding to $(a_{ij})$ to be
$R:=W\cdot \Pi$ and $R_+:=R\intersect Q_+$ (the positive roots of
$\g$).

\begin{lemma}
  The pairing $\pairing{\cdot,\cdot}:P\times Q\czek\mapto \Z$ is
  $W$-invariant:
  \begin{displaymath}
    \pairing{s_i(x),y\czek}= \pairing{x,s_i(y\czek)} \quad (\forall
    x\in P, y\czek\in Q\czek, i\in\range{1,r}).
  \end{displaymath}
\end{lemma}

\begin{proof}
  Let $x\in P$ and $y\czek\in Q\czek$.
  \begin{align*}
    \pairing{s_i(x),y\czek}
    &= \pairing{x- \pairing{x,\alpha\czek_i}\alpha_i, y\czek}\\ 
    &= \pairing{x,y\czek}- \pairing{x,\alpha\czek_i}
    \pairing{\alpha_i,y\czek}\\ 
    &= \pairing{x,y\czek- \pairing{\alpha_i,y\czek}\alpha\czek_i}\\ 
    &= \pairing{x,s_i(y\czek)}.
  \end{align*}
\end{proof}

\begin{corollary} \label{quea:W-invar-of-pairing}
  \begin{displaymath}
    \pairing{s_i(x),s_i(y\czek)}= \pairing{x,y\czek} \quad (\forall
    x\in P, y\czek\in Q\czek, i\in\range{1,r})
  \end{displaymath}
\end{corollary}

\subsection{Length function of $W$}
Let $w\in W$. The {\em length\/} of $w$ is defined to be the smallest
integer $n\geq0$ such that there exist
$i_1,i_2,\dots,i_n\in\range{1,r}$ with $w=s_{i_1}s_{i_2}\cdots
s_{i_n}$. I write $l(w)=n$ for the length of $w$. A shortest
expression $s_{i_1}s_{i_2}\cdots s_{i_n}$ of $w$ is called a {\em
  reduced expression\/} of $w$.  Note that a reduced expression is not
unique in general. The identity element $1\in W$ has length $l(1)=0$.
The generators of $W$ have length $l(s_i)=1$. There is a unique
longest element of $W$, which is denoted $w_0$.
It has the property that $w_0:R_+\mapto -R_+$ and ${w_0}^2=1$.

\section{The Hopf algebra $\Uq(\g)$}

\begin{notation}
  Denote by $\Cfree{q,q\inv}$ the ring of polynomials in the
  indeterminant $q$ and its inverse $q\inv$. Denote by $\C(q)$ the
  quotient field of $\Cfree{q,q\inv}$. Define $q_i:=q^{d_i}$. In
  $\C(q)$ I introduce the following standard notation
  \begin{alignat*}{2}
    \brackets{m}_q &:= \frac{q^m - q^{-m}}{q - q\inv} &\qquad &(m\in\Z),\\ 
    \brackets{m}! &:= \brackets{m}_q \brackets{m-1}_q \cdots
    \brackets{1}_q && (m>0),\\ 
    \brackets{0}_q! &:= 1, &&\\ 
    \comb{m}{n}_q &:= \frac{\brackets{m}_q!}{\brackets{n}_q!
      \brackets{m-n}_q!} && (m\geq n\geq 0).
  \end{alignat*}
\end{notation}

Let $m\in\N$, such that $m>1$. Note the following identity, which
holds in $\Zfree{q,q\inv}\subset \Cfree{q,q\inv}\subset \C(q)$
\begin{displaymath}
  q^m- q^{-m}= (q- q\inv)(q^{m-1}+ q^{m-3}+\dots + q^{3-m}+ q^{1-m}).
\end{displaymath}
{}From it follows that $\brackets{m}_q, \brackets{m}_q!\in
\Zfree{q,q\inv}$.  It is also well known~\cite{lusztig:reps} that
$\comb{m}{n}_q \in \Zfree{q,q\inv}$.


I come now to the definition of the quantum universal enveloping
algebra of a Lie algebra $\g$.

\begin{definition} \label{quea:Uqg-def}
  Let $\g$ be a simple finite Lie algebra with Cartan matrix $(a_{ij})$
  ($i,j\in \range{1,r}$). The {\em quantum universal enveloping
    algebra\/} $\Uq(\g)$ of $\g$ is the associative unital
  $\C(q)$-algebra with generators $\set{{k_i}^{\pm1}, e_i, f_i\mid i\in
    \range{1,r}}$ satisfying the relations
  \begin{alignat*}{2}
    k_i \cdot {k_i}\inv &= 1 = {k_i}\inv \cdot k_i, \qquad &
    k_i \cdot k_j &= k_j \cdot k_i,\\ 
    k_i \cdot e_j \cdot {k_i}\inv &= q_i^{a_{ij}} e_j, &
    k_i \cdot f_j \cdot {k_i}\inv &= q_i^{a_{ij}} f_j,\\ 
    e_i \cdot f_j - f_j \cdot e_i &= \delta_{ij} \frac{k_i -
      {k_i}\inv}{q_i-q_i\inv}, &&
  \end{alignat*}
  \begin{align*}
    \sum_{n=0}^{1-a_{ij}} (-1)^n \comb{1-a_{ij}}{n}_{q_i} {e_i}^n
    \cdot e_j \cdot {e_i}^{1-a_{ij}-n} &= \; 0 \qquad (i\neq j),\\
    \sum_{n=0}^{1-a_{ij}} (-1)^n \comb{1-a_{ij}}{n}_{q_i} {f_i}^n
    \cdot f_j \cdot {f_i}^{1-a_{ij}-n} &= \; 0 \qquad (i\neq j).
  \end{align*}
  The last two sets of relations are called the quantum Serre
  relations.
\end{definition}

\subsection{}
The quantum enveloping algebra $\Uq(\g)$ can be endowed with the
structure of a Hopf algebra.
\begin{proposition}
  The following maps, when extended homomorphically to the whole of
  $\Uq(\g)$, make $\structure{\Uq(\g),\cdot, \unity, \Delta, \epsilon,
    S; \C(q)}$ a Hopf algebra.

  The coproduct map $\Delta:\Uq(\g)\mapto \Uq(\g)\tensor \Uq(\g)$ is
  \begin{align*}
    \Delta:k_i &\mapsto k_i\tensor k_i,\\
    \Delta:e_i &\mapsto e_i\tensor \unity + k_i\tensor e_i,\\
    \Delta:f_i &\mapsto f_i\tensor {k_i}\inv+ \unity\tensor f_i.
  \end{align*}

  The antipode map $S:\Uq(\g)\mapto \Uq(\g)$ is
  \begin{displaymath}
    S:k_i \mapsto {k_i}\inv,\quad
    S:e_i \mapsto -{k_i}\inv e_i,\quad
    S:f_i \mapsto -f_i k_i.
  \end{displaymath}

  The counit map $\epsilon:\Uq(\g)\mapto \C(q)$ is
  \begin{displaymath}
    \epsilon: k_i \mapsto 1,\quad
    \epsilon: e_i \mapsto 0,\quad
    \epsilon: f_i \mapsto 0.
  \end{displaymath}
\end{proposition}

\subsection{}
There is a $\C$-algebra anti-automorphism $\w:\Uq(\g)\mapto \Uq(\g)$
\begin{alignat*}{2}
  \w:k_i &\mapsto {k_i}\inv, \qquad
  & \w:q &\mapsto q\inv,\\
  \w:e_i &\mapsto f_i,
  & \w:f_i &\mapsto e_i.
\end{alignat*}

Also there is a $\C$-algebra automorphism $\varphi:\Uq(\g)\mapto \Uq(\g)$
\begin{alignat*}{2}
  \varphi:k_i &\mapsto {k_i}, \qquad
  & \varphi:q &\mapsto q\inv,\\
  \varphi:e_i &\mapsto f_i,
  & \varphi:f_i &\mapsto e_i.
\end{alignat*}

\subsection{}
Let $\Uq(\n_+)$, $\Uq(\h)$ and $\Uq(\n_-)$ be the subalgebras of
$\Uq(\g)$ generated by $\set{e_i\mid i\in \range{1,r}}$,
$\set{k_i,{k_i}\inv\mid i\in\range{1,r}}$ and $\set{f_i\mid i\in
  \range{1,r}}$ respectively. From the defining relations of
$\Uq(\g)$, it follows that any element of $\Uq(\g)$ can be written as
a sum of monomials ordered such that elements of $\Uq(\n_-)$ appear on
the left, elements of $\Uq(\h)$ in the middle and elements of
$\Uq(\n_+)$ on right in each monomial.  Therefore $\Uq(\g)= \Uq(\n_-)
\Uq(\h) \Uq(\n_+)$,

\subsection{}
Denote by $\A$ the ring $\Cfree{q,q\inv}$. I define the elements $h_i$
in $\Uq(\g)$ to be
\begin{displaymath}
  h_i:=\frac{k_i- {k_i}\inv}{q_i- {q_i}\inv}.
\end{displaymath}
Define the $\A$-subalgebra $\UA(\g)$ of $\Uq(\g)$, to be the
subalgebra generated over $\A$ by the elements in $\set{{k_i}^{\pm1},
  e_i, f_i, h_i\mid i\in\range{1,r}}$. The generators $\set{e_i, f_i,
  {k_i}^{\pm1}\mid i\in\range{1,r}}$ satisfy the relations
in~\ref{quea:Uqg-def}, except that the relation between $e_i$ and $f_j$ is
replaced by
\begin{displaymath}
  e_i\cdot f_j- f_j\cdot e_i = \delta_{ij} h_i,
\end{displaymath}
and from the definition of $h_i$ there is the relation
\begin{displaymath}
  (q_i- q_i\inv) h_i = k_i- {k_i}\inv.
\end{displaymath}
Then $\UA(\g)$ is a Hopf subalgebra of $\Uq(\g)$. The coproduct,
antipode and counit maps act on $h_i$ as
\begin{align*}
  \Delta(h_i) &= h_i\tensor k_i+ {k_i}\inv\tensor h_i,\\
  S(h_i) &= -h_i,\\
  \epsilon(h_i) &= 0.
\end{align*}

\begin{notation}
  For $\e\in\Ccross$, define $\e_i:=\e^{d_i}$. Denote by
  $\brackets{n}_\e$ and $\comb{n}{m}_\e$ in $\C$, the numbers
  $\brackets{n}_q$ and $\comb{n}{m}_q$ in $\C(q)$ with $q$ replaced by
  $\e$.
\end{notation}

\subsection{}
Let $\e\in \Ccross$. Then I define the {\em specialisation\/} of
$\UA(\g)$ at $q=\e$ to be the $\C$-algebra $\Ue(\g):=\UA(\g)/
\ideal{q-\e}$.

\begin{proposition} \rom{(compare~\cite[1.5]{deconcini-kac:unity})}
  Let $\U_1(\g)$ be the specialisation at $\e=1$. The quotient algebra
  $\U_1(\g)/ \ideal{k_i-1, k_i\inv-1\mid i\in\range{1,r}}$ is Hopf
  algebra isomorphic to the universal enveloping algebra $\U(\g)$ of
  the Lie algebra $\g$ defined in~\ref{quea:g-assoc-to-aij}.
\end{proposition}

\begin{proof}
  The $\C$-algebra isomorphism is given by
  \begin{alignat*}{2}
    e_i &\mapsto \bar{e}_i, \qquad&
    f_i &\mapsto \bar{f}_i,\\
    h_i &\mapsto \bar{h}_i. &&
  \end{alignat*}
  From inspection of the Hopf algebra maps of $\U_1(\g)$, it is clear
  that in the quotient algebra they reduce to the standard Hopf
  algebra maps of a universal enveloping algebra
  (see~\ref{prel:Hopf-U(g)}).
\end{proof}

\section{Braid group automorphisms of $\Uq(\g)$}

\begin{notation}
  I define for each $n\in \N$ the following elements in $\Uq(\g)$
  \begin{displaymath}
    e_i^{(n)}:= \frac{e_i^n}{\brackets{n}_{q_i}!}\quad \text{and}\quad
    f_i^{(n)}:= \frac{f_i^n}{\brackets{n}_{q_i}!}.
  \end{displaymath}
\end{notation}

I quote the following well-known result, that appears for instance
in~\cite[2.2]{deconcini-kac-procesi:coadj}
and~\cite[2.1.2]{lusztig:book}
\begin{theorem}
  Let $B$ be the braid group and $W$ the Weyl group of $\g$. There is
  a unique map $\phi:W\mapto B$, such that $\phi:1\mapsto 1$,
  $\phi:s_i\mapsto T_i$ and $\phi(w w')=\phi(w)\phi(w')$ when
  $l(ww')=l(w)+l(w')$ ($w,w'\in W$).
\end{theorem}

\begin{corollary}
  Let $s_{i_1} s_{i_2}\cdots s_{i_n}$ and $s_{i'_1} s_{i'_2}\cdots
  s_{i'_n}$ be two reduced expressions of $w\in W$. Then
  $T_{i_1}T_{i_2}\cdots T_{i_n}= T_{i'_1}T_{i'_2}\cdots T_{i'_n}$ in
  $B$. Write $T_w$ for $\phi(w)$ $(w\in W)$. The map $\phi:w\mapsto
  T_w$ is well defined, since it is independent of the form of the
  reduced expression of $w$.
\end{corollary}

I introduce Lusztig's braid group automorphisms $T_i$ of $\Uq(\g)$
\cite{lusztig:reps,lusztig:jams,lusztig:geom-ded}.
\begin{theorem} \label{quea:B-action-on-Uq(g)}
  Let $B$ be the braid group associated to $\g$. There is a representation
  of $B$ on $\Uq(\g)$ as a group of automorphisms: 
  \begin{align*}
    T_i:k_i &\mapsto {k_i}\inv,\\
    T_i:k_j &\mapsto k_j {k_i}^{-a_{ij}} \quad (i\neq j),\\ 
    T_i:e_i &\mapsto -k_i f_i,\\ 
    T_i:e_j &\mapsto \sum_{n=0}^{-a_{ij}} (-1)^n q_i^{a_{ij}+n}
    e_i^{(n)} e_j e_i^{(-a_{ij}- n)} \quad (i\neq j),\\ 
    T_i:f_i &\mapsto -{k_i}\inv e_i,\\ 
    T_i:f_j &\mapsto \sum_{n=0}^{-a_{ij}} (-1)^n q_i^{-a_{ij}-n}
    f_i^{(-a_{ij}- n)} f_j f_i^{(n)} \quad (i\neq j).
  \end{align*}
\end{theorem}

\begin{proof}
  The proof of the theorem is rather lengthy. It can be found
  in~\cite{lusztig:book}.
\end{proof}

Note that the automorphisms commute with the anti-automorphism $\w$
\begin{displaymath}
  T_i\compose \omega = \omega\compose T_i.
\end{displaymath}
The action of the inverse elements is given by
\begin{displaymath}
  {T_i}\inv = \varphi\compose T_i\compose \varphi\inv.
\end{displaymath}

\begin{remark} \cite[1.6]{deconcini-kac:unity}
 Note that $T_i e_j= \ad(-e_i^{(-a_{ij})})e_j$ ($i\neq j$) and
 $T_i f_j= \w(T_i e_j)$.
\end{remark}

\subsection{} \label{quea:w0-ordered-R+}
Let $w_0\in W$ be the longest element of the Weyl group $W$. Fix a
reduced expression $s_{i_1} s_{i_2}\cdots s_{i_\ng}$ of $w_0$.  Then the
set
\begin{displaymath}
  \alpha_{i_1},\quad s_{i_1}\alpha_{i_2},\quad
  s_{i_1}s_{i_2}\alpha_{i_3},\quad\dots,\quad s_{i_1}s_{i_2}\cdots
  s_{i_{\ng-1}}\alpha_\ng
\end{displaymath}
is in bijective correspondence with the set of positive roots $R_+$ of
$\g$ and fixes an ordering of $R_+$. Define
$\beta_k:=s_{i_1}s_{i_2}\cdots s_{i_{k-1}}\alpha_k$.
So $R_+=\set{\beta_k\mid k\in\range{1,\ng}}$.

\begin{proposition}
  Let $\beta=w(\alpha_i)\in R_+$, for some $w\in W$ and
  $i\in\range{1,r}$. Then $T_w e_i\in \Uq(\n_+)$ and $T_w f_i\in \Uq(\n_-)$.
\end{proposition}

\begin{proposition} \cite[2.3]{deconcini-kac-procesi:coadj}
  Let $w\in W$ such that $w(\alpha_i)=\alpha_j$ ($i\neq j$). Then $T_w
  e_i= e_j$.
\end{proposition}

\section{A Basis of $\Uq(\g)$}
\label{quea:sec-Uq(g)-basis}

The following important result is due to Lusztig.
\begin{proposition} \cite[\S4]{lusztig:geom-ded}
  \label{quea:Uq(g)-basis}
  Let $s_{i_1} s_{i_2}\cdots s_{i_\ng}$ be a reduced expression of
  $w_0$, the longest element of $W$.  Let $\set{\beta_k\mid
    k\in\range{1,\ng}}$ be the ordered set of positive roots defined
  above. Then the positive root vectors defined by
  \begin{displaymath}
    e_{\beta_k}:= T_{i_1} T_{i_2} \cdots T_{i_{k-1}} e_{i_k} \quad
    (k\in\range{1,\ng}) 
  \end{displaymath}
  generate an ordered basis of $\Uq(\n_+)$ as a vector space
  \begin{displaymath}
    \Uq(\n_+)=\sum_{(m_1,\dots,m_\ng)\in \N^\ng} \C(q)\; e_{\beta_1}^{m_1}
    e_{\beta_2}^{m_2}\cdots e_{\beta_\ng}^{m_\ng}
    =: \sum_{m\in \N^\ng} \C(q) E^m.
  \end{displaymath}
  The corresponding negative root vectors are defined by
  $f_{\beta_k}:=\w(e_{\beta_k})$ and generate an ordered basis of
  $\Uq(\n_-)$
  \begin{displaymath}
    \Uq(\n_-)=\sum_{(m_1,\dots,m_\ng)\in \N^\ng} \C(q)\; f_{\beta_\ng}^{m_\ng}
    f_{\beta_{\ng-1}}^{m_{\ng-1}}\cdots f_{\beta_1}^{m_1}
    =: \sum_{m\in \N^\ng} \C(q) F^m.
  \end{displaymath}
\end{proposition}

The positive and negative root vectors form a linearly independent
basis of $\Uq(\n_+)$ and $\Uq(\n_-)$ respectively. This is
proved~\cite[proposition 1.10]{lusztig:jams} essentially by observing
that at the specialisation $q=1$, the induced $\U(\g)$-basis is
linearly independent.

\subsection{}
The set $\set{k_i, k_i\inv\mid i\in\range{1,r}}$ generates a basis of
$\Uq(\h)$
\begin{displaymath}
  \Uq(\h)=\sum_{(m_1,m_2,\dots,m_r)\in\Z^r} \C(q)\; k_1^{m_1}
  k_2^{m_2} \cdots k_r^{m_r}
  =: \sum_{m\in \Z^r} \C(q) K^m.
\end{displaymath}
So I have now a basis of $\Uq(\g)=\Uq(\n_-) \Uq(\h) \Uq(\n_+)$
\begin{displaymath}
  \Uq(\g)=\sum_{m_\pm\in \N^\ng,m_0\in \Z^r} \C(q) F^{m_-} K^{m_0} E^{m_+}.
\end{displaymath}
Note that different reduced expressions of the longest element $w_0$
of $W$ give inequivalent bases of $\Uq(\n_+)$ and $\Uq(\n_-)$.

\begin{remark}
  Let $\g$ be nonsimply-laced. Then the braid group action of $B$ on
  $\UA(\g)$ is not strictly well-defined, since the elements
  $e_i^{(n)}$ and $f_i^{(n)}$ ($i\in\range{1,r}$ and $n>1$) are not in
  $\UA(\g)$.  (Lusztig avoids this problem by considering a different
  $\A$-subalgebra of $\Uq(\g)$ which includes these elements as
  generators.) Another way to overcome this, is to extend the ring
  $\A$ to $\bar{\A}$ by elements $\frac{1}{q_i^n- q_i^{-n}}$
  ($n\in\set{1,2,3}$ (or $n\in\Zplus$), $i\in\range{1,r}$).  One
  should then `localise' $\bar{\A}$ at $q=\e$ (meaning remove from
  $\bar{\A}$ the elements that have poles at $q=\e$), so that one can
  consider the specialisation of $\U_{\bar{\A}}(\g)$ at $q=\e$. I
  thank Jonathan Beck for explaining to me the idea of the
  localisation of a ring in an indeterminant.
\end{remark}

\subsection{$\Ue(\g)$ basis}
The action of the braid group $B$ on $\Ue(\g)$ defined by the maps
in~\ref{quea:B-action-on-Uq(g)} with $q_i$ replaced by $\e_i$ is
well-defined. For each reduced expression of $w_0$ the action of $B$
on $\Ue(\g)$ gives an ordered basis of $\Ue(\n_+)$ (and
$\Ue(\n_-)$) over $\C$ as a vector space analogous to the basis of
$\Uq(\n_+)$ ($\Uq(\n_-)$) of~\ref{quea:Uq(g)-basis}. The basis of
$\Ue(\h)$ is given by
\begin{displaymath}
  \Ue(\h)=\sum
  \begin{Sb}
    (m_1,m_2,\dots,m_r)\in\Z^r\\
    (m'_1,m'_2\dots,m'_r)\in \N^r
  \end{Sb}
  \C\; k_1^{m_1} k_2^{m_2} \cdots k_r^{m_r} h_1^{m'_1}\cdots h_r^{m'_r}
  =: \sum_{m\in \Z^r,m'\in\N^r} \C K^m H^{m'}.
\end{displaymath}
Finally there is the corresponding basis of $\Ue(\g)$
\begin{displaymath}
  \Ue(\g)=\sum_{m_\pm,m'\in \N^\ng,m_0\in \Z^r} \C F^{m_-} K^{m_0}
  H^{m'} E^{m_+}.
\end{displaymath}

Observe that the above analysis is valid also at $\e=1$ and in the
quotient $\U_1(\g)/ \ideal{k_i-1}$. In the latter algebra the braid
action reduces to the classical braid group action of $B$ on $\U(\g)$
and allows the construction of a basis (a Poincar\'e-Birkhoff-Witt
basis), which in an obvious notation borrowed from above reads
\begin{displaymath}
  \U(\g)= \sum_{m_\pm,m_0\in \N^\ng} \C \bar{F}^{m_-} \bar{H}^{m_0}
  \bar{E}^{m_+}.
\end{displaymath}
Of course in this case the bases obtained from the different reduced
expressions of $w_0$ {\em are\/} all equivalent.

\subsection{Example}
$\sl_3$ is the easiest example of a simple Lie algebra with a
nontrivial root system. I briefly consider the construction of a basis
of $\Uq(\sl_3)$ using the techniques described above. The Cartan
matrix $(a_{ij})=\left( \begin{smallmatrix} 2& -1\\-1& 2 \end{smallmatrix}
\right)$ and $r=2$. Let $\set{\alpha_1, \alpha_2}$ be the simple roots of
$\sl_3$. The Weyl group $W$ is generated by $\set{s_1,s_2}$ and they
satisfy the braid relations $s_1 s_2 s_1= s_2 s_1 s_2$ and $s_i^2=1$
($i\in \set{1,2}$). $W$ contains the elements $\set{1, s_1,s_2,
  s_1s_2, s_2s_1, s_1s_2s_1}$. The longest element of $W$ is $w_0=
s_1s_2s_1= s_2s_1s_2$. The positive roots are
$R_+=\set{\alpha_1,\alpha_1+ \alpha_2, \alpha_2}$ and $\ng=3$.

Consider then the reduced expression $s_1s_2s_1$ of $w_0$.  Then
$\beta_1=\alpha_1$, $\beta_2=\alpha_1+\alpha_2$ and $\beta_3=
\alpha_2$. Straightforward calculations then show that the root
vectors of $\Uq(\sl_3)$ corresponding to this reduced expression of
$w_0$ are
\begin{align*}
  e_{\beta_1} &=e_1,\\
  e_{\beta_2} &=T_1 e_2= -e_1e_2+ q\inv e_2e_1,\\
  e_{\beta_3} &=T_1 T_2 e_1 = e_2,\\
  f_{\beta_1} &= \w(e_{\beta_1})= f_1,\\
  f_{\beta_2} &= \w(e_{\beta_2})= -f_2f_1+ qf_1f_2,\\
  f_{\beta_3} &= \w(e_{\beta_3})= f_2.
\end{align*}
If I choose instead the other reduced expression $s_2s_1s_2$.  Then
$\beta'_1=\alpha_2$, $\beta'_2=\alpha_2+\alpha_1$ and $\beta'_3=
\alpha_1$ and the root vectors obtained are
\begin{align*}
  e_{\beta'_1} &=e_2,\\
  e_{\beta'_2} &=T_2 e_1= -e_2e_1+ q\inv e_1e_2,\\
  e_{\beta'_3} &=T_1 T_2 e_1 = e_1,\\
  f_{\beta'_1} &= \w(e_{\beta'_1})= f_2,\\
  f_{\beta'_2} &= \w(e_{\beta'_2})= -f_1f_2+ qf_2f_1,\\
  f_{\beta'_3} &= \w(e_{\beta'_3})= f_1.
\end{align*}
Note that $e_{\beta_2}$ and $e_{\beta'_2}$ are not proportional, and
$f_{\beta_2}$ and $f_{\beta'_2}$ are not proportional to each other.
So inequivalent bases are obtained from the two reduced different
reduced expressions of $w_0$. However at the specialisation $q=1$,
$e_{\beta_2}= -e_{\beta'_2}$ and $f_{\beta_2}= -f_{\beta'_2}$.  The
basis of $\Uq(\sl_3)$ corresponding to the reduced expression
$s_1s_2s_1$ of $w_0$ reads
\begin{displaymath}
  \Uq(\sl_3)= \sum_{m^\pm\in\N^3,m^0\in\Z^2}
  f_{\beta_3}^{m^-_3} f_{\beta_2}^{m^-_2} f_{\beta_1}^{m^-_1}
  k_1^{m^0_1} k_2^{m^0_2}
  e_{\beta_1}^{m^+_1} e_{\beta_2}^{m^+_2} f_{\beta_3}^{m^+_3}.
\end{displaymath}

\section{Representations of $\Uq(\g)$}

\subsection{}
Let $P$ be the weight lattice, $Q$ the root lattice and
$(\cdot,\cdot):P\times Q\mapto \Z$ the bilinear form associated to the
Cartan matrix $(a_{ij})$ (as defined in~\ref{quea:lattices}). Let
$\Hom(Q,Z_2)$ be the group of homomorphisms of $Q$ into the group
$\set{1,-1}$.

\begin{lemma}
  Let $\sigma\in \Hom(Q,Z_2)$. Then the following map is an
  automorphism of $\Uq(\g)$
  \begin{align*}
    e_i &\mapsto \sigma(\alpha_i) e_i,\\
    k_i &\mapsto \sigma(\alpha_i) k_i,\\
    f_i &\mapsto f_i.
  \end{align*}
\end{lemma}

\begin{definition}
  Fix $\sigma\in\Hom(Q,Z_2)$ and $\lambda\in P$. Then the {\em
    Verma module\/} $M^\sigma(\lambda)$ over $\Uq(\g)$ with highest
  weight $\lambda$ and twisting $\sigma$ is defined to be the
  (unique) $\C(q)$-vector space generated by a vector $v_\lambda$ such
  that
  \begin{align*}
    \Uq(\n_+)\cdot v_\lambda &= 0,\\
    k_i\cdot v_\lambda &= \sigma(\alpha_i)
    q^{(\lambda,\alpha_i)} v_\lambda,\\ 
    M^\sigma(\lambda) :&= \Uq(\n_-)\cdot v_\lambda= \sum_{m\in \N^\ng}
    F^m\cdot v_\lambda.
  \end{align*}
  The vector $v_\lambda$ is called a highest weight vector of the
  module. 
\end{definition}

\subsection{Highest weight modules} \label{quea:Uq(g)-hw-modules}
I can construct the Verma module $M^\sigma(\lambda)$ in the following
way.  Consider the left ideal of $\Uq(\g)$
\begin{displaymath}
  J^\sigma(\lambda):= \sum_{i\in\range{1,r}}\Uq(\g) e_i
  + \sum_{i\in\range{1,r}} \Uq(\g) (k_i- \sigma(\alpha_i)q^{(\lambda,
    \alpha_i)}).
\end{displaymath}
Then the quotient $\Uq(\g)/ J^\sigma(\lambda)$ is isomorphic to 
$M^\sigma(\lambda)$ as a $\Uq(\g)$-module.

The $\Uq(\g)$-module $M^\sigma(\lambda)$ is a highest weight module.
Every quotient module of $M^\sigma(\lambda)$ is a
highest weight $\Uq(\g)$-module with highest weight $\lambda$.

In particular there exists a unique maximum submodule $M'$ of
$M^\sigma(\lambda)$. Therefore the quotient module $L^\sigma(\lambda):=
M^\sigma(\lambda)/ M'$ is unique and irreducible.

Let $V$ be a highest weight $\Uq(\g)$-module with highest weight
$\lambda$. Define the subspace $V_\mu$ of $V$ ($\mu\in P_+$) as
\begin{displaymath}
  V_\mu:=\set{v\in V\mid k_i\cdot v= \sigma(\alpha_i)
    q^{(\lambda-\mu,\alpha_i)} v}.
\end{displaymath}
Then $V$ admits the following weight space decomposition
($P_+$-gradation)
\begin{displaymath}
  V=\Directsum_{\mu\in P_+} V_\mu.
\end{displaymath}

\begin{definition}
  Let $V$ be a $\Uq(\g)$-module, with weight space decomposition. $V$
  is called {\em integrable\/}, if there exists a positive integer
  $m_0\in\Zplus$, such that $e_i^m\cdot v=0$ and $f_i^m\cdot v=0$ for
  all $m\geq m_0$, $i\in\range{1,r}$ and $v\in V$. (In this case the
  action of the Chevalley generators of $\Uq(\g)$ on $V$ is said to be
  locally nilpotent.)
\end{definition}

Integrable modules~\cite{lusztig:reps} of $\Uq(\g)$ can be constructed
as follows.

\begin{lemma} \label{quea:integrable-module-construction}
  Let $(m^+_i), (m^-_i)\in \N^r$ and $\lambda\in P$. Define a
  left ideal $I_\lambda$ of $\Uq(\g)$ by
  \begin{displaymath}
    I_\lambda:= \sum_{i\in\range{1,r}} \Uq(\g) e_i^{m^+_i+ 1}+
    \sum_{i\in\range{1,r}} \Uq(\g) f_i^{m^-_i+ 1}+
    \sum_{i\in\range{1,r}} \Uq(\g) (k_i- q^{(\lambda,\alpha_i)}).
  \end{displaymath}
  Then the quotient $\Uq(\g)/ I_\lambda$ is an integrable
  $\Uq(\g)$-module. 
\end{lemma}

\begin{proof}
  See~\cite[3.5.3]{lusztig:book}. I sketch the proof. The idea is to
  show using the defining relations of $\Uq(\g)$ that for every
  $x\in\Uq(\g)$, $f_i^N x= y f_i^{N'}$ ($y\in\Uq(\g)$) where $N'\geq
  N-c_x\in\N$ ($c_x\in\N$ depending only $x$). Then for
  $N\geq m^-_i+1+ c_x$ and $x\in \Uq(\g)/ I_\lambda$, $f_i^N x=0$.
  There is a corresponding result with $f_i$ replaced by $e_i$. This
  proves the lemma.
\end{proof}


\subsection{Trivial module}
As usual the counit map $\epsilon:\Uq(\g)\mapto \C(q)$ gives a trivial
(one dimensional) representation of $\Uq(\g)$ on $\C(q)$. (The
representation is integrable.)

\subsection{Tensor product module}
Let $V_1$ and $V_2$ be two $\Uq(\g)$-modules. Then the coproduct map
$\Delta:\Uq(\g)\mapto \Uq(\g)\tensor \Uq(\g)$ induces a
$\Uq(\g)$-module structure on $V_1\tensor V_2$. It can be
shown~\cite[3.5.2]{lusztig:book} that the tensor product of two
integrable $\Uq(\g)$-modules is integrable.

\subsection{Dual module} Let $V$ be a (left) $\Uq(\g)$-module and let $V^*$
be the $\C(q)$-vector space dual to $V$, under a pairing
$\pairing{\cdot, \cdot}:V^*\times V\mapto \C(q)$. Then the antipode
map $S$ induces a dual (left) $\Uq(\g)$-module
\begin{displaymath}
  \pairing{x\cdot v^*,v}:= \pairing{v^*,S(x)\cdot v} \quad
  (x\in\Uq(\g), v\in V, v^*\in V^*).
\end{displaymath}
The dual module to an integrable module is integrable.

\subsection{Right module}
Let $\pi:\Uq(\g)\mapto \End(V)$ be a left representation, so that
$\pi(xy)v= \pi(x)\pi(y)v$ ($x,y\in \Uq(\g)$, $v\in V$). Then the
composition of the anti-automorphism $\w$ of $\Uq(\g)$ with $\pi$ gives a
right representation $\pi^\w:=\pi\compose \w$ of $\Uq(\g)$, so that
$\pi^\w(xy) v= \pi^\w(y) \pi^\w(x) v$ ($x,y\in \Uq(\g)$, $v\in V$).
Conversely if $\pi'$ is a right representation then $\pi'\compose \w$
is a left representation. If $\pi$ is an integrable representation,
then $\pi\compose\w$ is an integrable representation.

\subsection{}
The Verma module $M_\A^\sigma(\lambda)$ over $\UA(\g)$ and the Verma module
$M_\e^\sigma(\lambda)$ over $\Ue(\g)$ are defined like
$M^\sigma(\lambda)$ in an obvious way.

\section{$\Ue(\g)$ at Roots of unity}

In this section I consider the specialisation of $\UA(\g)$ at $q=\e$,
a primitive root of unity.  First I present some identities, which
will be required shortly.

\begin{lemma} \label{quea:e^m-f-identity}
  \cite[\S3]{jimbo:lmp1} Let $m\in \Zplus$ and $i,j\in\range{1,r}$. The
  following identities hold in $\Uq(\g)$
  \begin{align*}
    \comm{e_i^m,f_j} &= \delta_{ij} \brackets{m}_{q_i} \frac{k_i
      q_i^{1-m}- k_i\inv q_i^{m-1}}{q_i- q_i\inv} e_i^{m-1},\\
    \comm{e_i,f_j^m} &= \delta_{ij} \brackets{m}_{q_j} f_j^{m-1}
    \frac{k_j q_j^{1-m}- k_j\inv q_j^{m-1}}{q_j- q_j\inv}.
  \end{align*}
\end{lemma}

\begin{proof}
  The first identity is proved by induction on $m$. The second then
  follows by applying the anti-automorphism $\w$.
\end{proof}

\begin{lemma}  \label{quea:e^m-e-identity}
  \cite[1.10]{deconcini-kac:unity}
  Define the following elements in $\Uq(\n_+)$:
  \begin{align*}
    e_{ij} &:= T_j e_i \qquad\qquad \text{if $a_{ij}< 0$},\\
    e_{iij} &:=
    \begin{cases}
      T_j T_i e_j & \text{if $a_{ij}=-2$}\\
      T_j T_i T_j e_i \quad & \text{if $a_{ij}=-3$},
    \end{cases}\\
    e_{iiij} &:= T_j T_i T_j T_i e_j \qquad \text{if $a_{ij}=-3$}.
  \end{align*}
  Let $m\geq -a_{ij}$. Then the following identity holds in
  $\Uq(\n_+)$
  \begin{displaymath}
    e_i^m e_j = \sum_{p=0}^{-a_{ij}} q_i^{(-a_{ij}- p)m+ p}
    \frac{\brackets{m}_{q_i}!}{\brackets{m-p}_{q_i}!} \;
    e_{\!\negthickspace \underbrace{{\scriptstyle i\ldots
        i}}_{\text{$p$ times}}\!\negthickspace j}
    e_i^{m-p}.
  \end{displaymath}
  A similar identity in the generators $f_i$ is obtained in
  $\Uq(\n_-)$ by applying the anti-automorphism $\w$ to the above
  identity.
\end{lemma}

\begin{proof}
  The lemma follows from direct calculations that can be found
  in~\cite[\S5]{lusztig:geom-ded}.
\end{proof}

\subsection{}
Fix a positive {\em odd\/} integer $\l\in \Zplus$, such that $\l>d_i$
($\forall i\in\range{1,r}$). Let $\e$ be a primitive (odd) $\l$-th
root of unity. 

\begin{proposition} \cite[3.1]{deconcini-kac:unity}
  Let $\set{e_\beta, f_\beta\mid \beta\in R_+}$ be the basis of the
  positive and negative root vectors of $\Ue(\g)$, introduced
  in~\ref{quea:Uq(g)-basis}.  In $\Ue(\g)$ at the odd root of unity
  $\e$ the following relations hold for all $\alpha,\beta\in R_+$,
  $i\in\range{1,r}$:
  \begin{alignat}{2}
    k_i^\l e_\beta &=  e_\beta k_i^\l,\qquad
    & k_i^\l f_\beta &=  f_\beta k_i^\l, \label{quea:k^l-relations}\\
    e_\alpha^\l f_\beta &= f_\beta e_\alpha^\l,
    & f_\alpha^\l e_\beta &= e_\beta f_\alpha^\l,\label{quea:e^lf,ef^l}\\
    e_\alpha^\l e_\beta &= e_\beta e_\alpha^\l,
    & f_\alpha^\l f_\beta &= f_\beta f_\alpha^\l \label{quea:e^le,ff^l}.
  \end{alignat}
\end{proposition}

\begin{proof}
  The relations in~\eqref{quea:k^l-relations} for $k_i^\l$ follow
  immediately from the corresponding defining relations of $\Uq(\g)$.
  Consider now~\eqref{quea:e^lf,ef^l} and~\eqref{quea:e^le,ff^l}. First
  let $\alpha=\alpha_i$ and $\beta=\alpha_j$. In this case the
  relations in~\eqref{quea:e^lf,ef^l} follow from
  lemma~\ref{quea:e^m-f-identity} and those in~\eqref{quea:e^le,ff^l}
  from lemma~\ref{quea:e^m-e-identity}. But then since
  $\set{e_{\alpha_j}\mid j\in\range{1,r}}$ generates $\Uq(\n_+)$ and
  $\set{f_{\alpha_j}\mid j\in\range{1,r}}$) generates $\Uq(\n_-)$, it
  follows that~\eqref{quea:e^lf,ef^l} and~\eqref{quea:e^le,ff^l} are true
  for all $\alpha_i\in\Pi$ and all $\beta\in R_+$. Next I apply a
  braid group automorphism $T_w$ ($w\in W$) to the relations
  in~\eqref{quea:e^lf,ef^l} and~\eqref{quea:e^le,ff^l} when
  $\alpha=\alpha_i$ and $\beta\in R_+$ and this proves that the
  relations are in fact true for all $\alpha=w(\alpha_i)$ ($\forall
  w\in W$) and $\beta\in R_+$. Therefore the relations hold for all
  $\alpha,\beta\in R_+$ and the proposition is proved.
\end{proof}

\begin{notation}
  Let $\Ze$ denote the centre of $\Ue(\g)$.
\end{notation}

\begin{corollary} \label{quea:centre-corol}
  At the root of unity $\e$, the following elements lie in the centre
  $\Ze$ of $\Ue(\g)$:
  \begin{displaymath}
    e_\alpha^\l, k_i^\l, f_\alpha^\l \qquad (\forall \alpha\in R_+,
    i\in\range{1,r}).
  \end{displaymath}
\end{corollary}

\subsection{Diagonal modules} \label{quea:diagonal-module}
Consider now a Verma module $M_\e^\sigma(\lambda)$ over $\Uq(\g)$ at
the root of unity $\e$, generated by $v_\lambda$. Let $\alpha\in R_+$.
Note that the vector $f_\alpha^\l\cdot v_\lambda$ is singular
(primitive) in $M_\e^\sigma(\lambda)$:
\begin{displaymath}
  \Ue(\n_+)\cdot f_\alpha^\l\cdot v_\lambda= 0 \quad (\forall \alpha\in
  R_+),
\end{displaymath}
since $f_\alpha^\l\in \Ze$. Therefore each vector
$f_\alpha^\l\cdot v_\lambda$ ($\alpha\in R_+$) generates a
$\Ue(\g)$-submodule of $M_\e^\sigma(\lambda)$. Define
\begin{displaymath}
  \bar{M}_\e^\sigma(\lambda):= M_\e^\sigma(\lambda)/ (\sum_{\alpha\in
    R_+} \Uq(\g)\cdot f_\alpha^\l\cdot v_\lambda).
\end{displaymath}
In~\cite[3.2]{deconcini-kac:unity} the $\Ue(\g)$-module
$\bar{M}_\e^\sigma(\lambda)$ is called diagonal.

\begin{proposition}
  Let $\Z_\l:=\Z/\l\Z$. The $\Ue(\g)$-module
  $\bar{M}_\e^\sigma(\lambda)$ is finite dimensional.  It has the
  following basis over $\C$
  \begin{displaymath}
      \bar{M}_\e^\sigma(\lambda)= \sum_{m\in \Z_\l^\ng} \C\; F^m v_\lambda.
  \end{displaymath}
\end{proposition}

\begin{remark}
  Like the Verma module $M_\e^\sigma(\lambda)$, the diagonal module
  $\bar{M}_\e^\sigma(\lambda)$ can of course be constructed as a
  quotient by a left ideal of $\Ue(\g)$ in a similar way
  to~\ref{quea:Uq(g)-hw-modules}.
\end{remark}

\subsection{The centre}
Let $x_\alpha:= e_\alpha^\l$, $y_\alpha:= f_\alpha^\l$ ($\alpha\in
R_+$) and $z_i^{\pm1}:= k_i^{\pm \l}$ ($i\in\range{1,r}$).  Define
$\Zo^+$ ($\Zo^-$) to be the central subalgebra of $\Ue(\n_+)$
($\Ue(\n_-)$) generated by $\set{x_\alpha\mid \alpha\in R_+}$
(respectively $\set{y_\alpha\mid \alpha\in R_+}$). Define $\Zo^0$ to
be the subalgebra of $\Ue(\h)$ generated by $\set{z_i^{\pm1}\mid
  i\in\range{1,r}}$. Let $\Zo$ be the subalgebra of $\Ze$ generated by
the subalgebras $\Zo^+$, $\Zo^-$ and $\Zo^0$. Note that $\Zo$ is a
proper subalgebra of $\Ze$, since $\Uq(\g)$ has one or more
$q$-analogue Casimir elements, that generate $\Ze$ when $\e$ is not a
root of unity.

\begin{lemma}
  The algebra $\Ue(\g)$ is finite dimensional over $\Zo$ and
  \begin{displaymath}
    \set{F^{m_-}K^{m_0}E^{m_+}\mid m_\pm\in Z_\l^\ng, m_0\in Z_\l^r}
  \end{displaymath}
  forms a basis of $\Ue(\g)$ over $\Zo$ as a free module. So $\dim_{\Zo}
  \Ue(\g)= \l^{\dim\g}$.
\end{lemma}


\begin{notation}
  Let $\alpha\in R_+$, such that $\alpha=\sum_{i\in\range{1,r}} a_i
  \alpha_i$. Define $k_\alpha:= \prod_{i\in\range{1,r}} k_i^{a_i}$.
\end{notation}

\subsection{Triangular modules}
Let $\lambda\in P$ and let $\nu$ be an algebra homomorphism
$\nu:\Zo^-\mapto \C$. Then define the {\em triangular module\/}
$\bar{M}_\e^\sigma(\lambda,\nu)$ over $\Ue(\g)$ to be the quotient of
$\Ue(\g)$ by the left ideal generated by $\set{e_i, k_i^{\pm1}-
  \sigma(\alpha_i)q^{\pm(\lambda,\alpha_i)}, y_\alpha-
  \nu(y_\alpha)\mid i\in\range{1,r}, \alpha\in R_+}$. Denote by
$v_\lambda$ the image of $1$ in $\bar{M}_\e^\sigma(\lambda,\nu)$. The
elements $F^m\cdot v_\lambda$ ($m\in \Z_\l^n$) form a (finite
dimensional) basis of $\bar{M}_\e^\sigma(\lambda,\nu)$.

Let $\alpha\in R_+$. When $\nu(y_\alpha)=0$, I say that $f_\alpha$
acts nilpotently in $\bar{M}_\e^\sigma(\lambda,\nu)$. In this case the
$\Ue(\sltwo)$-submodule generated by $\set{e_\alpha, f_\alpha,
  k_\alpha^{\pm1}}$ is called nilpotent. When $\nu(y_\alpha)\neq 0$, I
say that $f_\alpha$ acts cyclically (periodically). In this case the
$\Ue(\sltwo)$-submodule corresponding to the root $\alpha$ is called
semicyclic (semiperiodic), since $f_\alpha$ acts cyclically but
$e_\alpha$ acts nilpotently.

A $\Ue(\sltwo)$-module is called cyclic (periodic) when both its
Chevalley generators $e$ and $f$ act in it cyclically.  By definition
there are no cyclic $\Ue(\sltwo)$-submodules in a triangular module.

Note that the triangular $\Ue(\g)$-module
$\bar{M}_\e^\sigma(\lambda,\nu)$ contains the diagonal module
$\bar{M}_\e^\sigma(\lambda)$ as a special case: in the case $\nu=0$,
then $\bar{M}_\e^\sigma(\lambda,0)=\bar{M}_\e^\sigma(\lambda)$.

\subsection{Central characters}
Let $\structure{V,\pi}$ be an irreducible (finite dimensional)
representation of $\Ue(\g)$ (at the root of unity $\e$). Then by
Schur's lemma
\begin{displaymath}
  \pi(x)\cdot v= \chi^\pi(x) v \qquad (x\in\Ze, v\in V,
  \chi^\pi(x)\in \C).
\end{displaymath}
The map $\chi^\pi:\Ze\mapto \C$ is called the central
character of the representation $\pi$. Note in particular that 
$\chi^\pi(z_i^{\pm1})\neq 0$ ($i\in \range{1,r}$).

If a $\Ue(\g)$-module is diagonal, then $\chi:\Zo^\pm\mapto 0$. For
a triangular module the central character maps $\chi:\Zo^+\mapto 0$,
but $\chi(y_\alpha)\neq 0$ ($\alpha\in R_+$). In a completely cyclic
(periodic) module $\chi(x_\alpha)\neq 0$ and $\chi(y_\alpha)\neq 0$
($\alpha\in R_+$).

\section{$\Uq(\sltwo)$} \label{quea:Uq(sl2)-def}

In this section I consider as an example the case of $\sltwo$.

\subsection{}
The Lie algebra $\sltwo$ has a $1\times 1$ Cartan matrix $(2)$ and
only a single positive root $\alpha$. The weight lattice is $P=\Z$
($P_+= \N$) and the root lattice $Q=2\Z$ ($Q_+= 2\N$). The Weyl group
of $\sltwo$ has only two elements: the identity and the reflection
$\alpha\leftrightarrow -\alpha$. Therefore it is very simple to do
calculations in $\sltwo$, since it is the smallest simple Lie algebra
and it has a minimal root system ($\ng=1$).

The quantum group $\Uq(\sltwo)$ is generated over $\C(q)$ by
$\set{e,f,k^{\pm1}}$, satisfying the relations
\begin{alignat*}{2}
  k\cdot e\cdot k\inv &= q^2 e, \qquad\qquad&
  k\cdot f\cdot k\inv &= q^{-2} f,\\
  e\cdot f- f\cdot e &= \frac{k- k\inv}{q- q\inv}, &
  k\cdot k\inv &= 1= k\inv\cdot k.
\end{alignat*}
The algebra has the following basis
\begin{displaymath}
  \Uq(\sltwo):= \sum_{m_{\pm}\in \N, m_0\in \Z} \C(q) f^{m_-} k^{m_0}
  e^{m_+}.
\end{displaymath}

\subsection{}
The simplest nontrivial representation of $\Uq(\sltwo)$ is the
2-dimensional vector representation $\structure{\C(q)^2,\pi}$:
\begin{displaymath}
  \pi(e)= \twomatrix{0}{1}{0}{0},\quad
  \pi(k)= \twomatrix{q}{0}{0}{q\inv},\quad
  \pi(f)= \twomatrix{0}{0}{1}{0}.
\end{displaymath}

\subsection{Spin-$\frac{j}{2}$ representations}
Let $\sigma\in \set{1,-1}$ and $j\in P_+= \N$. Let the $\C(q)$-vector space
$V^\sigma(j):=\sum_{m=0}^j \C v_m$, with basis $\set{v_0,\dots,v_j}$.
Define a representation of $\Uq(\g)$ on $V^\sigma_j$ by:
\begin{align*}
  k\cdot v_m &= \sigma q^{j-2m} v_m,\\
  e\cdot v_m &= \sigma \brackets{j-m+1}_q v_{m-1},\\
  f\cdot v_m &= \brackets{m+1}_q v_{m+1},
\end{align*}
and $v_m:=0$ when $m\not\in\range{0,j}$.

\begin{lemma}
  The $\Uq(\sltwo)$-module $V^\sigma(j)$ is irreducible.
\end{lemma}

\begin{proof}
  The lemma is true since the coefficients of the $e$ and $f$ actions
  ($\brackets{m+1}_q$ and $\brackets{j-m+1}_q$) are nonzero for
  $m\in\range{0,j}$: so there are no singular vectors in
  $V^\sigma(j)$.
\end{proof}

For any irreducible $n$-dimensional representation $V$ of
$\Uq(\sltwo)$, $\sigma$ can be chosen in $\set{1,-1}$ such that $V$ is
equivalent to $V^\sigma(n-1)$ as a $\Uq(\sltwo)$-module.

\subsection{Verma module}
Consider the Verma module $M^\sigma(\lambda)$ of $\Uq(\sltwo)$
generated by the highest weight vector $v_\lambda$, with highest
weight $\lambda\in P$ ($k\cdot v_\lambda= \sigma q^\lambda v_\lambda$
and $e\cdot v_\lambda=0$). Then the vectors $v_m:=\frac{1}{\brackets{m}_q}
f^m\cdot v_\lambda$ ($m\in\Zplus$) and $v_0:=v_\lambda$ are a basis of
$M^\sigma(\lambda)$. The generators act on $M^\sigma(\lambda)$ as
\begin{align*}
  f \cdot v_m &= \brackets{m+1}_q v_{m+1},\\
  k\cdot v_m &= \sigma q^{\lambda-2m} v_m,\\
  e\cdot v_m &= \sigma \brackets{\lambda-m+1} v_{m-1}.
\end{align*}
Observe that if $\lambda\geq 0$, then $v':=v_{j'}$ with $j'=\lambda+1$
is a primitive (singular) vector in $M^\sigma(\lambda)$ ($e\cdot
v'=0$) and it is the first primitive vector in $M^\sigma(\lambda)$
below $v_\lambda$.  Therefore $L^\sigma(\lambda):=
M^\sigma(\lambda)/(\Uq(\sltwo)\cdot v')$ is a finite $j'$ dimensional
irreducible representation of $\Uq(\sltwo)$.  In fact
$L^\sigma(\lambda)$ is equivalent to the $\Uq(\sltwo)$-module
$V^\sigma_\lambda$ described above.  $L(\lambda)\isomorphic
\Uq(\sltwo)/ \ideal{e,f^{\lambda+1},k- \sigma q^\lambda}_L$.  (The
notation $\ideal{\cdot}_L$ denotes a left ideal in $\Uq(\sltwo)$.)
Clearly these irreducible finite dimensional representations are
integrable $\Uq(\sltwo)$-modules.

\subsection{Root of unity}
Let $\e\in \Ccross$. Define as before the $\A$-subalgebra
$\UA(\sltwo)$ and its specialisation $\Ue(\sltwo)$ at $q=\e$.

Fix a positive integer $\l\in \Zplus$, such that $\l>2$. Let $\e$ be
a primitive $\l$-th root of unity. Define
\begin{displaymath}
  \l'=
  \begin{cases}
    \l & \text{if $\l$ is odd},\\
    \frac{\l}{2} &\text{if $\l$ is even}.
  \end{cases}
\end{displaymath}

\begin{lemma}
  At the root of unity $\e$, the following elements are in the centre
  $\Ze$ of $\Ue(\sltwo)$
  \begin{displaymath}
    e^{\l'}, k^{\pm \l'}, f^{\l'}.
  \end{displaymath}
\end{lemma}

\subsection{}
Let $x:= e^{\l'}$, $z^{\pm 1}:= k^{\pm \l'}$ and $y:= f^{\l'}$. They
generate a subalgebra $\Zo$ of the centre $\Ze$.

\subsection{Spin-$\frac{j}{2}$ representations at a root of unity}
Consider now the $\Ue(\sltwo)$-module $V^\sigma(j)$ at the root of
unity $\e$. If $j< \l'$, then the module is irreducible. If $j\geq \l'$,
then the module contains one or more singular vectors $\set{v_p\mid
  p\in\set{j+1 \mod \l'}\intersect \range{0,j}}$ and is therefore
reducible. Let $\chi$ be the central character of the representation.
Observe that $\chi(x)=0$ and $\chi(y)=0$.

\subsection{Cyclic representations}
Every irreducible $\l'$-dimensional $\Ue(\sltwo)$-module, at the root
of unity $\e$, is isomorphic~\cite[4.2]{deconcini-kac:unity} for some
$(\zeta,a,b)\in \Ccross\times \C^2$ to the following representation
$V(\zeta,a,b)$ with basis $\set{v_m\mid m\in\range{0,\l'-1}}$
\begin{alignat*}{2}
  k\cdot v_m &= \zeta \e^{-2m} v_m \qquad&& (m\in\range{0,\l'-1}),\\
  f\cdot v_m &= v_{m+1} \qquad&& (m\in\range{0,\l'-2}),\\
  f\cdot v_{\l'-1} &= b v_0, &&\\
  e\cdot v_m &= \left(\frac{\zeta\e^{1-m}- \zeta\inv\e^{m-1}}{\e- \e\inv}
  \brackets{j}_\e+ ab\right) v_{m-1} \quad && (m\in\range{1,\l'-1}),\\
  e\cdot v_0 &= a v_{\l'-1}. &&
\end{alignat*}
Note that the $\Zo$ character $\chi$ is
\begin{align*}
  \chi(z) &= \zeta^{\l'},\\
  \chi(x) &=a \prod_{j=1}^{\l'-1} \left( \frac{\zeta\e^{1-m}-
    \zeta\inv\e^{m-1}}{\e- \e\inv} \brackets{j}_\e+ ab \right),\\
  \chi(y) &= b.
\end{align*}
In particular $\chi(x)=0$, if $a=0$.  Note that
$V(\sigma\e^{\l'-1},0,0)$ is equivalent to $V^\sigma(\l'-1)$. In this
case $(a,b)=(0,0)$ and the representation is called nilpotent.

In the case $(a,b)\in \Ccross\times\set{0}\union
\set{0}\times\Ccross$, the representation $V(\zeta,a,b)$ is called
semicyclic (semiperiodic).

In the case when $(a,b)\in\Ccross\times\Ccross$, the representation
$V(\zeta,a,b)$ is called cyclic (periodic).

The representation $V(\zeta,a,b)$ is fundamental, since it has minimal
dimension $\l'$.

\section{$\Upq(\gl_2)$}

\subsection{}
Let $\Upq(\gl_2)$ be the associative unital $\C(p,q)$-algebra with
generators
\begin{displaymath}
  \set{e,f,k^{\pm1},l^{\pm1}}
\end{displaymath}
that satisfy the relations
\begin{alignat*}{2}
  k\cdot k\inv &= 1 = k\inv k, &
  l\cdot l\inv &= 1 = l\inv \cdot l,\\
  k\cdot e\cdot k\inv &= p\; e, \qquad\qquad&
  k\cdot f\cdot k\inv &= p\inv f,\\
  l\cdot e\cdot l\inv &= q\; e, \qquad\qquad&
  k\cdot f\cdot k\inv &= q\inv f,\\
  e\cdot f- \frac{p}{q} f\cdot e &= \frac{k^2- l^{-2}}{p- q\inv}, &
  k\cdot l &= l\cdot k.
\end{alignat*}

The algebra $\Upq(\gl_2)$ is a Hopf algebra. The coproduct is
\begin{align*}
  \Delta(k) &= k\tensor k,\\
  \Delta(l) &= l\tensor l,\\
  \Delta(e) &= e\tensor k + l\inv\tensor e,\\
  \Delta(f) &= f\tensor k + l\inv\tensor f.
\end{align*}
Then the corresponding antipode map is
\begin{align*}
  S(k) &= k\inv,\\
  S(l) &= l\inv,\\
  S(e) &= -q k\inv l e,\\
  S(f) &= -q\inv k\inv l f.
\end{align*}
The counit is given by
\begin{alignat*}{2}
  \epsilon(k) &= 1,\qquad&
  \epsilon(l) &= 1,\\
  \epsilon(e) &= 0, &
  \epsilon(f) &= 0.
\end{alignat*}

\subsection{}
In the special case $p=q$, define $\Uq(\gl_2):=\Upq(\gl_2)/
\ideal{p-q}$, which is a $1$-parameter deformed quantum enveloping
algebra of $\gl_2$. The two invertible generators $\set{k,l}$ in its
Cartan subalgebra obey the same relations. Therefore the elements
$\set{kl\inv,k\inv l}$ are central in $\Uq(\gl_2)$.

Let $\U_P(\sl_2)$ be the {\em simply connected quantum group\/} of
$\sltwo$ in the sense of~\cite[0.3]{deconcini-kac-procesi:coadj}.
$\U_P(\sl_2)$ is a $\C(q)$-algebra with generators
$\set{e,f,l^{\pm1}}$ satisfying the relations
\begin{alignat*}{2}
  l\cdot e\cdot l\inv &= q\; e, &
  l\cdot f\cdot l\inv &= q\inv f,\\
  e\cdot f- f\cdot e &= \frac{l^2- l^{-2}}{q- q\inv}, \qquad\quad&
  l\cdot l\inv &= 1= l\inv\cdot l.
\end{alignat*}
$\U_P(\sl_2)$ is a quotient algebra of $\Upq(\gl_2)$:
\begin{displaymath}
  \U_P(\sltwo)\isomorphic \Upq(\gl_2)/ \ideal{p-q, k^{\pm1}-
    l^{\pm1}}.
\end{displaymath}

\part{$q$-oscillators}

%
%

\chapter{$q$-oscillators}
\label{chap:q-oscillators}

\section{Introduction}


In this chapter I discuss $q$-oscillator algebras (certain deformations of
the quantum mechanical oscillator algebra) which were brought into the
theory of quantum groups independently by Macfarlane, Hayashi and
Biedenharn in 1989. The $q$-oscillator algebras that I discuss are
not Hopf algebras. Nevertheless they have many intriguing properties
reminiscent of the finite simple quantum group $\Uq(\sltwo)$ discussed
in the last chapter.

In analogy with the quantum mechanical oscillator there is a natural
Fock module representation. Remarkably, for the $q$-oscillator that I
consider, the Fock module is essentially unique: in particular the
vacuum vector is fixed to have ``particle number'' zero.

At a root of unity, the $q$-oscillator algebra has an enlarged centre
and it is possible to construct finite dimensional irreducible
representations --- something which has no classical analogue: 
there exist cyclic, semicyclic~\cite{sun-ge} and nilpotent
representations.

The infinite dimensional Fock module (at generic specialisation) and
its irreducible quotient (at an even root of unity) are unitarisable.

The quantum enveloping algebra $\Uq(\sltwo)$ was first bosonised with
a pair of $q$-oscillators by Macfarlane and Biedenharn. At about the
same time Hayashi gave a bosonisation of each quantum enveloping
algebra corresponding to the Lie algebras of simple type $A$ and $C$,
and of affine type $A^{(1)}$ (and also $q$-spinor realisations of
simple Lie algebras of type $A$, $B$ and $D$). The bosonisation of
$\Uq(\g)$ can be used to construct representations of $\Uq(\g)$ on a
tensor product of $q$-oscillator representations.

I end this chapter with a short description of a 2-parameter
deformation of the oscillator algebra, a quotient of which degenerates
to the $q$-oscillator that I consider.

\section{Definitions}

I begin by recalling the definition of the quantum mechanical
oscillator algebra.

\begin{definition} \label{h4:classical-oscill}
  The {\em oscillator algebra} $\hfour$ (Heisenberg-Weyl algebra) is a
  non-semisimple Lie algebra with generators
  $\set{\bar{n},\bar{a}_+,\bar{a}_-,\bar{e}}$ that satisfy the
  following relations:
  \begin{alignat*}{2}
    \comm{\bar{n},\bar{a}_+} &= \bar{a}_+, \qquad&
    \comm{\bar{n},\bar{a}_-} &= -\bar{a}_-, \\
    \comm{\bar{a}_-,\bar{a}_+} &= \bar{e}, &
    & \text{$\bar{e}$ is central}.
  \end{alignat*}
\end{definition}

Since $\hfour$ is a Lie algebra, its universal enveloping algebra
$\U(\hfour)$ is a Hopf algebra with the usual maps
\begin{displaymath}
  \begin{aligned}
    \Delta(x) &= x\tensor 1 + 1\tensor x, \\
    \epsilon(x) &= 0, \\
    S(x) &= -x,
  \end{aligned}
  \qquad (x\in \set{\bar{a}_+,\bar{a}_-,\bar{n},\bar{e}}).
\end{displaymath}
For the unit element the Hopf maps are as usual: $\Delta(1)=1\tensor 1$,
$\epsilon(1)=1$ and $S(1)=1$.

\subsection{}
Usually physicists work with the quotient algebra $\U'(\hfour):=
\U(\hfour)/ \ideal{\bar{e}-1}$.  In $\U'(\hfour)$ the Heisenberg relation
is then $\comm{\bar{a}_-,\bar{a}_+}=1$. 

The definition of $\hfour$ has been made in terms of the step-up and
step-down operators, which appear also to be more important when the
generalisation to the $q$-oscillator is made.

\begin{definition}
  (compare~\cite{macfarlane:jpa,biedenharn:jpa,hayashi}) The {\em
    $q$-oscillator algebra} $\Uq'(\hfour)$ ($q$-Heisenberg-Weyl
  algebra) is the associative unital $\C(q)$-algebra with generators
  $\set{a_+,a_-,w, w\inv}$ with the relations:
  \begin{alignat*}{2}
    w \cdot w\inv &= 1 = w\inv\cdot w, &&\\ 
    w \cdot a_+ \cdot w\inv &= q\; a_+, \qquad&
    a_-\cdot a_+ - q\; a_+\cdot a_- &= w\inv, \\ 
    w \cdot a_- \cdot w\inv &= q\inv a_-, &
    a_-\cdot a_+ - q\inv a_+\cdot a_- &= w.
  \end{alignat*}
\end{definition}

\subsection{} \label{h4:Uq(h4')-automorphisms}
There are two $\C$-algebra anti-automorphisms $\w_c$ and $\w_r$ of
$\Uq'(\hfour)$ given by
\begin{alignat*}{4}
  \w_c(w) &:= w\inv,\quad&
  \w_c(a_+) &:= a_-, \quad&
  \w_c(a_-) &:= a_+, \quad&
  \w_c(q) &:= q\inv,\\
  \w_r(w) &:= w, &
  \w_r(a_+) &:= a_-, &
  \w_r(a_-) &:= a_+, &
  \w_r(q) &:= q.\\
  \intertext{There is a $\C$-algebra automorphism $\varphi$ of
    $\Uq'(\hfour)$ given by}
  \varphi(w) &:= w\inv, &
  \varphi(a_+) &:= a_+, &
  \varphi(a_-) &:= a_-, &
  \varphi(q) &:= q\inv.
\end{alignat*}

\begin{remark}
  The algebra $\Uq'(\hfour)$ has two different $q$-Heisenberg algebra
  relations. It actually suffices to define deformations of the
  oscillator algebra with only one of these relations and this is done
  by many authors.  But, as will be seen later in this chapter, both
  relations are needed in order to have an involutive
  anti-automorphism at a specialisation equal to a phase (necessary to
  construct a unitary representation of the $q$-oscillator algebra at
  $q$ specialised to an even root of unity) and also for the existence
  of the bosonisation homomorphisms $\Uq(\g)\mapto
  \Uq'(\hfour)^{\tensor n}$.
\end{remark}

\subsection{} \label{h4:Uq(h4')-"casimir"}
The two $q$-Heisenberg relations lead to the following relations in
$\Uq'(\hfour)$:
\begin{align*}
  a_+ \cdot a_- -\frac{w - w\inv}{q- q\inv} &= 0, \\
  a_- \cdot a_+ -\frac{qw - q\inv w\inv}{q - q\inv} &= 0,
\end{align*}
which can be compared to the the central element $\bar{a}_+ \cdot
\bar{a}_- -\bar{n}\cdot \bar{e}$ in $\U(\hfour)$, which acts as zero
on the standard vacuum representation.

\label{h4:Uq'(h4)-basis}
{}From this it follows that the $q$-oscillator algebra $\Uq'(\hfour)$
has the following basis as a vector space over $\C(q)$
\begin{displaymath}
  \Uq'(\hfour)= \sum_{m_-\in\Zplus, m_0\in\Z} \C(q) a_-^{m_-} w^{m_0}
  +\sum_{m_0\in\Z} \C(q) w^{m_0}
  +\sum_{m_+\in\Zplus, m_0\in\Z} \C(q) w^{m_0} a_+^{m_+}.
\end{displaymath}

\subsection{}
Let $\e\in\Ccross$. Let $\UA'(\hfour)$ be the
$\Cfree{q,q\inv}$-subalgebra of $\Uq'(\hfour)$ generated by $a_-$,
$a_+$, $w^{\pm1}$ and $n:=\frac{w^2- w^{-2}}{2(q- q\inv)}$. Note that
$n$ is fixed by the anti-automorphisms $\w_c$ and $\w_r$ and has the
commutation relations:
\begin{align*}
  n\cdot a_+- a_+\cdot n &= \half a_+(q w^2+ q\inv w^{-2}),\\
  a_-\cdot n- n\cdot a_- &= \half (q w^2+ q\inv w^{-2}) a_-.
\end{align*}
Note that the second relation can be obtained from the first by
applying the anti-automorphism $\w_c$ (or $\w_r$).  Define the
specialisation $\Ue'(\hfour)$ of $\UA'(\hfour)$ at $q=\e$ to be
$\Ue'(\hfour):= \UA'(\hfour)/ \ideal{q-\e}$.

\begin{proposition} \label{h4:classical-specialisation}
  The quotient algebra $\U_1'(\hfour)/ \ideal{w-1, w\inv-1}$ is
  $\C$-algebra isomorphic to $\U'(\hfour)$.
\end{proposition}

\begin{proof}
  Clearly in $\U_1'(\hfour)/ \ideal{w-1, w\inv-1}$ the two
  $q$-Heisenberg-Weyl relations degenerate into the usual
  Heisenberg-Weyl relation of $\U'(\hfour)$: $\comm{a_-, a_+}=1$.  From
  the commutation relation between $n$ and $a_\pm$, it follows that in
  the quotient algebra $\comm{n,a_+}=a_+$ and $\comm{a_-,n}=a_-$. The
  proposition is proved.
\end{proof}


\section{Other $q$-deformed oscillator algebras}

\subsection{} \label{h4:q-oscill-with-1-HeisenbergW-rel}
Clearly having two Heisenberg-Weyl relations is more restrictive than
just one. Consider the deformed oscillator algebra $\cs{A}_4$ with
generators $\set{A_+,A_-,W^{\pm1}}$ that satisfy the following
relations:
\begin{alignat*}{2}
  W \cdot W\inv &= 1 = W\inv\cdot W, &&\\
  W \cdot A_+ \cdot W\inv &= q A_+, \qquad&
  A_-\cdot A_+ - q^2 A_+\cdot A_- &= 1, \\
  W \cdot A_- \cdot W\inv &= q\inv A_-. &&
\end{alignat*}
Note that $\cs{A}_4$ has only one $q$-Heisenberg-Weyl type relation.

\begin{lemma}
  Let $n\in\Z$. The algebra $\cs{A}_4$ is $\Cfree{q,q\inv}$-algebra
  isomorphic to the algebra with the generators
  $\set{A'_+,A'_-,{W'}^{\pm1}}$ and the relations
  \begin{alignat*}{2}
    W' \cdot {W'}\inv &= 1 = {W'}\inv\cdot W', &&\\
    W' \cdot A'_+ \cdot {W'}\inv &= q A'_+, \qquad&
    A'_-\cdot A'_+ - q^{2-n} A'_+\cdot A'_- &= {W'}^{-n}, \\
    W' \cdot A'_- \cdot {W'}\inv &= q\inv A'_-. &&
  \end{alignat*}
\end{lemma}

\begin{proof}
  The isomorphism is given by the following map 
  \begin{displaymath}
    W\mapsto W',\qquad A_- \mapsto {W'}^n A'_-, \qquad A_+ \mapsto A'_+.
  \end{displaymath}
\end{proof}

\subsection{} \label{h4:CGST-oscillator}
Celeghini, Giachetti, Sorace and
Tarlini~\cite{celeghini-giachetti-sorace-tarlini:heisenberg}
introduced a $\C(q)$-algebra $\Uq(\hfour)$ with generators $\set{b_+,
  b_-, m, c, c\inv}$ that satisfy
\begin{displaymath}
  \begin{gathered}
    \text{$c$ is central},\\
    c\cdot c\inv = 1 = c\inv\cdot c,
  \end{gathered}
  \qquad
  \begin{aligned}
    b_-\cdot b_+ - b_+\cdot b_- &= \frac{c- c\inv}{q- q\inv},\\
    \comm{m, b_\pm} &= \pm b_\pm.
  \end{aligned}
\end{displaymath}
The algebra is a Hopf algebra:
\begin{alignat*}{2}
  \Delta(b_+) &= b_+\tensor 1+ c\tensor b_+,\qquad &
  S(b_+) &= -c\inv b_+,\\
  \Delta(b_-) &= b_-\tensor c\inv+ 1\tensor b_-, &
  S(b_-) &= -c b_-,\\
  \Delta(c) &= c\tensor c, &
  S(c) &= c\inv,\\
  \Delta(m) &= m\tensor 1+ 1\tensor m, &
  S(m) &= -m,\\
  \epsilon(b_+) &= 0, &
  \epsilon(b_-) &= 0,\\
  \epsilon(c) &= 1, &
  \epsilon(m) &= 0.
\end{alignat*}
This algebra has a nontrivial Hopf algebra structure and it is
known~\cite{gomez-sierra:oscillator} to be related to link invariants.

\begin{lemma} \label{h4:CGST-oscill-trivial}
  Let $\UA(\hfour)$ be the $\Cfree{q,q\inv}$-subalgebra of
  $\Uq(\hfour)$ generated by
  \begin{displaymath}
    \set{b_+, b_-, m, c^{\pm}, e:=\frac{c- c\inv}{q- q\inv}}
  \end{displaymath}
  (so that $\comm{b_-,b_+}=e$). Let $\e\in\Ccross$ and define
  $\Ue(\hfour):= \UA(\hfour)/ \ideal{q- \e}$.  The following map is a
  $\C$-algebra isomorphism $\Ue(\hfour)/ \ideal{c^{\pm1}- 1}\mapto
  \U(\hfour)$
  \begin{alignat*}{2}
    b_+ &\mapsto \bar{a}_+, \qquad&
    b_- &\mapsto \bar{a}_-,\\
    m &\mapsto \bar{n}, &
    e &\mapsto \bar{e}.
  \end{alignat*}
\end{lemma}

{}From the lemma it can be deduced that the basic representations of
$\Ue(\hfour)$ are equivalent ({\em even\/} at $\e$ a root of unity) to
those of $\U(\hfour)$. Therefore I will not discuss this algebra
further in this chapter.

\subsection{}
Schwenk and Wess have studied~\cite{schwenk-wess:q-oscill} another
interesting deformation of the
Heisenberg-Weyl algebra~\cite{manin:quantum-deRham} with generators
$\set{x,p}$ and relation
\begin{displaymath}
  p\cdot x- q\; x\cdot p= -i.
\end{displaymath}

\begin{remark}
  When considering the Hopf algebra structure of $\hfour$
  ($\Uq(\hfour)$), the generator $\bar{e}$ ($c$) plays an important
  role. The quotient algebra $\U'(\hfour)$ cannot enjoy a Hopf
  algebra structure because of the relation $\bar{e}=1$ (see for
  instance~\cite{palev:oscill-coproduct}):
  \begin{align*}
    \Delta(\bar{e}) &= \bar{e}\tensor 1 + 1\tensor \bar{e} \equiv
    2(1\tensor1)\\ 
    &\neq \Delta(1) = 1\tensor 1.
  \end{align*}
  It follows that $\Uq'(\hfour)$ does not have a canonical coproduct
  (with a well defined classical limit).
\end{remark}

\section{A Fock Representation of $\Uq'(\hfour)$}
\label{h4:Uq(h4)-Fock-module}

\begin{lemma} \label{h4:Uq(h4')-relations}
  Let $n\in\Zplus$. The follow identities hold in $\Uq'(\hfour)$:
\begin{align*}
  a_- a_+^n &\equiv q^{\pm n} a_+^n a_- + \brackets{n}_q
  a_-^{n-1}w^{\mp1},\\ 
  a_-^n a_+ &\equiv q^{\pm n} a_+ a_-^n + \brackets{n}_q
  a_+^{n-1} w^{\mp1}.
\end{align*}
\end{lemma}

\subsection{}
Let $F$ be a $\C(q)$-vector space with basis $\set{u_n\mid n\in\N}$
\begin{displaymath}
  F:=\sum_{n\in\N} \C(q) u_n.
\end{displaymath}
The following action of $\Uq'(\hfour)$ on $F$, makes $F$ a
$\Uq'(\hfour)$-module
\begin{align*}
  a_+ \cdot u_n &:= u_{n+1},\\
  w \cdot u_n &:= q^n u_n, \\
  a_- \cdot u_n &:= \brackets{n}_q u_{n-1}.
\end{align*}
For $n\in\Zminus$, I set $u_n=0$.

The vector $u_0$ is a vacuum vector (lowest weight vector)
\begin{align*}
  a_- \cdot u_0 &= 0, \\
  w \cdot u_0 &= u_0,\\
  F &= \Uq'(\hfour) \cdot u_0.
\end{align*}
Note that the weight ($w$-eigenvalue) of $u_0$ is essentially fixed
(up to a factor of $-1$) by the two $q$-Heisenberg-Weyl relations.

\begin{remark}
  Consider for a moment the construction of a Fock module $\tilde{F}$
  over the deformed oscillator algebra $\cs{A}_4$ defined
  in~\ref{h4:q-oscill-with-1-HeisenbergW-rel}. The weight of the
  vacuum vector $v_0$ of $\tilde{F}$ is not fixed by the single
  $q$-Heisenberg-Weyl relation. In fact there are many different
  vacuum states that can be defined: $w\cdot v_0 = q^{n_0} v_0$
  ($n_0\in\Z$).
\end{remark}

\begin{proposition}
  The $\Uq'(\hfour)$-module $F$ is irreducible.
\end{proposition}

\begin{proof}
  There is not a (singular) vector $u'$ in $F$, such that $a_-\cdot
  u'=0$, except $u'= u_0$.
\end{proof}


\section{The Centre of $\Uq'(\hfour)$ at a root of unity}

Let $\l$ be a positive integer, such that $\l>2$. Fix $\e$ to be a
primitive $\l$-th root of unity (then $\e^\l=1$).

\begin{notation}
  Define $\l'$ to be
  \begin{displaymath}
    \l':=
    \begin{cases}
      \l & \text{if $\l$ is odd},\\
      \frac{\l}{2} & \text{if $\l$ is even.}
    \end{cases}
  \end{displaymath}
\end{notation}


\begin{proposition}
  At the root of unity $\e$ the following elements lie in the centre
  of $\Ue'(\hfour)$
  \begin{displaymath}
    \set{a_+^\l, a_-^\l, w^\l, w^{-\l}}.
  \end{displaymath}
\end{proposition}

\begin{proof}
  The proof of the proposition is straightforward using the defining
  relations of $\Uq'(\hfour)$ and lemma~\ref{h4:Uq(h4')-relations}.
\end{proof}

At the root of unity $\e$ the algebra $\Ue'(\hfour)$ is finite
dimensional (finitely generated as a free module) over its centre.

\section{The Fock module at a root of unity}

Let $F_\A$ be the $\UA'(\hfour)$-submodule of $F$ over
$\Cfree{q,q\inv}$.  Define the specialisation $F_\e$ of the
$\UA'(\hfour)$-module $F_\A$ at $q=\e$, to be the
$\Ue'(\hfour)$-submodule of $F_\A$ over $\C$.

I consider now the Fock module $F_\e$ over $\Ue'(\hfour)$ at the root
of unity $\e$.

\begin{proposition}
  The $\Ue'(\hfour)$-module $F_\e$ at the root of unity $\e$ is
  reducible and has a finite number of weight spaces.
\end{proposition}

\begin{proof}
  The elements in $\set{u_{k\l'} \mid k\in\Zplus}$ are all singular
  vectors in $F_\e$, since 
  \begin{displaymath}
    a_-\cdot u_{k\l'} = \brackets{k\l'}_\e u_{k\l'-1} \equiv 0.
  \end{displaymath}
  Therefore $F_\e$ is reducible. Consider now the action of $w$ on
  $F_\e$
  \begin{displaymath}
    w\cdot u_n = \e^n u_n.
  \end{displaymath}
  The possible weights of $w$ are then
  $\set{1,\e,\e^2,\ldots,\e^{\l-1}}$. So there are only $\l$ weight
  spaces: $F_\e=\directsum_{n=0}^{\l-1} F_{(n)}$ ($F_{(n)}:=\set{v\in
    F_\e\mid w\cdot v= \e^n v}$).
\end{proof}

\subsection{} \label{Uq(h4')-sub&quotient-modules}
Let $k\in\N$. Denote by $F'_{k\l'}$ the submodule generated by the
singular vector $u_{k\l'}$. Note that $F'_0\equiv F_\e$. Denote by
$L_\e$ the irreducible quotient module $F_\e/ F'_{\l'}$.

\begin{lemma} \label{h4:quotient-modules}
  Let $m,n\in\N$, such that $m<n$.
  \begin{enumerate}
  \item \label{h4:quotient-reducible} The quotient module
    $F'_{m\l'}/F'_{n\l'}$ is irreducible if and only if $n=m+1$.
  \item \label{h4:quotient-irred} The quotient module
    $F'_{m\l'}/F'_{(m+1)\l'}$ is equivalent to $L_\e$.
  \end{enumerate}
\end{lemma}

\begin{proof}
  \eqref{h4:quotient-reducible}~If $n> m+1$ then clearly the quotient
  module contains (the image of) the singular vector $u_{(m+1)\l'}$,
  so it is reducible. On the other hand if $n=m+1$ then $a_+$ and
  $a_-$ act transitively and the module is irreducible and has
  dimension $\l'$. \eqref{h4:quotient-irred}~is obvious since
  $\brackets{m+k\l'}_\e= (-1)^{\frac{k\l}{\l'}} \brackets{m}_\e$.
\end{proof}

\section{Semicyclic Representations}

\subsection{}
Let $(\lambda,\mu)\in \C\times\set{0}\union \set{0}\times\C$. Let
$V_{\lambda,\mu}:= \sum_{n=0}^{\l-1} \C v_n$ be a $\C$-vector space.
At the root of unity $\e$, define the following $\Ue'(\hfour)$-module
structure on $V_{\lambda,\mu}$
\begin{align*}
  a_+\cdot v_n &:=
  \begin{cases}
    v_{n+1} & \text{if $n\in\range{0,\l-2}$}\\ 
    \lambda v_0 & \text{if $n=\l-1$}
  \end{cases} \\
  a_-\cdot v_n &:=
  \begin{cases}
    \brackets{n}_\e v_{n-1} & \text{if $n\in\range{1,\l-1}$}\\ 
    \mu v_{\l-1} & \text{if $n=0$}
  \end{cases} \\
  w \cdot v_n &:= \e^n v_n.
\end{align*}
Note that $V_{0,0}$ is equivalent to the quotient $F_\e/F'_\l$.

\begin{lemma}[odd case] \label{h4:odd-lem}
  Let $\l$ be odd.
  \begin{enumerate}
  \item \label{h4:odd-irrep} $V_{\lambda,\mu}$ is an irreducible
    $\Ue'(\hfour)$-module.
  \item \label{h4:odd-semicyclic} If $(\lambda,\mu)\in
    \Ccross\times\set{0}\union \set{0}\times\Ccross$, then
    $V_{\lambda,\mu}$ is a semicyclic (semiperiodic)
    $\Ue'(\hfour)$-module.
  \item \label{h4:odd-nilpotent} $V_{0,0}$ is a nilpotent
    $\Ue'(\hfour)$-module.
  \end{enumerate}
\end{lemma}

\begin{proof}
  \eqref{h4:odd-irrep} At an odd root of unity, $V_{\lambda,\mu}$ does
  not contain any $\Ue'(\hfour)$-submodules. Therefore it is
  irreducible. \eqref{h4:odd-semicyclic} If $\lambda$ (respectively
  $\mu$) is zero, then $a_+$ ($a_-$) acts nilpotently and $a_-$
  ($a_+$) injectively.  So the module is semicyclic: $a_+^\l\cdot v_n=
  \lambda v_n$ and $a_-^\l\cdot v_n= \mu \brackets{\l-1}_\e! v_n$
  ($n\in\range{0,\l-1}$). \eqref{h4:odd-nilpotent}~From
  $(\lambda,\mu)=(0,0)$, it is clear that $V_{0,0}$ is nilpotent.
\end{proof}

\begin{lemma}[even case] \label{h4:Ue(h4')-even-semicyclic}
  Let $\l$ be even.
  \begin{enumerate}
  \item \label{h4:even-nilpotent}
    $a_-$ acts nilpotently in $V_{\lambda,\mu}$.
  \item \label{h4:even-reducible} $V_{0,0}$ is a reducible nilpotent
    $\Ue'(\hfour)$-module. The irreducible quotient of $V_{0,0}$ is
    equivalent as a $\Ue'(\hfour)$-module to $L_\e$.
  \item \label{h4:even-new-nilpotent} Let
    $(\lambda,\mu)\in\set{0}\times\Ccross$. $V_{0,\mu}$ is an
    reducible nilpotent $\Ue'(\hfour)$-module.
  \item \label{h4:even-semicyclic} Let
    $(\lambda,\mu)\in\Ccross\times\set{0}$. $V_{\lambda,0}$ is an
    irreducible semicyclic $\Ue'(\hfour)$-module.
  \end{enumerate}
\end{lemma}

\begin{proof}
  \eqref{h4:even-nilpotent}~For even $\l$, $\brackets{\l'}_\e=0$ and
  $a_-$ acts nilpotently, annihilating the vector $v_{\l'}$.
  \eqref{h4:even-reducible}~$(\lambda,\mu)=(0,0)$, so $V_{0,0}$ is
  nilpotent. $V_{0,0}$ contains a submodule generated by $v_{\l'}$.
  Hence it is reducible.  \eqref{h4:even-new-nilpotent}~$a_+$ acts
  nilpotently since $\lambda=0$ and $a_-$ acts nilpotently
  by~\eqref{h4:even-nilpotent}.  $V_{0,\mu}$ is generated by $v_0$ and
  contains a submodule generated by $v_{\l'}$.
  \eqref{h4:even-semicyclic}~follows since $a_+$ acts cyclically,
  whereas $a_-$ acts nilpotently by~\eqref{h4:even-nilpotent}.
\end{proof}

\begin{remark}
  It can be checked that the space of the parameters $\lambda$ and
  $\mu$ of the module $V_{\lambda,\mu}$ cannot be extended to
  $\C\times \C$, since an action of $\Ue'(\hfour)$ on $V_{\mu,\nu}$
  with $(\lambda,\mu)\in \Ccross\times\Ccross$ does not form a
  consistent (cyclic) representation.
\end{remark}

\section{Cyclic Representations}
\label{h4:sec-cyclic-reps}

\begin{proposition}
  Let $\l$ be even. There does not exist an $\l$ (or $\l'$)
  dimensional (minimal) cyclic (periodic) representation of
  $\Ue'(\hfour)$.
\end{proposition}

\begin{proof}
  There is no $\l'$-dimensional cyclic representation, since
  $a_+^{\l'}$ and $a_-^{\l'}$ are not central in $\Ue'(\hfour)$. In
  fact from~\ref{h4:Uq(h4')-relations} it is clear that $a_\pm^{\l'}$
  anti-commutes with $a_\mp$, $w$ and $w\inv$. So it is not possible
  to have a cyclic representation in this case. There is no
  $\l$-dimensional cyclic representation, since
  by~\ref{h4:Ue(h4')-even-semicyclic}\eqref{h4:even-nilpotent} the
  action of $a_-$ would be nilpotent.
\end{proof}

\begin{proposition}
  Let $(\lambda,\mu,\zeta)\in \C\times\C\times\Ccross$. Consider the
  left quotient
  \begin{displaymath}
    M_{\lambda,\mu,\zeta}:= \Ue'(\hfour)/ \ideal{a_+^\l-
    \lambda, a_-^\l- \mu, w- \zeta}_L.
  \end{displaymath}
  It is a $\Ue'(\hfour)$-module of dimension $2\l-1$. Let $v$ denote
  the image of $1$ in $M_{\lambda,\mu,\zeta}$. For odd $\l$ and
  $(\lambda,\mu)\in \Ccross\times\Ccross$, it is an
  irreducible cyclic (periodic) $\Ue'(\hfour)$-module. For $\l$ even,
  if $\zeta\not\in\set{1,-1}$ then $M_{\lambda,\mu,\zeta}$ is an
  irreducible cyclic $\Ue'(\hfour)$-module,
\end{proposition}

\begin{proof}
  The dimension of the module follows from~\ref{h4:Uq'(h4)-basis}.
  For odd $\l$, the module is cyclic and irreducible, since
  by~\ref{h4:Uq(h4')-relations} $a_+$ and $a_-$ act injectively. For
  even $\l$, $a_+$ and $a_-$ also act injectively, unless $\zeta\in
  \set{1,-1}$ in which case the module is nilpotent and contains a
  submodule generated by $a_+^{\l'} v$.
\end{proof}

\begin{remark}
  Most of the statements that have been made on the centre and
  representations of $\Ue'(\hfour)$ at the root of unity $\e$ are also
  true for the algebra $\cs{A}_4$ at the root of unity $\e$: the
  elements $\set{A_+^\l, A_-^\l, W^\l, W^{-\l}}$ are central, there
  exists a similar set of representations and so on (see also the
  following chapter on this point).
\end{remark}


\section{Unitary Representations}

\subsection{}
Let $F$ be the $\C(q)$-Fock module of $\Uq'(\hfour)$ defined
in~\ref{h4:Uq(h4)-Fock-module}. Let $\w$ be one of the involutive
($\w^2=\id$) $\C(q)$-algebra anti-automorphisms of $\Uq'(\hfour)$
defined in~\ref{h4:Uq(h4')-automorphisms} and require that
$\w(\alpha):= \alpha^*$ ($\alpha\in\C$). Define on $F$ a scalar
product $(\cdot,\cdot): \Uq'(\hfour)\times \Uq'(\hfour)\mapto \C(q)$,
contravariant with respect to $\w$, such that
\begin{align*}
  (u_m,u_n) &:= \delta_{m,n} \brackets{m}_q! \qquad (m,n\in\N),\\
  (x\cdot v,w) &= (v,\w(x)\cdot w) \qquad (x\in\Uq'(\hfour); v,w\in F).
\end{align*}
Denote also by $(\cdot,\cdot)$ the corresponding induced scalar
products on $F_\A$ and $F_\e$.

\begin{lemma}
  The triple $\structure{F,(\cdot,\cdot),\w}$ is a
  $\star$-representation of $\Uq'(\hfour)$.
\end{lemma}

\begin{proposition}
  Let $\w_r$ be the anti-automorphism of $\Ue'(\hfour)$ defined
  in~\ref{h4:Uq(h4')-automorphisms} (with $\w_r(\alpha):= \alpha^*$
  ($\alpha\in\C$)). If $\e\in \Rplus$, then the
  $\star$-representation $\structure{F_\e,(\cdot,\cdot),\w_r}$ of
  $\Ue'(\hfour)$ is unitary.
\end{proposition}

\begin{proof}
  When $\e$ is real and positive, then $\brackets{n}_\e\in \Rplus$ (for
  all $n\in\Zplus$). Therefore in this case the scalar product is
  positive definite and the $\star$-representation is unitary.
\end{proof}

\begin{lemma}
  Let $\e\in\Ccross$ be such that $\modulus{\e}=1$. Then the triple
  $\structure{F_\e, (\cdot,\cdot), \w_c}$ is a $\star$-representation
  of $\Ue'(\hfour)$.
\end{lemma}

\begin{proposition} \label{h4:unitary-rep-root-of-1}
  Let $\l$ be an {\em even\/} positive integer ($\l\geq4$). Let
  $\l':=\frac{\l}{2}$ and let $\e$ be a primitive $\l$-th root of unity.
  Consider the irreducible quotient $\Ue'(\hfour)$-module $L_\e$ at the
  root of unity $\e$, defined in~\ref{Uq(h4')-sub&quotient-modules}
  and denote again by $(\cdot,\cdot):L_\e\times L_\e\mapto \C$ the
  induced scalar product on $L_\e$.  The $\star$-representation
  $\structure{L_\e, (\cdot,\cdot), \w_c}$ is a unitary representation
  of $\Ue'(\hfour)$.
\end{proposition}

\begin{proof}
  The scalar product on $L_\e$ gives $(u_m,u_m)= \brackets{m}_\e!$.
  Since $\brackets{m}_\e=\frac{sin(\frac{m\pi
      i}{\l'})}{sin(\frac{i\pi}{\l'})}$, $\brackets{m}_\e>0$ for $m\in
  \range{1,\l'-1}$. Therefore $\brackets{m}_\e!>0$ ($m\in
  \range{1,\l'-1}$) and the scalar product on $L_\e$ is positive
  definite.
\end{proof}

\begin{remark}
  There does not exist a unitary representation at an odd root of
  unity, because $\brackets{m}_\e<0$ for
  $m\in\range{\frac{\l+1}{2},\l-1}$, so the scalar product is not
  positive definite.
\end{remark}

\section{Bosonisation of $\Uq(\sln)$}

The simple quantum groups corresponding to Cartan matrices of type
$A_n$ and $C_n$ can been bosonised~\cite{hayashi} with
$\Uq'(\hfour)$. Here I present only the bosonisation of $\Uq(\sln)$.

\begin{notation}
  For $x,y\in \Uq'(\hfour)$, define
  \begin{displaymath}
    x\overset{i}{\tensor} y:=
    \underbrace{1\tensor\cdots \tensor 1}_{\text{$i-1$ times}}\tensor
    x\tensor y\tensor 1\cdots 1\in \Uq'(\hfour)^{\tensor n}.
  \end{displaymath}
\end{notation}

\medskip

\begin{theorem}~\cite[3.2]{hayashi}
  The following map $\Uq(\sln)\mapto \Uq'(\hfour)^{\tensor n}$ is an
  $\C(q)$-algebra homomorphism
  \begin{align*}
    e_i &\mapsto a_-\overset{i}{\tensor} a_+,\\
    f_i &\mapsto a_+\overset{i}{\tensor} a_-,\\
    k_i &\mapsto w\inv \overset{i}{\tensor} w.
  \end{align*}
\end{theorem}

\begin{proof}
  The homomorphism is easily verified on the relations $k_ie_jk_i\inv=
  q^{a_{ij}}e_j$ and $k_if_jk_i\inv= q^{-a_{ij}}f_j$. Similarly the
  map is checked on the relation $\comm{e_i,f_j}= \delta_{ij}\frac{k_i-
    k_i\inv}{q- q\inv}$ using the relations
  in~\ref{h4:Uq(h4')-"casimir"}. The Serre relations require a little
  more work.
\end{proof}

\subsection{}
Using the theorem, an infinite dimension (unitary) representation of
$\Uq(\sln)$ can be constructed on $F^{\tensor n}$. At the root of
unity $\e$, a finite dimensional (unitary) representation of
$\Ue(\sln)$ can be constructed on $L_\e^{\tensor n}$ and semicyclic
(cyclic) representations of $\Ue(\sln)$ can be constructed on
$V_{\lambda,\mu}^{\tensor n}$ ($M_{\lambda,\mu}^{\tensor n}$
respectively). Interestingly a mixture of different representations
can also be taken.

\section{2-parameter deformed $pq$-oscillator}

Consider the associative unital $\C(p,q)$-algebra $\Upq'(\hfour)$ with
generators
\begin{displaymath}
  \set{a_+, a_-, w^{\pm1}, x^{\pm1}},
\end{displaymath}
that satisfy the relations
\begin{alignat*}{2}
  w \cdot w\inv &= 1 = w\inv\cdot w, &
  x \cdot x\inv &= 1 = x\inv\cdot x,\\ 
  w \cdot a_+ \cdot w\inv &= p a_+, \qquad&
  w \cdot a_- \cdot w\inv &= p\inv a_-,\\
  x \cdot a_+ \cdot x\inv &= q a_+, \qquad&
  x \cdot a_- \cdot x\inv &= q\inv a_-,\\
  a_-\cdot a_+ - p a_+\cdot a_- &= x\inv,&
  a_-\cdot a_+ - q\inv a_+\cdot a_- &= w.
\end{alignat*}
It is a two parameter deformation of $\U'(\hfour)$.

\begin{lemma}
  The following  map $\Upq'(\hfour)\mapto \Upq'(\hfour)$ is an
  anti-automorphism of $\Upq'(\hfour)$
  \begin{alignat*}{3}
    a_+ &\mapsto a_-, \qquad&
    a_- &\mapsto a_+, \qquad&
    w &\mapsto x\inv,\\
    x &\mapsto w\inv, &
    p &\mapsto q\inv, &
    q &\mapsto p\inv.
  \end{alignat*}
\end{lemma}

\subsection{}
The quotient algebra $\Upq'(\hfour)/ \ideal{p-q, w- x}$ is
isomorphic to $\Uq'(\hfour)$.

\chapter[$qr$-oscillator]{A Quadratic 2-parameter deformation of\\
the Oscillator Algebra}
\label{chap:qr-oscillator}
\section{Introduction}

In this chapter I consider a new 2-parameter
deformation~\cite{petersen:qr-oscillator} of the oscillator algebra
$\U(\hfour)$, with a natural $q$-Heisenberg-Weyl subalgebra.

\subsection{Quadratic deformations}
In the theory of (classical) differential geometry of a Lie group $G$,
its Lie algebra $\g$ appears naturally as the set of left invariant
vector fields $\cs{L}(G)$ on $G$ with a commutator operation.  There
is an elegant theory of noncommutative differential geometry on a
matrix
quantum group $\funq(G)$ (see~\cite{woronowicz:cmp2,manin:cmp,%
  wess-zumino:npbproc}), from which it turns out that the quantum Lie
algebra that arises from the set of invariant vector fields on
$\funq(G)$, is not the quantum enveloping algebra $\Uq(\g)$, but
rather a quadratic Hopf algebra $\cs{L}_q(G)$. The quantum calculus of
the simplest case, the compact quantum group $\funq(SU_2)$, was first
studied by Woronowicz~\cite{woronowicz:rims} and he found the
quadratic quantum Lie algebra $\cs{L}_q(SU_2)$. (Sklyanin had already
introduced a quadratic deformation of $\U(\sltwo)$
in~\cite{sklyanin:uspekhi}: though it was not a Hopf algebra.) Later a
deformation of $\U(\sltwo)$ similar to the one found by Woronowicz was
uncovered~\cite{witten:vertex-models} in the context of vertex models.
Fairlie~\cite{fairlie:su2} and Curtright and
Zachos~\cite{curtright-zachos} generalised this algebra to a
2-parameter deformation $\Uqr(\sltwo)$ of $\U(\sltwo)$.

\subsection{}
An interesting problem is the study of twisted deformations of
multidimensional oscillator algebras. Deformed multi-oscillators,
covariant under the action of a quantum group, have been studied by
Pusz and Woronowicz~\cite{pusz-woronowicz:twisted-SU(n)-oscill,%
  pusz:twisted-fermionic-oscill} and Zumino~\cite{zumino:mpla}.
Twisted multiparameter deformed multi-oscillators have been studied
in~\cite{fairlie-zachos:multiparameter-oscill,%
  fairlie-nuyts:multiparam-oscill}.
Another challenge is the construction of $q$-quantum mechanical
systems~\cite{schwenk-wess:q-oscill,chodos-caldi}.

\subsection{}
I describe the contents of this chapter.  Following an idea I got from
the 2-parameter quadratic deformation $\Uqr(\sltwo)$ of $\U(\sltwo)$
that I mentioned above, I introduce a 2-parameter deformation
$\Uqr(\hfour)$ of the oscillator algebra $\U(\hfour)$ and its one
parameter deformed Heisenberg-Weyl subalgebra $\Ur(\hthree)$.  It has
a family of Fock modules parametrised by the $q$-number operator
eigenvalue of the vacuum vector and the central charge of the central
element.  It is possible to unitarise these Fock modules over
$\Uqr(\hfour)$ when the parameters are specialised to positive real
numbers.

At an $\l$-th root of unity the specialisation $\Uef(\hfour)$ of
$\Uqr(\hfour)$ has an enlarged centre: the step-up and step-down
operators to the $\l'$-th power are in the centre of $\Uef(\hfour)$.
There is a finite dimensional irreducible quotient of the Fock module,
which may have some connection with paragrassmann
algebras~\cite{filippov-isaev-kurdikov}.  There also exist semicyclic
and cyclic modules.

One of the important applications of $q$-oscillators, mentioned in the
last chapter, is the bosonisation of other quadratic algebras and
quantum groups. The $q$-oscillator discussed in the previous chapter
was ideally suited for bosonising quantum enveloping algebras. The
1-parameter quadratic deformation of $\U(\sltwo)$ can be bosonised by
the quadratic $r$-Heisenberg algebra $\Ur(\hthree)$.  It is also
possible to bosonise the centreless $q$-Virasoro ($q$-Witt) algebra
with an extension of $\Ur(\hthree)$. From this realisation at a root
of unity, some finite dimensional representations of the $q$-Witt algebra can
be constructed (compare~\cite{narganes}). Kassel has
found~\cite{kassel:q-virasoro} a full $q$-analogue of the Virasoro
algebra with central extension.

In~\cite{celeghini-giachetti-sorace-tarlini:contraction}
Wigner-I\"onu contractions of quantum groups were first
performed. The quantum group $\Uq(\sltwo)$ (defined
in~\ref{quea:Uq(sl2)-def}) was contracted to Heisenberg and Euclidean
quantum groups.  A contraction of $\Uqr(\sltwo)$, at the
specialisation $q=1$, leads to $\Ur(\hthree)$. Another contraction gives a
2-parameter quadratic deformed Euclidean algebra.

The quadratic algebra $\Ur(\hthree)$ is covariant under left and right
coactions of $\fun_{r^2}(\SL_2)$.

\section{A Quadratic deformation of $\sltwo$}
\label{qr:quad-sltwo}

\begin{notation}
  Define the $q$-commutator $\comm{X,Y}_q:= q X\cdot Y- q\inv Y\cdot
  X$.
\end{notation}

\begin{definition}~\cite{fairlie:su2,curtright-zachos}
  Define $\Uqr(\sltwo)$ to be the associative unital algebra over
  $\Cfree{q,q\inv,r,r\inv}$ with generators $\set{W_+,W_-,W_0}$ that
  satisfy the following relations
  \begin{align*}
    \comm{W_0,W_+}_q &= W_+,\\ 
    \comm{W_-,W_0}_q &= W_-,\\ 
    \comm{W_+,W_-}_{r\inv} &= W_0.
  \end{align*}
\end{definition}

\subsection{}
The algebra $\Uqr(\sltwo)$ is an example of what is called a quadratic
algebra: it has defining relations that are quadratic in its
generators. It seems that $\Uqr(\sltwo)$ is only a Hopf algebra for
certain values of $q$ and $r$: specifically when $q=r^2$ or $r=q^2$
(the latter corresponding to the quantum algebra
of~\cite{witten:vertex-models}).  This agrees with results in the
literature that the number of deformation parameters of a Hopf algebra
deformation of $\U(\g)$ ($\g$ a simple Lie algebra of rank $r$) is
bounded by $r$.

\section{Definition of $\Uqr(\hfour)$}

\subsection{} 
Denote by $\Uqr(\hfour)$ the associative unital
$\Cfree{q,q\inv,r,r\inv}$-algebra with generators $\set{A_+, A_-, N,
  E}$ that satisfy the relations
\begin{align*}
  \text{$E$ is central},\\
  \comm{N,A_+}_q &= A_+,\\
  \comm{A_-,N}_q &= A_-,\\
  \comm{A_-,A_+}_r &= E.
\end{align*}
It is easy to check that the quantum algebra $\Uqr(\hfour)$ is well
defined by these relations. In particular it does not matter in which
way a cubic monomial $X_1X_2X_3$ in the generators of the quantum
algebra is re-ordered: the result should be the same answer either
way. For example in $\Uqr(\hfour)$ the re-orderings
\begin{align*}
  N A_+ A_- &= r^{2} N A_- A_+ -rEN \\
  &= q^2 r^2 A_- N A_+ - qr^2 A_-A_+ - rEN \\
  &= r^2 A_- A_+ N - rEN\\
  \intertext{and}
  N A_+ A_- &= q^{-2} A_+ N A_- + q^{-1} A_+ A_- \\
  &= A_+ A_- N \\
  &= r^2 A_- A_+ N - rEN
\end{align*}
give the same expression.

\subsection{}
Let $\e,\f\in \Ccross$. Define the partial specialisation
$\Uqf(\hfour)$ of $\Uqr(\hfour)$ at $r=\f$ to be the
$\Cfree{q,q\inv}$-algebra $\Uqf(\hfour):=\Uqr(\hfour)/ \ideal{r- \f}$.
The full specialisation $\Uef(\hfour)$ of $\Uqr(\hfour)$ at $(q,r)=
(\e,\f)$ is defined to be the $\C$-algebra $\Uef(\hfour):=
\Uqr(\hfour)/ \ideal{q-\e, r-\f}$.

\begin{lemma}
  The specialisation $\U_{1,1}(\hfour)$ of $\Uqr(\hfour)$ is
  isomorphic to the universal enveloping algebra $\U(\hfour)$ defined
  in~\ref{h4:classical-oscill}. The quotient $\U_{1,1}(\hfour)
  /\ideal{E-1}$ is isomorphic to $\U'(\hfour)$.
\end{lemma}

\subsection{}
The algebra $\Uqr(\hfour)$ has a subalgebra $\Ur(\hthree)$ generated
by $\set{A_+, A_-, E}$. I call it the $r$-Heisenberg-Weyl subalgebra
of the $qr$-oscillator algebra $\Uqr(\hfour)$. Let $\f\in \Ccross$. I
define the specialisation $\Uf(\hthree)$ of $\Ur(\hthree)$ at $r=\f$
to be $\Uf(\hthree):= \Ur(\hthree)/ \ideal{r-\f}$. The specialisation
at $r=1$ is isomorphic to the subalgebra $\U(\hthree)$ of $\U(\hfour)$
generated by $\set{\bar{a}_+, \bar{a}_-, \bar{e}}$.

\begin{lemma}
  Let $\Ur'(\hthree):= \Ur(\hthree)/ \ideal{E-1}$. Let $\cs{A}_3$ be
  the $\Cfree{q,q\inv}$-algebra (the Heisenberg-Weyl subalgebra of
  $\cs{A}_4$ defined in~\ref{h4:q-oscill-with-1-HeisenbergW-rel}) with
  generators $\set{c_+,c_-}$ and the relation
  \begin{displaymath}
    c_-\cdot c_+ -q^2 c_+\cdot c_-= 1.
  \end{displaymath}
  $\Ur'(\hthree)$ is $\C$-algebra isomorphic to $\cs{A}_3$.
\end{lemma}

\begin{proof}
  The isomorphism is given by
  \begin{displaymath}
    A_- \mapsto q c_-, \qquad
    A_+ \mapsto c_+, \qquad
    r \mapsto q\inv.
  \end{displaymath}
\end{proof}

\subsection{} \label{h4:anti-automorphism}
There is an anti-automorphism $\w$ of $\Uqr(\hfour)$ given by
\begin{alignat*}{3}
  \w(N) &= N, \qquad&
  \w(A_+) &= A_-, \qquad&
  \w(A_-) &= A_+,\\
  \w(E) &= E, &
  \w(q) &= q, &
  \w(r) &= r.
\end{alignat*}
The following maps $\phi$ and $\iota$ are automorphisms of
$\Uqr(\hfour)$
\begin{alignat*}{3}
  \phi(N) &= -N, \qquad&
  \phi(A_+) &= -A_+, \qquad&
  \phi(A_-) &= -A_-,\\
  \phi(E) &= E, &
  \phi(q) &= q\inv, &
  \phi(r) &= r,\\
  \iota(N) &= N, &
  \iota(A_+) &= -A_+, &
  \iota(A_-) &= A_-,\\
  \iota(E) &= -E, &
  \iota(q) &= q, &
  \iota(r) &= r.
\end{alignat*}

\subsection{} \label{qr:power-m-relations}
Let $m\in\N$. The following identities hold in $\Uqr(\hfour)$ 
\begin{align*}
  \comm{N,{A_+}^m}_{q^m} &\equiv \brackets{m}_q {A_+}^m,\\
  \comm{{A_-}^m,N}_{q^m} &\equiv \brackets{m}_q {A_-}^m,\\
  \comm{A_-,{A_+}^m}_{r^m} &\equiv \brackets{m}_r E\cdot {A_+}^{m-1}.
\end{align*}
The proof is by induction on $m$.

\section{Fock module Representations}

\begin{notation}
  Let $m\in\N$. Define $(m)_q:= q^{-m} \brackets{m}_q\equiv
  \sum_{i=1}^{m} q^{1-2i}$. Note that $(0)_q= \brackets{0}_q= 0$.
\end{notation}

\subsection{}
Let $F$ be an infinite dimensional module over
$\Cfree{q,q\inv,r,r\inv}$, with basis $\set{u_n \mid n\in\N}$
\begin{displaymath}
  F:=\sum_{n\in\N} \Cfree{q,q\inv,r,r\inv} u_n.
\end{displaymath}
Let $c,j\in\C$. Then the following action of $\Uqr(\hfour)$ on $F$ makes
$F$ a $\Uqr(\hfour)$-module
\begin{align*}
  A_+\cdot u_k &= u_{k+1},\\
  A_-\cdot u_k &= c(k)_r u_{k-1},\\
  N\cdot u_k &= (q^{-2k}j + (k)_q) u_k,\\
  E\cdot u_k &= cu_k.
\end{align*}
Note that for the vacuum vector $u_0$:
\begin{align*}
  A_-\cdot u_0 &= 0,\\
  N \cdot u_0 &= j u_0,\\
  \Uqr(\hfour)\cdot u_0 &= F.
\end{align*}
Write $F_{j,c}$ for $F$, when it is necessary to emphasise the
dependence on $j,c$.

\begin{lemma}
  If $c\in\Ccross$ then the Fock module $F$ is irreducible.
\end{lemma}

\begin{proof}
  If $c$ is nonzero, then for each $v\in F$ there exists $n\in\N$ such
  that $A_-^n\cdot v= \alpha_v u_0$ ($\alpha_v\in \Ccross$). Since
  $u_0$ generates $F$, any other vector in $F$, can then be reached
  from the vacuum vector by applying a suitable polynomial in $A_+$.
\end{proof}

\begin{remark}
  Let $j,j',c,c'\in \C$ be such that $j'\neq j$ and/or $c'\neq c$. The
  Fock module $F_{j,c}$ is inequivalent $F_{j',c'}$.
\end{remark}

\begin{remark}
  A new representation, can be obtained from $F$ by
  twisting by the automorphism $\iota$. It is inequivalent to $F$ as a
  $\Uqr(\hfour)$-module.
\end{remark}

\subsection{}
Since $\Ur(\hthree)$ is a subalgebra of $\Uqr(\hfour)$, many of the
results on $\Uqr(\hfour)$ that I mention in this chapter lead to
corresponding results for the subalgebra $\Ur(\hthree)$. For example
the restriction to $\Ur(\hthree)$ of the $\Uqr(\hfour)$-action on $F$,
gives immediately a $\Ur(\hthree)$ Fock module. To avoid repetition I
will generally only write down the corresponding results for
$\Ur(\hthree)$ if they differ from those of $\Uqr(\hfour)$, otherwise
they will be taken as understood.

\subsection{}
In the quotient algebra $\Uqr(\hfour)/ \ideal{q-r}$ there exists a
quadratic central element
\begin{displaymath}
  A_+\cdot A_- -E\cdot N.
\end{displaymath}
It coincides in form with the quadratic central element of $\hfour$
(mentioned in~\ref{h4:Uq(h4')-"casimir"}).

\section{The centre at roots of unity}

\subsection{}
Let $\l$ be a positive integer such that $\l>2$. Let $\e=\f$ be a primitive
$\l$-th root of unity. As usual define $\l'$ as
\begin{displaymath}
  \l'=
  \begin{cases}
    \l & \text{if $\l$ is odd},\\
    \frac{\l}{2} &\text{if $\l$ is even}.
  \end{cases}
\end{displaymath}

\begin{lemma}
  The following elements are central in $\Uef(\hfour)$
  ($\Uf(\hthree)$) at the root of unity $\e=\f$
  \begin{displaymath}
    \set{A_+^{\l'}, A_-^{\l'}}.
  \end{displaymath}
\end{lemma}

\begin{proof}
  The result follows directly from the relations
  in~\ref{qr:power-m-relations}.
\end{proof}

$\Uf(\hthree)$ is finite dimensional over its centre, but
$\Uef(\hfour)$ is not since $N$ does not generate a central element.

\section{Fock module at roots of unity}

\subsection{}
Let $F_\f$ denote the Fock module over $\Uqf(\hfour)$ obtained by
partial specialisation of the $\Uqr(\hfour)$ Fock module $F$ at
$r=\f$.

\begin{lemma}
  Consider the $\Uqf(\hfour)$ Fock module $F_\f$ at the root of unity
  $\f$. The vectors $\set{u_{k\l'}\mid k\in \Zplus}$ are singular in
  $F_\f$. The module $F_\f$ is reducible.
\end{lemma}

\subsection{}
Let $k\in\Zplus$. Denote by $F'_{k\l'}$ the $\Uqf(\hfour)$-submodule
generated by the singular vector $u_{k\l'}$.

\begin{lemma}
  The quotient module $L_\f:=F_\f/ F'_{\l'}$ is irreducible.
\end{lemma}

\begin{lemma}
  Let $m,n\in\N$, such that $m<n$.

  (a) The quotient module $F'_{m\l'}/F'_{n\l'}$ is irreducible if and
  only if $n=m+1$.

  (b) The quotient module $F'_{m\l'}/F'_{(m+1)\l'}$ is equivalent to
  $L_\f$.
\end{lemma}

\begin{proof}
  The proof is almost entirely the same as
  for~\ref{h4:quotient-modules}. 
\end{proof}

\subsection{}
The irreducible quotient $\Uqf(\hfour)$-module $L_\f$ is an example of
a nilpotent representation of $\Uqf(\hfour)$: $A_+$ and $A_-$
act on it nilpotently.

\subsection{}
In the specialisation $\U_{1,r}(\hfour)$ of $\Uqr(\hfour)$ at $q=1$,
the $N$ has real eigenvalues and so it has an interpretation as a
standard quantum mechanical number operator (rather than a $q$-number
operator).

The algebra $\U_{1,\f}$ and with its representation $L_\f$ could be
interpreted abstractly as a $\f$-parafermionic ($\f$-paragrassmann)
oscillator.

\section{Semicyclic and Cyclic representations}

\subsection{}
Let $(\lambda,\mu)\in\C\times\set{0}\union \C\times\set{0}$. Let
$V_{\lambda,\mu}:= \sum_{n=0}^{\l'-1} \C v_n$ be a $\C$-vector space.
At the root of unity $\e=\f$, define the following
$\Uef(\hfour)$-module structure on $V_{\lambda,\mu}$
\begin{align*}
  A_+\cdot v_n &:=
  \begin{cases}
    v_{n+1} & \text{if $n\in\range{0,\l'-2}$}\\ 
    \lambda v_0 & \text{if $n=\l'-1$},
  \end{cases} \\
  A_-\cdot v_n &:=
  \begin{cases}
    c (n)_\f v_{n-1} & \text{if $n\in\range{1,\l'-1}$}\\ 
    \mu v_{\l'-1} & \text{if $n=0$},
  \end{cases} \\
  N \cdot v_n &:= (\e^{-2k}j + (k)_\e) v_n,\\
  E\cdot v_n &= cv_n.
\end{align*}
Note that $V_{0,0}$ ($\lambda=0$ and $\mu=0$) is equivalent to the
quotient module $L_{\e\f}$.

\begin{proposition}
  The $\Uef(\hfour)$-module $V_{\lambda,\mu}$ is irreducible.
\end{proposition}

\begin{notation}
  Let $k\in\Zplus$. Define $(k)_r!:= \prod_{m\in\range{1,k}}(m)_r$ and
  $(0)_r!:=1$.
\end{notation}

\begin{proposition}[semicyclic]
  If $(\lambda,\mu)\in\Ccross\times\set{0}\union \set{0}\times\Ccross$
  then $V_{\lambda,\mu}$ is a semicyclic (semi-periodic)
  $\Ue'(\hfour)$-module.
\end{proposition}

\begin{proof}
  The module is semicyclic, since $A_+^{\l'}\cdot v_n= \lambda v_n$,
  $A_-^{\l'}\cdot v_n= \mu c^{\l'-1} (\l'-1)_\f!  v_n$ ($n\in\range{0,\l'-1}$):
  so only one of $A_+$ and $A_-$ acts cyclicly.
\end{proof}

\begin{remark}
  In the module $V_{1,0}$ the restriction of the $\Uef(\hfour)$-action
  to the generator $A_+$ gives a representation of the cyclic group
  $Z_{\l'}$. Hence the action of $A_+$ is called cyclic.
\end{remark}

\subsection{Cyclic}
Let $(\lambda,\mu)\in \C\times\C$. Consider the following quotient of
$\Uef(\hfour)$ by a left ideal:
\begin{displaymath}
  M_{\lambda,\mu,j}:= \Uef(\hfour)/ \ideal{A_+^{\l'}- \lambda,
    A_-^{\l'} -\mu, N- j}_L,
\end{displaymath}
on which $\Uef(\hfour)$ acts naturally on the left, giving a left
$\Uef(\hfour)$-module.  $M_{\lambda,\mu,j}$ has the decomposition
\begin{displaymath}
  M_{\lambda,\mu,j}= \sum_{n_\pm\in \range{0,\l'-1}} \C A_-^{m_-}
  A_+^{n_+}.
\end{displaymath}
$M_{\lambda,\mu,j}$ is an irreducible ${\l'}^2$ dimensional cyclic
$\Uef(\hfour)$-module. If $(\lambda,\mu)\in (0,0)$, then
$M_{\lambda,\mu,j}$ is nilpotent. If $(\lambda,\mu)\in (\Ccross\times
\set{0})\union (\set{0}\times\Ccross$), then $M_{\lambda,\mu,j}$ is
semicyclic. If $(\lambda,\mu)\in \Ccross\times\Ccross$, then
$M_{\lambda,\mu,j}$ is cyclic.

\section{Unitary representations}

In this section I describe a (complex) unitary representation of
$\Uqr(\hfour)$.


\subsection{}
Define the sesquilinear scalar product $(\cdot,\cdot)$ on the
$\Uqr(\hfour)$ Fock module $F$, to be contravariant with respect to
the involutive $\Uqr(\hfour)$ anti-automorphism $\w$ defined
in~\ref{h4:anti-automorphism} (with $\w(\alpha):= \alpha^*$
($\alpha\in\C$))
\begin{align*}
  (u_k,u_l) &:= \delta_{k,l} (k)_r! \qquad (k,l\in\N),\\
  (x\cdot u_k,u_l) &= (u_k,\w(x)\cdot u_l) \qquad (x\in\Uqr(\hfour)).
\end{align*}

\begin{lemma}
  The triple $\structure{F, (\cdot,\cdot), \w}$ is a
  $\star$-representation of $\Uqr(\hfour)$.
\end{lemma}

\begin{proposition}
  Let $\e\in \Rcross$ and $\f\in\Rplus$. Denote again by
  $(\cdot,\cdot)$ the specialisation of the scalar product
  $(\cdot,\cdot)$ on $F$ to $F_{\e,\f}$. The triple
  $\structure{F_{\e,\f}, (\cdot,\cdot), \w}$ is a unitary
  representation of $\Uef(\hfour)$.
\end{proposition}

\begin{proof}
  If $\f\in\Rplus$, then $\brackets{m}_\f$ is also real and positive.
  Therefore in this case the scalar product on $F_{\e\f}$ is positive
  definite. The condition on $\e$ ensures that $(N\cdot v, N\cdot
  v)>0$ ($v\in F$).
\end{proof}

\begin{remark}
  It appears that there is no anti-automorphism of
  $\Uqr(\hfour)$ compatible with specialising $q$ and $r$ to a phase.
  So it is not possible to construct a unitary representation
  of $\Uef(\hfour)$ at a root of unity
  (\cf~\ref{h4:unitary-rep-root-of-1}).
\end{remark}

\section{Bosonisations of some quantum algebras with $\Us(\hthree)$}

In this section I construct $q$-bosonisation homomorphisms of some
well known finite and infinite dimensional quadratic quantum algebras
into the algebra $\Ur'(\hthree)$, defined by
\begin{displaymath}
  \Ur'(\hthree):= \Ur(\hthree)/ \ideal{E-1}.
\end{displaymath}
(In other words I construct realisations of these algebras in terms of
the generators of $\Ur'(\hthree)$.)

\subsection{}
Let $\alpha\in\C$. The following map
\begin{displaymath}
  \U_{rr}'(\hfour):=\Uqr(\hfour)/ \ideal{q-r,E-1}\mapto
  \Ur'(\hthree)
\end{displaymath}
is a $\Cfree{r,r\inv}$-algebra homomorphism
\begin{align*}
  A_+ &\mapsto A_+,\\ 
  A_- &\mapsto A_-,\\ 
  N &\mapsto \half(A_+\cdot A_- + A_-\cdot A_+ + \alpha).
\end{align*}

\subsection{}
Let $\Uqr(\sltwo)$ denote the quantum algebra introduced
in~\ref{qr:quad-sltwo} and consider it here as a $\C(q,r)$-algebra.
The following map
\begin{displaymath}
  \U_{r^2,r^4}(\sltwo):= \U_{q,r^4}(\sltwo)/ \ideal{q- r^2}\mapto
  \Ur'(\hthree)
\end{displaymath}
is a $\C(r)$-algebra homomorphism
\begin{align*}
  W_0 &\mapsto \left(\brackets{2}_{r^2} \brackets{2}_r \right)\inv
  (r^2 A_- \cdot A_+ +r^{-2}A_+ \cdot A_-),\\ 
  W_\pm &\mapsto \pm \left((\brackets{2}_{r^2})^{\half} \brackets{2}_r
\right)\inv A_\pm \cdot A_\pm.
\end{align*}

\subsection{} 
The subset $\set{A_-,{A_+}^k \mid k\in\N}$ of $\Ur'(\hthree)$ generates
a subalgebra of $\Ur'(\hthree)$ with the relation
$\comm{A_-,{A_+}^k}_{r^k}=\brackets{k}_r {A_+}^{k-1}$.  This
subalgebra has a natural interpretation as the algebra of polynomials
in one variable ($A_+$) with a $q$-derivation ($A_-$). From this
point of view $\Ur'(\hthree)$ is the algebra of polynomials in a variable and
a $q$-differential operator.

\subsection{Contractions}
Let $\eta\in\R$. Consider the following elements of the quantum
algebra $\U_{1,r\inv}(\sltwo)$ (defined in~\ref{qr:quad-sltwo})
\begin{align*}
  X_0 &:= \eta^2 W_0,\\
  X_\pm &:= \eta W_\pm.
\end{align*}
They satisfy the relations
\begin{align*}
  \comm{X_0,X_+} &= \eta^2 X_+,\\ 
  \comm{X_-,X_0} &= \eta^2 X_-,\\ 
  \comm{X_+,X_-}_r &= X_0.
\end{align*}
In the limit $\eta\rightarrow 0$, the subalgebra generated by
$\set{X_0,X_+,X_-}$ (a {\em contraction\/} of $\U_{1,r\inv}(\sltwo)$) is
isomorphic to $\Ur(\hthree)$. The isomorphism is given by
\begin{displaymath}
  X_0 \mapsto E,\qquad
  X_+ \mapsto A_-, \qquad
  X_- \mapsto A_+.
\end{displaymath}

A different contraction of $\U_{q,r\inv}(\sltwo)$ is also possible.
Let $\eta\in\Rcross$. Consider the following elements in
$\U_{q,r\inv}(\sltwo)$
\begin{align*}
  Y_0 &:= W_0,\\
  Y_\pm &:= \eta\inv W_\pm.
\end{align*}
They satisfy the relations
\begin{align*}
  \comm{Y_0,Y_+}_q &= Y_+,\\ 
  \comm{Y_-,Y_0}_q &= Y_-,\\ 
  \comm{Y_+,Y_-}_r &= \eta Y_0.
\end{align*}
In the limit as $\eta\rightarrow 0$, the contracted algebra generated
by $\set{Y_0,Y_+,Y_-}$ becomes a quadratic 2-parameter deformation
of the
Euclidean algebra.

\subsection{$r$-Witt algebra}
The $r$-Witt algebra $\Ur(\cs{V}_0)$ (deformed centreless Virasoro
algebra) is defined to be the $\Cfree{r,r\inv}$-algebra with
generators $\set{L_m\mid m\in\Z}$ that satisfy the relations
\begin{displaymath}
  \comm{L_m,L_n}_{r^{n-m}} = \brackets{m-n}_r L_{m+n},
\end{displaymath}
and certain complicated associativity relations
(compare~\cite{polychronakos:q-vir-consistency}), which I will not
write down here. (I thank Cosmas Zachos for bringing these to my
attention.) $\Ur(\cs{V}_0)$ is a deformation of the algebra of
vector fields on the circle.

I define an extension $\Ur(\hthree\spbreve)$ of $\Ur'(\hthree)$ by
adjoining an (formal) generator $A_+\inv$ with the relations
\begin{align*}
  A_+\inv \cdot A_+  &= 1=A_+ \cdot A_+\inv,\\
  \comm{A_+\inv,A_-}_r &= A_+\inv\cdot A_+\inv.
\end{align*}
The following map $\Ur(\cs{V}_0)\mapto \Ur(\hthree\spbreve)$ is a
$\Cfree{r,r\inv}$-algebra homomorphism
\begin{displaymath}
  L_m \mapsto -(A_+)^{1+m} \cdot A_- \qquad (m\in\Z).
\end{displaymath}
Note that the associativity relations are automatically satisfied,
since $\Ur(\hthree\spbreve)$ is associative.

\begin{remark}
  It is also possible to construct a centreless $q$-deformed
  $W_\infty$ algebra~\cite{chaichian-kulish-lukierski} (with
  generators $W^{(k+1)}_n:= {A_+}^{n+k} \cdot {A_-}^k$, $k\in\Zplus$).
  This was done in~\cite{gelfand-fairlie} using a $q$-Heisenberg
  algebra.
\end{remark}

\subsection{}
Note the following identity in $\Ur(\hthree\spbreve)$
\begin{displaymath}
  \comm{A_+^{-m},A_-}_{r^m}= \brackets{m}_r A_+^{-2m-1}.
\end{displaymath}
Hence it follows that at the root of unity $\f$, $A_+^{-\l'}$ lies in
the centre of $\Uf(\hthree\spbreve)$.

The following map $\varphi:\Uf(\cs{V}_0)\mapto \Uf(\hthree\spbreve)/
\ideal{A_+^{\l'}-1, \A_+^{-\l'}-1,N}_L$
\begin{displaymath}
  \varphi(L_m)= -(A_+)^{1+m} \cdot A_- \qquad (m\in\Z).
\end{displaymath}
gives a finite representation of $\Uf(\cs{V}_0)$ with a cyclic property:
 $\varphi(L_{m+k\l'})= \varphi(L_m)$ for all
$m,k\in\Z$.

Similarly the map $\vartheta:\Uf(\cs{V}_0)\mapto
\Uf(\hthree\spbreve)/ \ideal{A_+^{\pm \l'}}$ given by
\begin{displaymath}
  \vartheta(L_m):= -(A_+)^{1+m} \cdot A_- \qquad (m\in\Z).
\end{displaymath}
gives a finite dimensional representation $\vartheta(\Uf(\cs{V}_0))$
with a nilpotent property.

\section{A Symmetry of $\Ur(\hthree)$}

Let $T:= \twomatrix{a}{b}{c}{d}$ be a matrix, whose entries generate a
quadratic $\Cfree{r,r\inv}$-bialgebra $\frak{A}$, with relations that
will be given presently and coproduct $\Delta(T):= T\dottensor T$.

\begin{proposition} 
  Consider the left coaction $\Delta_L:\Ur(\hthree)\mapto
  \frak{A}\tensor \Ur(\hthree)$ of $\frak{A}$ on $\Ur(\hthree)$:
  \begin{displaymath}
    \begin{pmatrix}
      A_+ \\ A_- 
    \end{pmatrix}
    \mapsto
    T \dottensor
    \begin{pmatrix}
      A_+ \\ A_-
    \end{pmatrix},
    \qquad
    E \mapsto E':=1\tensor E.
  \end{displaymath}
  If the elements of $T$ satisfy the following relations, then the left
  coaction $\Delta_L$ is a $\Cfree{r,r\inv}$-algebra homomorphism
  \begin{alignat*}{2} 
    ac &=r^2 ca, \qquad&
    ad-r^2cb &=1,\\
    bd &= r^2 db, &
    da- r^{-2}bc &=1.
  \end{alignat*}
\end{proposition}


\begin{proposition}
  The right coaction $\Delta_R:\Ur(\hthree)\mapto \Ur(\hthree)\tensor
  \frak{A}$ of $T$ on $\Ur(\hthree)$, given by:
  \begin{displaymath}
    \begin{pmatrix}
      A_+ & A_-
    \end{pmatrix}
    \mapsto
    \begin{pmatrix}
      A_+ & A_-
    \end{pmatrix}
    \dottensor T, \qquad
    E\mapsto E\tensor 1
  \end{displaymath}
  is an algebra homomorphism, if the elements of $T$ satisfy the
  relations
  \begin{alignat*}{2}
    ab &= r^2 ba, \qquad &
    ad- r^2 bc &= 1,\\
    cd &= r^2 dc, &
    da- r^{-2} cb &= 1.
  \end{alignat*}
\end{proposition}


\begin{proposition}
  If $\Ur(\hthree)$ is covariant with respect to both the left and
  right $\frak{A}$-coactions $\Delta_L$ and $\Delta_R$, then the
  generators of $\frak{A}$ satisfy the relations of
  $\fun_{r^2}(\SL_2)$.
\end{proposition}

\begin{proof}
  Combining the relations from the two previous propositions forces
  the additional relation $bc=cb$ in $\frak{A}$. Then all the
  relations given coincide with those of $\fun_{r^2}(\SL_2)$.
\end{proof}

\subsection{}
By construction the maps $\Delta_L$ and $\Delta_R$ are
compatible with the coproduct of $\frak{A}$:
\begin{align*}
  (\Delta_L \tensor \id) \compose \Delta_L &= (id \tensor \Delta)\compose
  \Delta_L,\\
  (\id \tensor \Delta_R) \compose \Delta_R &= (\Delta \tensor id)\compose
  \Delta_R.
\end{align*}
and the two coactions cocommute
\begin{displaymath}
  (\id \tensor \Delta_R)\compose \Delta_L= (\Delta_L \tensor
  \id)\compose \Delta_R.
\end{displaymath}


\chapter{Quantum affine algebras}
\label{chap:kac-moody}

\section{Introduction}

To every affine Cartan matrix of an affine (Kac-Moody) Lie algebra
$\kmg$, Drinfeld and Jimbo have associated a Hopf algebra $\Uq(\kmg)$, the
quantum affine (Kac-Moody) algebra or quantum group of $\kmg$.
In this chapter I describe the quantum affine algebra $\Uq(\kmg)$,
when $\kmg$ is an untwisted affine Kac-Moody algebra.

In the classical theory of Kac-Moody algebras and their applications,
a very important role is played by their centrally extended loop
algebra presentation. It has a number of advantages over the Chevalley
presentation. In particular the loop algebra presentation $\kmg$ makes
it manifest that the algebra is infinite dimensional and that it has a
central element. In the case of quantum affine Kac-Moody algebras such
a loop algebra presentation was given by
Drinfeld~\cite{drinfeld:presentation}. This presentation has been
extensively utilised in applications of quantum affine algebras and in
their representation theory (see for
 example~\cite{davies-foda-jimbo-miwa-nakayashiki:XXZ-diag,%
  chari-pressley:eval-reps}). Unfortunately Drinfeld did not give a
proof that his loop algebra presentation is isomorphic to the usual
Chevalley presentation. Another related problem is the construction of
a basis of $\Uq(\kmg)$ in terms of the underlying root system, like
the one described for $\Uq(\g)$ in~\ref{quea:sec-Uq(g)-basis}. This
problem is much harder in the affine case because the root system and
affine Weyl group $\hat{W}$ of $\kmg$ are both infinite dimensional:
in particular there is no unique longest element of $\hat{W}$ that
gives a natural basis of root vectors in $\Uq(\kmg)$ as there is in
the finite case and also because of the existence of imaginary root
vectors.

The isomorphism between the two presentations was proved recently by
Beck~\cite{beck:qkm-braid-group}, and Khoroshkin and
Tolstoy~\cite{khoroshkin-tolstoy:drinfeld-realisation} and their
work~\cite{khoroshkin-tolstoy:cartan-weyl-basis,beck:qkm-convex-bases}
also led to Poincar\'e-Birkoff-Witt type bases of $\Uq(\kmg)$. Beck's
work in particular gave an extension of Lusztig's braid group
action~\cite{lusztig:geom-ded,lusztig:book} on $\Uq(\kmg)$ to the
group of automorphisms $\hat{B}$ associated to the extended affine
Weyl group of $\kmg$. The action of $\hat{B}$ on the Chevalley
generators, corresponding to the underlying $\Uq(\g)$ subalgebra,
generates all the real root vectors of $\Uq(\kmg)$. There is a
subgroup $\cs{P}$ of $\hat{B}$ corresponding to the lattice of
translations in the extended affine Weyl group of $\kmg$. The
imaginary root vectors on the other hand cannot be reached by the
action $\hat{B}$. The imaginary root vectors are fixed under the
action of $\cs{P}$.

As in the classical case, $\Uq(\kmg)$ has a Heisenberg subalgebra
$\Uq(\H)$ generated by its imaginary root vectors. $\Uq(\H)$ has a
similar representation theory to the classical Heisenberg subalgebra.
It has highest weight Fock modules which are irreducible. At a real
positive specialisation the Fock modules are unitarisable.

\section{Preliminaries}

I start by recalling the definition of a quantum affine Kac-Moody
algebra due to Jimbo~\cite{jimbo:lmp1} and
Drinfeld~\cite{drinfeld:icm}.

\subsection{} Let $\g$ be a simple Lie algebra of rank $r$
with Cartan matrix $(a_{ij})_{i,j\in \range{1,r}}$. Let
$(a_{ij})_{i,j\in \range{0,r}}$ be the extended affine Cartan matrix
of the (untwisted) Kac-Moody algebra $\kmg$ of $\g$, in the sense of
Kac~\cite{kac:book}, so that $a_{ii}=2$ and $a_{ij}\leq 0$ for $i\neq
j$.  Fix $r+1$ positive coprime integers $(d_i)_{i\in\range{0,r}}$,
such that $(d_i a_{ij})$ is a symmetric matrix. The Cartan matrix
$(a_{ij})_{i,j\in\range{0,r}}$ of $\kmg$ has rank $r$:
$\det((a_{ij})_{i,j\in\range{0,r}})=0$.

\begin{definition}
  The quantum affine (Kac-Moody) algebra $\Uq(\kmg)$ is the associative
  $\C(q)$-algebra with $\unity$ and generators $\set{e_i, f_i,
    {k_i}^{\pm1}, \deriv \mid i\in \range{0,r}}$ with the relations:
  \begin{alignat*}{2}
    k_i \cdot {k_i}\inv &= 1 = {k_i}\inv \cdot k_i, \qquad &
    k_i \cdot k_j &= k_j \cdot k_i,\\ 
    k_i \cdot e_j \cdot {k_i}\inv &= q_i^{a_{ij}} e_j, &
    k_i \cdot f_j \cdot {k_i}\inv &= q_i^{-a_{ij}} f_j,\\ 
    e_i \cdot f_j - f_j \cdot e_i &= \delta_{ij} \frac{k_i -
      {k_i}\inv}{q_i-q_i\inv}, &
    \deriv \cdot k_i &= k_i \cdot \deriv, \\ 
    \deriv \cdot e_i \cdot\deriv\inv &= q^{\delta_{i0}} e_i, &
    \deriv \cdot f_i \cdot \deriv\inv &= q^{-\delta_{i0}} f_i,
  \end{alignat*}
  \begin{align*}
    \sum_{n=0}^{1-a_{ij}} (-1)^n \comb{1-a_{ij}}{n}_{q_i} {e_i}^n
    \cdot e_j \cdot {e_i}^{1-a_{ij}-n} &= \; 0 \qquad (i\neq j),\\
    \sum_{n=0}^{1-a_{ij}} (-1)^n \comb{1-a_{ij}}{n}_{q_i} {f_i}^n
    \cdot f_j \cdot {f_i}^{1-a_{ij}-n} &= \; 0 \qquad (i\neq j).
  \end{align*}
  The generators $\set{e_i,f_i\mid i\in\range{0,r}}$ are called the
  Chevalley generators. 
\end{definition}

\subsection{} The quantum enveloping algebra $\Uq(\kmg)$ is a Hopf
algebra. The Hopf algebra structure is as follows.

The coproduct map $\Delta:\Uq(\kmg)\mapto \Uq(\kmg)\tensor \Uq(\kmg)$ is
\begin{align*}
  \Delta:k_i &\mapsto k_i\tensor k_i,\\ 
  \Delta:e_i &\mapsto e_i\tensor \unity + k_i\tensor e_i,\\ 
  \Delta:f_i &\mapsto f_i\tensor {k_i}\inv+ \unity\tensor f_i,\\
  \Delta:\deriv &\mapsto \deriv\tensor \deriv.
\end{align*}

The antipode map $S:\Uq(\kmg)\mapto \Uq(\kmg)$ is
\begin{displaymath}
  S:k_i \mapsto {k_i}\inv,\quad
  S:e_i \mapsto -{k_i}\inv e_i,\quad
  S:f_i \mapsto -f_i k_i, \quad
  S:\deriv \mapsto \deriv\inv.
\end{displaymath}

The counit map $\epsilon:\Uq(\kmg)\mapto \C(q)$ is
\begin{displaymath}
  \epsilon: k_i \mapsto 1,\quad
  \epsilon: e_i \mapsto 0,\quad
  \epsilon: f_i \mapsto 0, \quad
  \epsilon: \deriv \mapsto 1.
\end{displaymath}

\subsection{Triangular decomposition}
The Cartan subalgebra $\Uq(\kmh)$ is generated by $\set{k_i,
  \deriv}_{i \in \range{0,r}}$. The (positive) Chevalley generators
$\set{e_i}_{i\in\range{0,r}}$ generate the positive roots subalgebra
$\Uq(\kmn_+)$ and the (negative) Chevalley generators
$\set{f_i}_{i\in\range{0,r}}$ generate the negative roots subalgebra
$\Uq(\kmn_-)$. As in the finite case~\cite{rosso:cmp}, $\Uq(\kmg)$ has
a triangular decomposition $\Uq(\kmg)\isomorphic\Uq(\kmn_-) \tensor
\Uq(\kmh) \tensor \Uq(\kmn_+)$ (vector space isomorphism by
multiplication).

\subsection{Derived algebra}
The derived quantum affine algebra $\Uq(\kmg')$, which is the quantum
enveloping algebra of the derived affine algebra $\kmg'$, is the
subalgebra of $\Uq(\kmg)$ generated by $\set{e_i, k_i^{\pm1}, f_i}_{i
  \in \range{0,r}}$ (\ie~excluding $\deriv^{\pm1}$). Let
$\Uq(\kmn_+')$, $\Uq(\kmh')$ and $\Uq(\kmn_-')$ be the positive roots
subalgebra, Cartan subalgebra and negative roots subalgebra of
$\Uq(\kmg')$, generated respectively by $\set{e_i}$, $\set{k_i}$ and
$\set{f_i}$. There is a corresponding triangular decomposition
$\Uq(\kmg')= \Uq(\kmn_-')\tensor \Uq(\kmh')\tensor \Uq(\kmn_+')$.
Clearly $\Uq(\kmn_+') \isomorphic \Uq(\kmn_+)$ and $\Uq(\kmn_-')
\isomorphic \Uq(\kmn_-)$, but $\Uq(\kmh)$ is an extension of
$\Uq(\kmh')$ by the grading generator $\deriv$. $\deriv$ is called the
{\em derivation\/} or scaling element of $\Uq(\kmg)$.

\begin{notation}
  For $\e\in\Ccross$, define $\e_i:=\e^{d_i}$. Denote by
  $\brackets{n}_\e$ and $\comb{n}{m}_\e$ in $\C$ the specialisation of
  the $q$-numbers $\brackets{n}_q$ and $\comb{n}{m}_q$ in
  $\Cfree{q,q\inv}$ at $q=\e$.
\end{notation}

\begin{definition}[specialisation]
  Define the following elements in $\Uq(\kmg)$
  \begin{displaymath}
    h_i:=\frac{k_i-{k_i}\inv}{q_i-q_i\inv} \qquad \text{and}\qquad
    d:=\frac{\deriv-\deriv\inv}{q-q\inv}.
  \end{displaymath}
  Define $\U_{\cs{A}}(\kmg)$ to be the $\Cfree{q,q\inv}$-subalgebra of
  $\Uq(\kmg)$ generated by
  \begin{displaymath}
    \set{e_i, f_i, {k_i}^{\pm1}, \deriv^{\pm1}, h_i, d}.
  \end{displaymath}
  For $\epsilon\in\Ccross$, I define the specialisation of $\UA(\kmg)$
  at $q=\e$ to be $\Ue(\kmg):= \U_{\cs{A}}(\kmg)/ \ideal{q-\epsilon}$.
\end{definition}

\subsection{}
The affine Lie algebra $\kmg$ associated to the affine Cartan matrix
$(a_{ij})_{i,j\in\range{0,r}}$ has generators $\set{\bar{h}_i,
  \bar{e}_i, \bar{f}_i, \bar{d} \mid i\in\range{0,r}}$, which satisfy
the following relations (Chevalley presentation)
\begin{alignat*}{2}
  \comm{\bar{h}_i,\bar{h}_j}&=0, \qquad&
  \comm{\bar{e}_i,\bar{f}_j} &= \delta_{ij} \bar{h}_i,\\ 
  \comm{\bar{h}_i,\bar{e}_j} &= a_{ij} \bar{e}_j, &
  \comm{\bar{h}_i,\bar{f}_j} &= -a_{ij} \bar{f}_j,\\ 
  \comm{\bar{d},\bar{h}_i} &= 0, &
  \comm{\bar{d},\bar{e}_i} &= \delta_{i0}\bar{e}_i,\\
  \comm{\bar{d},\bar{f}_i} &= -\delta_{i0}\bar{f}_i, &&\\
  \ad(\bar{e}_i)^{1-a_{ij}} &(\bar{e}_j) = 0, &
  \ad(\bar{f}_i)^{1-a_{ij}} &(\bar{f}_j) = 0.
\end{alignat*}
Let $\U(\kmg)$ be the universal enveloping algebra $\kmg$.

\begin{proposition} \cite{lusztig:reps,deconcini-kac:unity}
  \label{km:chevalley-classical-limit}
  Let $\U_1(\kmg)$ be the specialisation of $\UA(\kmg)$ at $q=1$. The
  following map $\U_1(\kmg)/\ideal{k_i-1,\deriv-1 \mid
    i\in\range{0,r}}\longrightarrow \U(\kmg)$ is a $\C$-algebra
  isomorphism
  \begin{alignat*}{2}
    e_i &\mapsto \bar{e}_i, & \qquad f_i &\mapsto \bar{f}_i,\\
    h_i &\mapsto \bar{h}_i, & d &\mapsto \bar{d}.
  \end{alignat*}
\end{proposition}


\begin{remark} \label{km:sl2,g-embedding-in-kmg}
   Fix $i\in\range{0,r}$. The elements $\set{e_i,{k_i}^{\pm1},f_i}$ in
   $\Uq(\kmg)$ generate a $\Uq(\sltwo)$ subalgebra of $\Uq(\kmg)$.

   The generators $\set{e_i, (k_i)^{\pm1}, f_i| i\in \range{1,r}}$
   generate a $\Uq(\g)$ subalgebra of $\Uq(\kmg)$.
\end{remark}

\subsection{}
Let $(a_i)_{i\in\range{0,r}}\in \Zplus^{r+1}$ be the unique vector of
coprime positive integers that satisfy $\sum_{j=0}^r a_{ij}
a_j=0$ (with $a_0=1$). 
Then $\set{a_0,\ldots, a_r}$ are the Coxeter labels (or Kac numbers) of
the Dynkin diagram of $\kmg$. Recall that the canonical central
element of $\kmg$ is given by $\sum_{i=0}^r a_i\czek \bar{h}_i$, where
$\set{a_0\czek,\ldots, a_r\czek}$ are the dual Coxeter labels (dual
Kac numbers) of $\kmg$ satisfying $\sum_{i=0}^r a_i\czek a_{ij}=0$
($a_i\czek= d_i a_i$).

\begin{lemma} \label{km:centre}
  The canonical central element of $\Uq(\kmg)$ is
  \begin{displaymath}
    C:=\prod_{i\in\range{0,r}} {k_i}^{a_i}
  \end{displaymath}
\end{lemma}

\begin{proof}
  It is easily checked that $C$ commutes with all the Chevalley
  generators of $\Uq(\kmg)$. I show that $C$ commutes with
  $e_j$. A similar simple calculation shows that $C$ and $f_j$
  commute. 
  \begin{align*}
    C\cdot e_i &= \prod_{i=0}^r {k_i}^{a_i} \cdot e_j\\
    &= q^{\sum_{i=0}^r a_i d_i a_{ij}} e_j \cdot \prod_{i=0}^r
    {k_i}^{a_i}\\
    &= e_j \cdot C.
  \end{align*}
  The last step follows from the symmetry of the symmetrised Cartan
  matrix: $d_i a_{ij}=d_j a_{ji}$ and the definition of the Coxeter
  labels.
\end{proof}

\subsection{Gradations}
Let $\structure{M,+}$ be an abelian group. Then an $M$-gradation of
$\Uq(\kmg)$ is a decomposition with respect to $M$:
\begin{displaymath}
  \Uq(\kmg):=\Directsum_{\alpha\in M} \Uq(\kmg)_{(\alpha)},
\end{displaymath}
such that $\forall \alpha,\beta\in M$, $\Uq(\kmg)_{(\alpha)}\cdot
\Uq(\kmg)_{(\beta)} \subseteq \Uq(\kmg)_{(\alpha+\beta)}$. An element
$x\in \Uq(\kmg)_{(\alpha)}$ is said to have degree $\alpha$ ($\deg x=
\alpha$).

Let $s:=(s_0,\dots,s_r)\in \Z^{r+1}$. Then
\begin{alignat*}{2}
  \deg e_i &:= s_i, \qquad& \deg f_i &:= -s_i,\\
  \deg k_i &:= 0, & \deg \deriv &:= 0, 
\end{alignat*}
defines a $\structure{\Z,+}$-gradation of $\Uq(\kmg)$, (which in the
classical case, Kac calls a $\Z$-gradation of type~$s$).  For example
choosing $s_i=1$ ($i\in \range{0,r}$) gives the principal gradation of
$\Uq(\kmg)$. Choosing instead $s_0=1$ and $s_i=0$ ($i\in \range{1,r}$)
gives the homogeneous gradation of $\Uq(\kmg)$.

Let $\hat{Q}:= \sum_{i\in\range{0,r}} \Z \alpha_i$ be the root lattice
of $\kmg$ (see~\ref{km:^Q-root-lattice}). There is a
$\hat{Q}$-gradation of $\Uq(\kmg)$ given by
\begin{alignat*}{2}
  \deg e_i &:= \alpha_i, \qquad& \deg f_i &:= -\alpha_i,\\
  \deg k_i &:= 0, & \deg \deriv &:= 0.
\end{alignat*}



\begin{lemma}
  The map $\w:\Uq(\kmg)\mapto \Uq(\kmg)$
  \begin{alignat*}{3}
    k_i &\mapsto {k_i}\inv, \qquad&
    \deriv &\mapsto \deriv\inv, \qquad &
    q &\mapsto q\inv, \\
    e_i &\mapsto f_i, &
    f_i &\mapsto e_i, &&
  \end{alignat*}
  extends as an $\C$-algebra anti-automorphism to the whole of
  $\Uq(\kmg)$. It is called the Cartan involution of $\Uq(\kmg)$.

  The following map $\varphi:\Uq(\kmg)\mapto \Uq(\kmg)$ is a
  $\C$-algebra automorphism of $\Uq(\kmg)$
  \begin{alignat*}{3}
    k_i &\mapsto {k_i}, \qquad&
    \deriv &\mapsto \deriv, \qquad &
    q &\mapsto q\inv, \\
    e_i &\mapsto f_i, &
    f_i &\mapsto e_i. &&
  \end{alignat*}
\end{lemma}


\section{Lattices and Weyl group}

\subsection{}
Recall from~\ref{quea:lattices} the definitions of the following
structures associated to $\g$. The weight lattice $P$ of $\g$ over
$\Z$ has generators $\omega_i$ ($i\in\range{1,r}$)
\begin{displaymath}
  P:= \sum_{i\in\range{1,r}} \Z \omega_i.
\end{displaymath}
The simple roots of $\g$ are $\alpha_i:=\sum_{j\in\range{1,r}} a_{ij}
\omega_j$ ($i\in\range{1,r}$) and the root lattice of $\g$ is the
sub-lattice of $P$ generated over $\Z$ by $\set{\alpha_i\mid
  i\in\range{1,r}}$
\begin{displaymath}
  Q:= \sum_{i\in\range{1,r}} \Z \alpha_i.
\end{displaymath}
The coroot lattice is the lattice dual to $P$ over $\Z$:
$Q\czek:=\Hom(P,\Z)$, with a basis $\alpha\czek_i$ ($i\in\range{1,r}$)
with the pairing $\pairing{\cdot,\cdot}:P\times Q\czek\mapto \Z$ given
by
\begin{displaymath}
  \pairing{\omega_i,\alpha\czek_j}= \delta_{ij}.
\end{displaymath}
Then $\pairing{\alpha_i,\alpha\czek_j}= a_{ij}$.

Let $W$ be the Weyl group associated to $\g$ with generators $s_i$ (the
fundamental reflections) as defined in~\ref{quea:g-Weyl-group}.
The Weyl group $W$ acts on $P$ as
\begin{displaymath}
  s_i:x\mapsto x- \pairing{x, \alpha_i\czek}\alpha_i \quad (x\in P).
\end{displaymath}
$Q$ is invariant under the action of $W$.  The set of simple roots is
$\Pi:=\set{\alpha_i\mid i\in\range{1,r}}$. The root system $R$ of $\g$
is the subset of $Q$ given by the $W$-orbit of $\Pi$, $R:=W\cdot\Pi$.
$P_+:= \sum_{i\in\range{1,r}} \N \omega_i$ and $Q_+:=
\sum_{i\in\range{1,r}} \N \alpha_i$. The set of positive roots of $\g$
is $R_+:= R\intersect Q_+$.

The coroot lattice $Q\czek$ can be embedded in the coweight lattice
$P\czek$, which is the dual lattice to $Q$, $P\czek:=\Hom(Q,\Z)$, with
basis $\set{\omega\czek_i\mid i\in\range{1,r}}$ and pairing
\begin{align*}
  \pairing{\cdot,\cdot}': Q\times P\czek &\mapto \Z,\\
  \pairing{\alpha_i,\omega\czek_j}' &= \delta_{ij}.
\end{align*}
The embedding is given by $\alpha\czek_i=\sum_{j\in\range{1,r}}
a_{ij}\omega\czek_j$. Note that the two pairings coincide on $Q\times
Q\czek$:
\begin{displaymath}
  \pairing{\cdot,\cdot}_{\vert Q\times Q\czek}=
  \pairing{\cdot,\cdot}'_{\vert Q\times Q\czek}.
\end{displaymath}
Let $\Pi\czek:= \set{\alpha\czek_i\mid i\in\range{1,r}}$ be the set of
simple coroots and $R\czek:= W\cdot \Pi\czek$ be the coroot system of
$\g$. The correspondence $\alpha_i \leftrightarrow \alpha\czek_i$
($i\in\range{1,r}$) extends to $R \leftrightarrow R\czek$.
The Weyl group acts on
$P\czek$ by
\begin{displaymath}
  s_i:x\czek \mapsto x\czek- \pairing{\alpha_i,x\czek}'\alpha\czek_i
  \quad(x\czek\in P\czek).
\end{displaymath}
The coroot lattice $Q\czek$ is invariant under the action of $W$:
\begin{displaymath}
  s_i:\alpha\czek_j\mapsto \alpha\czek_j-
  \pairing{\alpha_i,\alpha\czek_j}\alpha\czek_j.
\end{displaymath}
{}From~\ref{quea:W-invar-of-pairing} it follows that $\pairing{\alpha,
  \alpha\czek}=2$ for all $\alpha\in R$.

\subsection{Affine Weyl group}
The affine Weyl group of $\kmg$ associated to $(a_{ij})_{i,j\in
  \range{0,r}}$ has generators $\set{s_i\mid i\in\range{0,r}}$
and the relations of~\ref{quea:g-Weyl-group} (with $m_{ij}=\infty$ if
$\frac{a_{ij} a_{ji}}{4}\geq 1$).

\label{km:^Q-root-lattice}
Let $\hat{P}:= \sum_{i\in\range{0,r}} \Z \omega_i$ be the affine
weight lattice of $\kmg$ with basis $\set{\omega_i\mid
  i\in\range{0,r}}$. Let $\set{\alpha_i:= \sum_{j\in\range{0,r}}
  a_{ij} \omega_j\mid i\in\range{0,r}}$ be the simple roots of $\kmg$
($\alpha_0$ is the extra root).  Define the affine root lattice of
$\kmg$ to be $\hat{Q}:= \sum_{i\in\range{0,r}} \Z \alpha_i$. 

\subsection{Symmetric form} \label{km:symmetric-form}
I define a bilinear map $(\cdot,\cdot): \hat{P}\times \hat{Q} \mapto \Z$
\begin{displaymath}
  (\omega_i,\alpha_j):= d_i \delta_{ij}.
\end{displaymath}
Note that the restriction of this pairing to $\hat{Q}\times \hat{Q}$
defines a symmetric $\Z$-valued bilinear form on $\hat{Q}$, since
\begin{displaymath}
  (\alpha_i,\alpha_j)= d_i a_{ij}.
\end{displaymath}

\subsection{Extended affine Weyl group} 
Following~\cite{beck:qkm-braid-group}, I make the following
definitions.

Define the extended affine Weyl group $\kmW:= W\ltimes P\czek$, where
the semi-direct product is defined using the action of $W$ on $P\czek$
to be
\begin{displaymath}
  (w_1,x_1)\cdot (w_2,x_2):= (w_1 w_2, w_2\inv(x_1)+ x_2)\quad
  (\forall w_1,w_2\in W; x_1,x_2\in P\czek).
\end{displaymath}
The identity element in $\kmW$ is $(1,0)$ and for each $(w,x)\in\kmW$
the inverse is $(w\inv,-w\inv(x))$ ($w\in W$, $x\in P\czek$).
%
Write $w$ for $(w,0)\in \kmW$ ($w\in W$). Also 
write $x$ for $(1,x)$ ($x\in P\czek$).

\begin{notation}
  Let $\alpha\in R$. Define $s_\alpha$ to act on $x\in P$ as
  $s_\alpha(x)= x- \pairing{x,\alpha\czek}\alpha$ and to act on
  $x\czek\in P\czek$ as $s_\alpha(x\czek)= x\czek-
  \pairing{\alpha,x\czek}'\alpha\czek$.
\end{notation}

Denote by $\theta:=\sum_{i\in\range{1,r}} a_i \alpha_i$ the highest
root in $R_+$ of $\g$ and $\theta\czek$ the highest coroot of $\g$.
Let $s_0=(s_\theta,\theta\czek)\in\kmW$. The set $\set{s_i\mid
  i\in\range{0,r}}$ of $\kmW$ generates the affine Weyl group
$\tilde{W}$. Therefore $\tilde{W}$ is a subgroup of $\kmW$: in fact
$\tilde{W}$ is a normal (Coxeter) subgroup of $\kmW$. The quotient
group $\cs{T}:= \kmW/\tilde{W}$ is a finite group (a certain subgroup
of Dynkin diagram automorphisms of
$\kmg$)~\cite[VI]{bourbaki:Lie-456}. A diagram automorphism $\tau\in
\cs{T}$ acts on $\tilde{W}$ as $\tau s_i \tau\inv= s_{\tau(i)}$. Then
I can identify $\kmW= \cs{T}\ltimes \tilde{W}$.  In $\cs{T}\ltimes
\tilde{W}$ the product is given by $(\tau_1,w_1)\cdot (\tau_2,w_2):=
(\tau_1 \tau_2, \tau_2\inv(w_1)w_2)$.

Define the canonical imaginary root of $\kmg$ to be
\begin{displaymath}
  \delta:= \alpha_0+ \theta= \sum_{i\in\range{0,r}} a_i \alpha_i.
\end{displaymath}


The pairing $\pairing{\cdot,\cdot}:P\times Q\czek\mapto \Z$ extends naturally
to a pairing $\pairing{\cdot,\cdot}:\hat{P}\times \hat{Q}\czek\mapto \Z$ and
likewise the pairing $\pairing{\cdot,\cdot}':Q\times P\czek$ to
$\hat{Q}\times \hat{P}\czek\mapto \Z$.

The extended affine Weyl group $\kmW$ acts on $\hat{Q}$: in particular
the elements of $P\czek$ act on $\hat{Q}$ as
\begin{displaymath}
  x(\beta)= \beta- \pairing{\beta,x}' \delta \quad (\beta\in Q, x\in
  P\czek).
\end{displaymath}

\subsection{Root system}
Let $R$ be the (finite) root system of $\g$. Recall that the root
system $\hat{R}$ of $\kmg$ is infinite dimensional and has the
following structure.

There are real roots and imaginary roots. The set of real roots is
\begin{displaymath}
  \hat{R}^{\text{re}}:= \set{\alpha+ n\delta\mid \alpha\in R, n\in
    \Z}.
\end{displaymath}
The set of imaginary roots is
\begin{displaymath}
  \hat{R}^{\text{im}}:= \set{n\delta\mid n\in \Zcross}. 
\end{displaymath}
The real roots are generated by the action of $\kmW$ on the simple roots,
$\hat{R}^{\text{re}}= \kmW\cdot \Pi$. The imaginary roots are fixed by
$\tilde{W}$: $s_i(n\delta)=n\delta$ (for all $i\in\range{0,r}$ and
$n\in\Z$), since $\pairing{\delta, \alpha\czek_i}=0$ 
($i\in\range{0,r}$). The positive real roots are
\begin{displaymath}
  \hat{R}^{\text{re}}_+:= \hat{R}^{\text{re}}\intersect \hat{Q}_+=
  \set{\beta+ n\delta\mid \beta\in R, n\in\Zplus}\union
  \set{\alpha\mid \alpha\in R_+}.
\end{displaymath}
The positive imaginary roots are
\begin{displaymath}
  \hat{R}^{\text{im}}_+=\hat{R}^{\text{im}}\intersect Q_+=
  \set{n\delta\mid n\in \Zplus}.
\end{displaymath}
Finally the root system and positive roots are
\begin{align*}
  \hat{R} &:= \hat{R}^{\text{re}}\union \hat{R}^{\text{im}},\\
  \hat{R}_+ &:= \hat{R}\intersect Q_+= \hat{R}^{\text{re}}_+\union
  \hat{R}^{\text{im}}_+.
\end{align*}

\subsection{Ordering of $\hat{R}_+$}\label{km:ordering-of-kmR+}
Recall~\ref{quea:w0-ordered-R+} that a reduced expression of the
longest element $w_0\in W$ gives an ordering in $R_+$ and a
corresponding ordering of $-R_+$.  Defining also $\alpha+ m\delta>
\alpha+ n\delta$, if $m>n$ ($\alpha\in R\union \set{0}$, $m,n\in\Zplus$) and
$\beta+ n\delta> m\delta> -\beta+ p\delta$ ($\beta\in R_+$,
$m,n,p\in\Zplus$), gives then a ordering of $\hat{R}_+$.

\section{The braid group action}
\label{km:affine-braid-group}

The results from this section will be used
in~\ref{qkm-1:centre-of-Uqkmg} when a quantum affine algebra at a root
of unity is considered.

\begin{notation}
  I define for each $n\in \N$ and $i\in\range{0,r}$ the following
  elements in $\Uq(\kmg)$
  \begin{displaymath}
    e_i^{(n)}:= \frac{e_i^n}{\brackets{n}_{q_i}!}\quad \text{and}\quad
    f_i^{(n)}:= \frac{f_i^n}{\brackets{n}_{q_i}!}.
  \end{displaymath}
\end{notation}

\subsection{}
Recall from~\ref{quea:B-action-on-Uq(g)} that the braid group of $W$
associated to $\g$ acts as a group of automorphisms on $\Uq(\g)$. This
result extends~\cite{lusztig:book} to an action of the braid group
$\tilde{B}$ associated to $\tilde{W}$ on $\Uq(\kmg)$ as a group of
automorphisms. The generators of $\tilde{B}$ are $\set{T_i^{\pm1}\mid
  i\in\range{0,r}}$. Let $l$ be the length function on $\tilde{W}$.
Let $s_{i_1}s_{i_2}\cdots s_{i_n}$ ($i_k\in\range{0,r}$) be a reduced
expression of $w\in\tilde{W}$ ($l(w)=n$). Then define
$T_w:=T_{i_1}T_{i_2}\cdots T_{i_n}$. The map $\tilde{W}\mapto
\tilde{B}$, that maps $w\mapsto T_w$ such that $T_{w_1}T_{w_2}=
T_{w_1w_2}$ if $l(w_1)+ L(w_2)= l(w_1w_2)$, is unique and
well-defined. The action of the braid group generators on the
generators of $\Uq(\kmg)$ is given by
\begin{align*}
  T_i:k_i &\mapsto {k_i}\inv,\\ 
  T_i:k_j &\mapsto k_j {k_i}^{-a_{ij}} \quad (i\neq j),\\ 
  T_i:e_i &\mapsto -k_i f_i,\\ 
  T_i:e_j &\mapsto \sum_{n=0}^{-a_{ij}} (-1)^n q_i^{a_{ij}+n}
  e_i^{(n)} e_j e_i^{(-a_{ij}- n)} \quad (i\neq j),\\ 
  T_i:f_i &\mapsto -{k_i}\inv e_i,\\ 
  T_i:f_j &\mapsto \sum_{n=0}^{-a_{ij}} (-1)^n q_i^{-a_{ij}-n}
  f_i^{(-a_{ij}- n)} f_j f_i^{(n)} \quad (i\neq j).
\end{align*}
The action of the inverse elements is given by $T_i\inv=
\varphi\compose T_i\compose \varphi\inv$ ($i\in\range{0,r}$). Note
also that $T_i\compose \w= \w\compose T_i$ ($i\in\range{0,r}$). The
braid group $\tilde{B}$ and its action on $\Uq(\kmg)$ can be
extended~\cite{beck:qkm-braid-group} to include also the finite group
$\cs{T}$ with a natural action on $\Uq(\kmg)$, so giving the group
$\hat{B}$ of automorphisms of $\Uq(\kmg)$ associated to
$\kmW=\cs{T}\ltimes \tilde{W}$. For $\tau\in \cs{T}$, define $T_\tau$
which acts on $\Uq(\kmg)$ as
\begin{alignat*}{2}
  T_\tau(e_i) &= e_{\tau(i)}, \qquad&
  T_\tau(f_i) &= f_{\tau(i)},\\
  T_\tau(k_i) &= k_{\tau(i)}. &&
\end{alignat*}
The group $\hat{B}$ is then generated by $\set{T_i^{\pm1}, T_\tau\mid
  i\in\range{0,r}, \tau\in\cs{T}}$.

\subsection{Lattice of translations}
The subgroup $P\czek$ of $\kmW$, induces a subgroup $\cs{P}$ (the
lattice of translations) of $\hat{B}$, whose generators are
$\set{(T_{\omega\czek_i})^{\pm1}\mid i\in\range{1,r}}$. 

\begin{remark}
  Note the action of $\hat{B}$ on $\Uq(\kmg)$ is only known in terms
  of the generators $\set{T_i,T_\tau\mid i\in\range{0,r}, \tau\in
    \cs{T}}$ that correspond to the $\cs{T}\ltimes \tilde{W}$ form of
  $\kmW$: calculating explicitly the bijection between $W\ltimes
  P\czek$ and $\cs{T}\ltimes \tilde{W}$ is somewhat complicated.  From
  here on this bijection will be assumed and braid group generators
  $\set{T_i, T_{\omega\czek_i}\mid i\in\range{1,r}}$ corresponding to
  the $W\ltimes P\czek$ form of $\kmW$ will be used.
\end{remark}

\begin{notation}
  Define an orientation (colouring) of the vertices of the Dynkin
  diagram of $\g$, by a map $o:\range{1,r}\mapto \set{1,-1}$, such
  that if $a_{ij}<0$ then $o(i)o(j)=-1$. Then define
  $\hat{T}_{\omega\czek_i}:= o(i) T_{\omega\czek_i}$.
\end{notation}

\section{Drinfeld's presentation}

In~\cite{drinfeld:presentation} Drinfeld introduced an important loop-like
presentation of $\Uq(\kmg)$ which I now describe. It turns out to be very
useful for many calculations and applications of quantum affine
Kac-Moody algebras.


\begin{definition} \cite{drinfeld:presentation}
  \label{km:drinfeld-presentation}
  Let $\g$ be a simple, rank~$r$ Lie algebra, with Cartan matrix
  $(a_{ij})_{i,j\in\set{1,\ldots,r}}$. The {\em centrally extended
    quantum loop algebra\/} $\Uq(\hatL{\g})$ of $\g$ with derivation
  is defined to be the associative $\C(q)$-algebra with unity $1$ and
  generators $\set{E_m^{+,i}, E_m^{-,i}, H_n^i, K_i^{\pm1},
    \gamma^{\pm \half}, \deriv^{\pm1} \mid i \in \range{1,r}, \: m\in
    \Z, \: n\in \Zcross}$ ($\gamma^{\pm \half}$ is central) satisfying
  the relations:
  \begin{alignat*}{2}
    \gamma \cdot \gamma\inv &= 1 = \gamma\inv \cdot \gamma, \qquad &
    \deriv \cdot \deriv\inv &= 1 = \deriv\inv \cdot \deriv,\\ 
    K_i \cdot K_j &= K_j \cdot K_i, &
    K_i \cdot K_i\inv &= 1 = K_i\inv \cdot K_i,\\ 
    \deriv \cdot K_i &= K_i \cdot \deriv, &&\\ 
    K_i \cdot H_n^j &= H_n^j \cdot K_i, \qquad &
    \deriv \cdot H_n^i \cdot \deriv\inv &= q^n H_n^i,\\ 
    K_i \cdot E_m^{\pm,j} \cdot K_i\inv &= q_i^{\pm a_{ij}}
    E_m^{\pm,j}, &
    \deriv \cdot E_m^{\pm,i} \cdot \deriv\inv &= q^m E_m^{\pm,i},
  \end{alignat*}
  \begin{align*}
    \comm{H_m^i,H_n^j} &= \delta_{m+n,0}
    \frac{\brackets{a_{ij}m}_{q_i}}{m} \cdot \frac{\gamma^m -
      \gamma^{-m}}{q_j-q_j\inv},\\ 
    \comm{H_m^i,E_n^{\pm,j}} &= \pm \frac{\brackets{a_{ij}m}_{q_i}}{m}
    \gamma^{\mp \frac{\modulus{m}}{2}} E_{m+n}^{\pm,j},\\ 
    \comm{E_m^{+,i},E_n^{-,j}} &= \delta^{ij}\frac{\gamma^{\half(m-n)}
      \psi_{m+n}^{+,i} - \gamma^{-\half(m-n)}
      \psi_{m+n}^{-,i}}{q_i-q_i\inv},
  \end{align*}
  \begin{displaymath}
    E_{m+1}^{\pm,i} \cdot E_n^{\pm,j} - q_i^{\pm a_{ij}} E_n^{\pm,j}
    \cdot E_{m+1}^{\pm,i} = q_i^{\pm a_{ij}} E_m^{\pm,i} \cdot
    E_{n+1}^{\pm,j} - E_{n+1}^{\pm,j} \cdot E_m^{\pm,i},
  \end{displaymath}
  \begin{multline*}
    \sum_{n=0}^{1-a_{ij}} \sum_{\sigma\in S_{1-a_{ij}}} (-1)^n
    \comb{1-a_{ij}}{n}_{q_i} E_{m_{\sigma(1)}}^{\pm,i} \cdots
    E_{m_{\sigma(n)}}^{\pm,i} \cdot E_p^{\pm,j} \cdot
    E_{m_{\sigma(n+1)}}^{\pm,i} \cdots \;
    E_{m_{\sigma(1-a_{ij})}}^{\pm,i}\\ 
    = \quad 0 \qquad (i\neq j).
  \end{multline*}
  The last two relations are the $q$-Serre relations. I call the
  former the quadratic Serre relations and the latter simply the Serre
  relations. The $\set{\psi_m^{+,i},\psi_m^{-,i} \mid i\in\range{1,r},
    \: m\in\Z}$ are defined by equating powers of $z$ in the formal series:
  \begin{align*}
    \sum_{m\in\Z} \psi_m^{+,i} z^{- m} &:= K_i \exp
    \left( (q_i-q_i\inv) \sum_{n=1}^\infty H_n^i z^{- n} \right),\\
    \sum_{m\in\Z} \psi_{-m}^{-,i} z^m &:= K_i\inv \exp
    \left( -(q_i-q_i\inv) \sum_{n=1}^\infty H_{-n}^i z^n \right).
  \end{align*}
\end{definition}

\begin{lemma}
  There is a family of $\Uq(\g)$ subalgebras of $\Uq(\hatL{\g})$ 
  parametrised by the elements $m\in\Z$. Let $\set{k_i^{\pm1}, e_i,
    f_i}$ be the standard generators of $\Uq(\g)$. The embedding of each
  subalgebra $\Uq(\g) \embed \Uq(\hatL{\g})$ is given by the
  $\C(q)$-algebra homomorphism:
  \begin{displaymath}
    k_i \mapsto \gamma^m K_i, \qquad
    e_i \mapsto E_m^{+,i}, \qquad f_i \mapsto E_{-m}^{-,i}.
  \end{displaymath} 
  The `horizontal' subalgebra corresponds to the case with $m=0$.
\end{lemma}

\begin{proof}
  The homomorphism is easily checked explicitly.
\end{proof}

Recently Beck gave a constructive proof of the following theorem by
Drinfeld.

\begin{theorem} \label{km:drinfeld-thm}
  \cite{drinfeld:presentation} The quantum affine Kac-Moody algebra
  $\Uq(\kmg)$ extended by a central element $C^{\half}$ (satisfying
  $C=\prod_{i=0}^r {k_i}^{a_i}$) is isomorphic as an algebra to the
  centrally extended quantum loop algebra $\Uq(\hatL{\g})$.
\end{theorem}

\begin{proof}
  The theorem is proved in~\cite{beck:qkm-braid-group}, where the
  isomorphism $\Uq(\kmg)\overset{\isomorphic}{\longrightarrow}
  \Uq(\hatL{\g})$ is shown to be:
  \begin{gather*}
    \begin{alignedat}{3}
      \gamma^{\half} &= C^{\half}, &
      K_i &= k_i, \qquad&& (i\in\range{1,r})\\ 
      E_m^{+,i} &= (\hat{T}_{\omega\czek_i})^{-m} e_i, \qquad &
      E_m^{-,i} &= (\hat{T}_{\omega\czek_i})^m f_i, \qquad&& (m\in \Z)\\ 
      \deriv &= \deriv, &&\\ 
      \psi_0^{+,i} &= k_i, &
      \psi_0^{-,i} &= k_i\inv, \qquad\quad && (i\in\range{1,r})
    \end{alignedat}
    \\
    \begin{alignedat}{2}
      \psi_m^{+,i} &= (q_i- q_i\inv) C^{\frac{m}{2}}
      \comm{e_i,\hat{T}_{\omega\czek_i}^m f_i} \qquad&& (m\in\Zplus),\\ 
      \psi_{-m}^{-,i} &= \w(\psi_m^{+,i})= (q_i- q_i\inv)
      C^{-\frac{m}{2}} \comm{f_i,\hat{T}_{\omega\czek_i}^m e_i}
      \qquad&& (m\in\Zplus),\\ 
      H_m^i &= k_i\inv \left( \frac{\psi_m^{+,i}}{q_i- q_i\inv}-
      \frac{1}{m} \sum_{p=1}^{m-1} p\psi_{m-p}^{+,i} H_p^i \right)
      \qquad&& (m\in\Zplus),\\ 
      H_{-m}^i &= \w(H_m^i) \qquad&& (m\in\Zplus).
    \end{alignedat}
  \end{gather*}
  The elements in $\Uq(\hatL{\g})$ on the left-hand side of each
  equation are the image under the map of the elements in $\Uq(\g)$ on
  the right-hand side.
\end{proof}

\begin{corollary}
  The derived quantum Kac-Moody algebra $\Uq(\kmg')$ is isomorphic to
  $\Uq(\hatL{\g}')$, the subalgebra of $\Uq(\hatL{\g})$ without the
  derivation $\deriv$.
\end{corollary}

The elements $H_m^i$ and $\psi_m^{\pm,i}$ are fixed under the action
of the translations $\hat{T}_{\omega\czek_i}$ and $\gamma
\hat{T}_{\omega\czek_i}$ respectively
(see~\cite[\S4]{damiani:PBW-kmsl2}):
\begin{corollary}
  Let $m\in\Zplus$. Then
  \begin{align*}
    \hat{T}_{\omega\czek_i} \psi_m^{+,i} &= \gamma\inv \psi_m^{+,i},\\
    \hat{T}_{\omega\czek_i} H_m^i &= H_m^i.
  \end{align*}
\end{corollary}

\begin{remark}
  In Drinfeld's presentation of $\Uq(\kmg)$, the Cartan subalgebra
  $\Uq(\kmh)$ is generated by $\set{K_i^{\pm1}, \gamma^{\pm\half},
    \deriv^{\pm1}}$. The set $\set{E_0^{+,i}, E_m^{\pm,i}, H_m^i \mid
    m\in \Zplus}$ generates $\Uq(\kmn_+)$ and $\set{E_0^{-,i},
    E_m^{\pm,i}, H_m^i \mid m\in \Zminus}$ generates $\Uq(\kmn_-)$.
\end{remark}

\subsection{Basis} \label{km:Uq(kmg)-basis}
For fixed $m\in\Zcross$, the generators $\set{H_m^i\mid i\range{1,r}}$
are the imaginary root vectors associated to the imaginary root
$m\delta$.  The generators $\set{H_m^i\mid i\in\range{1,r},
  m\in\Zplus}$ are the positive imaginary root vectors and
$\set{H_{-m}^i\mid i\in\range{1,r}, m\in\Zplus}$ the negative
imaginary root vectors.  The generator $E_m^{\pm,i}$ corresponds to
the real root $\pm\alpha_i+ m\delta$.

Let $s_{i_1}s_{i_2}\dots s_{i_n}$ be a reduced expression of the
longest element $w_0\in W$. Then the ordered set $\set{\beta_k:=
  s_{i_1}s_{i_2}\cdots s_{i_{k-1}}\alpha_{i_k}\mid k\in \range{1,n}}$
is $R_+$ with an ordering. For each $m\in\Z$, I define
\begin{align*}
  E_m^{+,\beta_k} &:= T_{i_1} T_{i_2}\cdots T_{i_{k-1}} E_m^{+,i_k},\\
  E_{-m}^{-,\beta_k} &:= \w(E_m^{+,\beta_k}).
\end{align*}
Then the set $\set{E_0^{+,\beta_k}, E_m^{\pm,\beta_k}, H_m^i\mid
  m\in\Zplus, k\in\range{1,n}}$ generates a basis of $\Uq(\kmn_+)$ in
an appropriate order (such as the one in~\ref{km:ordering-of-kmR+}).
Similarly $\set{E_0^{-,\beta_k}, E_m^{\pm,\beta_k}, H_m^i\mid
  m\in\Zminus, k\in\range{1,n}}$ generates a basis of $\Uq(\kmn_-)$.

The basis of $\Uq(\kmh)$ is simply
\begin{displaymath}
  \Uq(\kmh)= \sum_{m_i,m_c,m_d\in\Z} K_1^{m_1}\cdots K_r^{m_r}
  \gamma^{\frac{m_c}{2}} \deriv^{m_d}.
\end{displaymath}


\subsection{}
The Hopf algebra structure of $\Uq(\hatL{\g})$ is not known explicitly
in terms of Drinfeld's generators. However Chari and Pressley have
given some partial results for the coproduct of
$\Uq(L(\sltwo))$~\cite{chari-pressley:eval-reps}.  Beck recently
presented a construction of the coproduct for the real root vectors
$\set{E_m^{\pm,i}|m\in\Z}$ in $\Uq(\hatL{\g})$ using the
$\cs{R}$-matrix of $\Uq(\kmg)$. From this the coproduct for the
imaginary root vectors can also be deduced.

\subsection{} \label{km:heisenberg-subalgebra} 
The set $\set{K_i, H_m^i, \gamma^{\pm\half} \mid
  m\in\Z}_{i\in\range{1,r}}$ generates a subalgebra $\Uq(\H)$ of
$\Uq(\kmg)$ called the {\em Heisenberg subalgebra}. If I adjoin the
formal central elements $\ln(q)$ and $\ln(q)\inv$ to the algebra, then
the generators $\set{H_m^i,K_i,x^i\mid i\in\range{1,r}}$ ($x^i$
conjugate to the zero mode momenta $K_i$:
$\comm{K_i,x^j}=\ln(q)\delta^{ij}K_i$) are Fourier modes of
$q$-analogue 2d chiral bosonic quantum fields $\set{\phi^i(z)}$:
\begin{align*}
  \phi^i(z) &= x^i + \half(K_i-K_i\inv)\frac{\ln(z)}{\ln(q_i)} +
  \sum_{n\in\Zcross}
  \frac{\gamma^{\half\modulus{n}}}{\brackets{n}_{q_i}} H_n^i z^{-n}\\ 
  \partial_{q_i} \phi^i(z) &= \frac{K_i-K_i\inv}{q_i-q_i\inv}
  z\inv + \sum_{n\in\Zcross} \gamma^{\half \modulus{n}} H_n^i z^{-n-1}.
\end{align*}
Here the finite difference operator acts as $\partial_q f(z):=
\frac{f(qz)-f(q\inv z)}{(q-q\inv)z}$ on any complex function $f(z)$.
These $q$-bosonic fields play a crucial role in the constructions of
bosonisations of quantum affine Kac-Moody algebras and their vertex
operators (see the references cited in~\ref{intro:qkm-generic} and
references therein).


\begin{remark}
  Setting $\gamma^{\half}$ to $1$ in $\Uq(L(\g))$, corresponds to
  having zero central extension in the classical limit. I define the
  quantum loop algebra of $\g$ (with zero central extension) to be
  $\Uq(L(\g)):= \Uq(\hatL{\g})/ \ideal{\gamma^{\half} -1}$.  The
  subalgebra $\Uq(\H)/ \ideal{\gamma^{\half} -1}$ of $\Uq(L(\g))$
  forms a commutative subalgebra. So the Cartan subalgebra of
  $\Uq(L(\g))$ is generated by $\set{K_i^{\pm1},
    \deriv^{\pm1}}\union \Uq(\H)/ \ideal{\gamma^{\half}-1}$.
\end{remark}


\begin{lemma} \label{km:quad-Serre-reln}
  Let $m,n\in\Z$ be two integers such that $m>n$.

  (a) If $m=n+1$
  \begin{displaymath}
    E_m^{\pm,i} E_n^{\pm,i} - q_i^{\pm2} E_n^{\pm,i} E_m^{\pm,i} =
    0.
  \end{displaymath}

  (b) For $i,j\in\range{1,r}$
  \begin{multline*}
    E_m^{\pm,i} E_n^{\pm,j} - q_i^{\pm a_{ij}} E_n^{\pm,j}
    E_m^{\pm,i} =
    (q_i^{\pm 2a_{ij}}- 1) \sum_{p=1}^{\ell} q_i^{\pm(p-1)a_{ij}}
    E_{n+p}^{\pm,j} E_{m-p}^{\pm,i}\\ 
    + q_i^{\pm \ell a_{ij}} (q_i^{\pm a_{ij}}
    E_{m-\ell-1}^{\pm,i} E_{n+\ell+1}^{\pm,j} - E_{n+\ell+1}^{\pm,j}
    E_{m-\ell-1}^{\pm,i}).
  \end{multline*}

  (c) If $m>n+2$ and $m-n$ is odd, then

  \begin{displaymath}
    E_m^{\pm,i} E_n^{\pm,i} - q_i^{\pm2} E_n^{\pm,i} E_m^{\pm,i} =
    (q_i^{\pm4}- 1) \sum_{j=1}^{m-n-1} q_i^{\pm2(j-1)} E_{n+j}^{\pm,i}
    E_{m-j}^{\pm,i}.
  \end{displaymath}

  (d) If $m>n+1$ and $m-n$ is even,
  \begin{multline*}
    E_m^{\pm,i} E_n^{\pm,i} - q_i^{\pm2} E_n^{\pm,i} E_m^{\pm,i} =
    (q_i^{\pm4}- 1) \sum_{j=1}^{m-n-2} q_i^{\pm2(j-1)} E_{n+j}^{\pm,i}
    E_{m-j}^{\pm,i}\\ 
    + (q_i^{\pm2}-1)q_i^{m-n-2} E_{\frac{m+n}{2}}^{\pm,i}
    E_{\frac{m+n}{2}}^{\pm,i}.
  \end{multline*}
\end{lemma}

\begin{proof}
  (a) The $m=n+1$ case follows immediately from the quadratic Serre
  relations. The proof of (b) is by induction for fixed
  $n$ using the quadratic Serre relations recursively. (c) and (d) are
  special cases of (b).
\end{proof}

\begin{remark}
  Note that in the case of $\Uq(\hatL{\sltwo})$ there are no Serre
  relations just as in the classical case, but there are the quadratic
  Serre relations. So lemma~\ref{km:quad-Serre-reln} allows us to
  order the sets of root vectors $\set{E_m^+}_{m\in\Z}$ and
  $\set{E_m^-}_{m\in\Z}$ in $\Uq(\hatL{\sltwo})$ and within the
  $\Uq(\kmsltwo)$ subalgebras of $\Uq(\hat{\g})$. For $i,j$ such
  that $a_{ij}=0$, the Serre relation tells us that $E_m^{\pm,i}$ and
  $E_n^{\pm,j}$ commute for all $m,n\in\Z$ as for a classical affine
  Kac-Moody algebra, so the quadratic Serre relation is trivially
  satisfied. On the other hand for $i,j$ such that $a_{ij}<0$, the
  quadratic Serre relation gives additional relations to
  the Serre relation.
\end{remark}


\subsection{} \label{km:classical-extended-loop-algebra}
Let $\set{\bar{H}_m^i, \bar{E}_m^{\pm,i}, \bar{k}, \bar{d} \mid
  m\in\Z}$ denote the usual loop algebra presentation of $\U(\kmg)$ with
central element $\bar{k}$ and derivation $\bar{d}$). The generators
satisfy the relations:
\begin{align*}
  \comm{\bar{H}_m^i,\bar{H}_n^j} &= \delta_{m+n} a_{ij} m\bar{k},\\ 
  \comm{\bar{H}_m^i,\bar{E}_n^{\pm,j}} &= \pm a_{ij}
  \bar{E}_{m+n}^{\pm,j},\\ 
  \comm{\bar{E}_m^{+,i},\bar{E}_n^{-,j}} &= \delta^{ij}
  (\bar{H}_{m+n}^i + m\bar{k}),\\ 
  \comm{\bar{E}_m^{\pm,i},\bar{E}_n^{\pm,i}} &= 0,\\
  \comm{E_{m_1}^{\pm,i},\comm{E_{m_2}^{\pm,i},\comm{\ldots,
        \comm{E_{m_{1-a_{ij}}}^{\pm,i},E_{n}^{\pm,j}}}\ldots}} &= 0
  \qquad (i\neq j).
\end{align*}

\begin{lemma}[classical limit] \label{km:drinfeld-classical-lim}
  Define $\UA(\hatL{\g})$ to be the $\Cfree{q,q\inv}$-subalgebra of
  $\Uq(\hatL{\g})$ generated by
  \begin{displaymath}
    H_0^i:=\frac{K_i - K_i\inv}{q_i-q_i\inv},\quad
    d:=\frac{\deriv- \deriv\inv}{q-q\inv},\quad
    k:=\frac{\gamma- \gamma\inv}{q- q\inv},
  \end{displaymath}
  and all the generators of $\Uq(\hatL{\g})$. Let $\e\in \Ccross$.
  Define the specialisation of $\UA(\hatL{\g})$ at $q=\e$ to be
  $\Ue(\hatL{\g}):= \UA(\hatL{\g})/\ideal{q-\e}$. The quotient algebra
  $\U_1(\kmg)/ \ideal{\deriv-1, \gamma^{\half} -1, K_i-1| i\in
    \range{1,r}}$ is $\C$-algebra isomorphic to $\U(\kmg)$:
  \begin{alignat*}{2}
    H_m^i &\mapsto \bar{H}_m^i, \qquad&
    E_m^{\pm,i} &\mapsto \bar{E}_m^{\pm,i}, \qquad(m\in\Z)\\ 
    k &\mapsto \bar{k}, & d &\mapsto \bar{d}.
  \end{alignat*}
\end{lemma}


\begin{remark}
  From lemma~\ref{km:quad-Serre-reln}, it follows that the quadratic
  Serre relations with $i=j$ (and in particular for $\Uq(\hatL{\sltwo})$)
  become trivial at the specialisation $q=1$, as would be expected for
  consistency. The quadratic Serre relations have no classical
  analogue with $i=j$: $\comm{\bar{E}_m^{\pm,i}, \bar{E}_n^{\pm,i}}=0$
  in $\U(\hatL{\g})$.
\end{remark}

\subsection{} \label{km:drinfeld-cartan-involution}
In Drinfeld's presentation the $\C$-algebra anti-automorphism $\w$
(Cartan involution) takes the form
  \begin{alignat*}{2}
    K_i &\mapsto K_i\inv, \qquad&
    q &\mapsto q\inv,\\
    \gamma^{\half} &\mapsto \gamma^{-\half}, &
    \deriv &\mapsto \deriv\inv,\\
    E_m^{\pm,i} &\mapsto E_{-m}^{\mp,i}, &
    H_m^i &\mapsto H_{-m}^i.
  \end{alignat*}

\begin{definition}
  There is a natural homogeneous $\Z$-gradation of $\Uq(\hatL{\g})$ given
  by:
  \begin{alignat*}{3}
    \deg K_i&=0, \qquad&
    \deg \gamma^{\half}&=0, \qquad&
    \deg \deriv &=0,\\
    \deg H_m^i&=z^m, & \deg E_m^{\pm,i}&=z^m. &&
  \end{alignat*}
\end{definition}

\subsection{The evaluation map}
Let $n>1$ and let $\kmg$ be a affine Lie algebra of type
$A_{n-1}^{(1)}\isomorphic \widehat{\sl}_{n}$.  The evaluation map of
the quantum affine Kac-Moody algebra $\Uq(\kmsln)\mapto
\Uq(\sln)\tensor \Cfree{z,z\inv}$ was first constructed by
Jimbo~\cite{jimbo:toda}.  Chari and
Pressley~\cite{chari-pressley:eval-reps} studied the evaluation
representations (corresponding to the homogeneous gradation) of
$\Uq(\kmsltwo)$ in detail. Apparently it is not possible to construct
evaluation representations of $\Uq(\kmg)$, for $\kmg$ of affine type
other than $A_{n-1}^{(1)}$.

The evaluation map $\ev_z$ is useful because it allows one to
construct loop modules of $\Uq(\kmsln')$ over $\Uq(\sln)$-modules and
also to construct parameter dependent solutions of the Yang-Baxter
equation (see also~\cite{bracken-delius-gould-zhang:vertex-models}).
When the formal parameter $z$ is taken to be a non-zero complex number
$a$ then $\ev_a:\Uq(\kmsln') \mapto \Uq(\sln)$ and irreducible finite
dimensional representations of $\Uq(\kmsln')$ can be obtained.  Note
that an evaluation map of $\Uq(\kmsln')$ onto a finite dimensional
highest weight $\Uq(\kmsln)$-module is not in general a highest weight
representation of $\Uq(\kmsln')$, since $e_0$ and $f_0$ are mapped
into the $\Uq(\n_-)$ and $\Uq(\n_+)$ subalgebras of $\Uq(\sln)$
respectively. The evaluation map is {\em not\/} a Hopf algebra map, so
in particular the tensor product structure of $\Uq(\kmsln)$-loop
modules is different from that of the corresponding
$\Uq(\sln)$-modules. The evaluation representations of
$\Uq(\kmsltwo')$ have been classified by Chari and
Pressley~\cite{chari-pressley:eval-reps}.


\begin{remark}
  Recently an interesting two parameter (elliptic) deformation of
  $\kmsltwo$ was
  introduced~\cite{foda-iohara-jimbo-kedem-miwa-yan:elliptic-kmsl2-notes}.
\end{remark}

\section{Heisenberg subalgebra}
\label{km:heisenberg-section}

In this section I consider the Heisenberg subalgebra $\Uq(\H)$ of
$\Uq(\kmg)$, which has the generators
$\set{H_n^i,K_i^{\pm1},\gamma^{\pm\half}\mid
  n\in\Zcross}_{i\in\range{1,r}}$. The generators $K_i^{\pm1}$ and
$\gamma^{\pm\half}$ are central in $\Uq(\H)$ and
\begin{displaymath}
  \comm{H_m^i,H_n^j}= \delta_{m+n,0}
  \frac{\brackets{a_{ij}m}_{q_i}}{m} \cdot \frac{\gamma^m-
    \gamma^{-m}}{q_j-q_j\inv}
  \qquad (m,n\in\Zcross;i,j\in\range{1,r}).
\end{displaymath}
Let $\Uq(\H^+)$ and $\Uq(\H^-)$ denote the commutative subalgebras of
$\Uq(\H)$ generated by $\set{H_n^i\mid n\in\Zplus}$ and
$\set{H_n^i\mid n\in\Zminus}$ respectively.

\begin{remark}
  The Heisenberg algebra of $\Uq(\H)$ of $\Uq(\kmsltwo)$ is
  essentially an infinite dimensional generalisation of the Heisenberg
  subalgebra of the $q$-oscillator algebra mentioned
  in~\ref{h4:CGST-oscillator}.
\end{remark}

\subsection{}
Let $\UA(\H)$ denote the $\Cfree{q,q\inv}$-subalgebra of $\Uq(\H)$
generated by $\set{\gamma^{\pm\half}, k:=\frac{\gamma- \gamma\inv}{q-
    q\inv}, H_0^i:=\frac{K_i- K_i\inv}{q- q\inv},
  H_m^i|m\in\Z,i\in\range{1,r}}$. Let $\e\in\Ccross$. The
specialisation of $\UA(\H)$ at $q=\e$ is $\Ue(\H):=
\UA(\H)/\ideal{q-\e}$.  Let $\U(\H)$ denote the Heisenberg enveloping
subalgebra of $\U(\kmg)$ introduced
in~\ref{km:classical-extended-loop-algebra}, generated by
$\set{\bar{k}, \bar{H}^i_m\mid m\in\Z}$.
{}From~\ref{km:drinfeld-classical-lim} it follows that $\U_1(\H)/
\ideal{\gamma-1, K_i-1| i\in\range{1,r}}\isomorphic \U(\H)$. The
isomorphism is given by $H^i_m\mapsto \bar{H}^i_m$ and $k\mapsto
\bar{k}$.

\subsection{}
Let $F$ be a Fock module over $\C(q)$ of $\Uq(\H)$, generated by an
element $v_0\in F$, called the vacuum vector, such that $\Uq(\H^+)\cdot
v_0=0$ and $F= \Uq(\H^-)\cdot v_0$.

\begin{lemma}[Irreducibility]
  $F$ is an irreducible representation of $\Uq(\H)$, if and only if
  $\gamma^2\cdot v\neq v$ ($v\in F$).
\end{lemma}

\begin{proof}
  The proof is similar to the classical
  case~\cite[2.2]{kac-raina:book}. From any element of $F$ one can
  reach the vacuum vector~$v_0$ by appropriate applications of elements
  of $\Uq(\H^+)$. Then acting with elements of $\Uq(\H^-)$ any other
  element of $F$ can reached. If $\gamma^2\cdot v=v$ ($v\in F$), then
  clearly $\comm{H_m^i,H_n^j}$ acts on $F$ as~$0$, which would give a
  trivial representation.
\end{proof}

\begin{notation}
  A nonzero complex number is called {\em generic\/}, if it is real or
  if it is not equal to a (nontrivial) root of unity.
\end{notation}

\begin{lemma}
  Let $F$ be a $\C(q)$-Fock module of $\Uq(\H)$ with vacuum vector
  $v_0$. $F_\A:= \UA(\H)\cdot v_0$ is a $\UA(\H)$ submodule of $F$.
  If $F$ is an irreducible Fock module of $\Uq(\H)$, then the Fock
  module $F_\A$ over $\UA(\H)$ is also irreducible.

  Let $\e\in\Ccross$. Then $F_\e:=F_\A/\ideal{q-\e}$ is a $\Ue(\H)$
  submodule of $F_\A$.  If $F_\A$ is an irreducible Fock module of
  $\UA(\H)$, then its specialisation $F_\e$ over $\Ue(\H)$ at generic
  $q=\e$ is also irreducible.
\end{lemma}

\subsection{}
Let $F$ be an irreducible Fock module of $\Uq(\H)$.  When the central
elements $\gamma$ and $K_i$ ($i\in\range{1,r}$) of $\Uq(\H)$ act on
$F$ with eigenvalues $q^c$ ($c\in\Z$) and $q^{\alpha_i}$
($\alpha_i\in\Z$, $i\in\range{1,r}$), I call $c$ the {\em level\/} and
$\alpha:=(\alpha_i)_{i\in\range{1,r}}$ the {\em charge\/} (or
momentum) respectively of the Fock module $F$. Write $F_\alpha$
for $F$, when it is necessary to emphasise the charge of $F$.

In the following I will only consider (non-trivial) representations with
non-zero level $c\in\Zcross$.

\begin{lemma}[Intertwiners]
  Let $x^i$ $(i\in\range{1,r})$ be the zero mode coordinate conjugate
  to $K_i$ introduced in~\ref{km:heisenberg-subalgebra}.  Let
  $F_\alpha$ and $F_\beta$ be two level $c$ Fock modules over
  $\Uq(\H)$ with charge $\alpha$ and $\beta$.  Then the map
  $e^{\sum_{i=1}^r (\beta_i -\alpha_i) x^i}:
  F_{\alpha} \mapto F_{\beta}$ is a $\Uq(\H)$-module intertwiner from
  $F_\alpha$ to $F_\beta$.
\end{lemma}

\begin{proof}
  Follows straightforwardly from $K_i\cdot e^{x^j}
  =q^{\delta^{ij}}e^{x^j} \cdot K_i$ and $\comm{H_n^i,x^j}=0$.
\end{proof}


\subsection{}
Recall from~\ref{km:drinfeld-cartan-involution} that the Cartan
involution $\w$ of $\Uq(\H)$ is
\begin{alignat*}{2}
  q &\mapsto q\inv, \qquad&
  H_n^i &\mapsto H_{-n}^i,\\
  K_i &\mapsto K_i\inv, &
  \gamma &\mapsto \gamma\inv,
\end{alignat*}
which extends as an anti-automorphism to all of $\Uq(\H)$. Note that
the elements $\frac{K_i- K_i\inv}{q- q\inv}$ and $\frac{\gamma-
  \gamma\inv}{q- q\inv}$ in $\Uq(\H)$ are invariant under $\w$. Hence
the elements $H_0^i$ and $k$ in $\UA(\H)$ and $\Ue(\H)$ are $\w$
invariant. Clearly $\w$ is also an anti-automorphism of $\UA(\H)$ and
$\Ue(\H)$.

\subsection{}
Let $(\cdot,\cdot):F \times F \mapto \C(q)$ denote the unique scalar
product on $F$, which is contravariant with respect to $\omega$:
\begin{displaymath}
  \begin{aligned}
    (x\cdot v,w) &= (v,\omega(x)\cdot w),\\ 
    (v,x\cdot w) &= (\omega(x)\cdot v,w),
  \end{aligned}
  \qquad \forall x\in \H, v,w\in F
\end{displaymath}
and normalised so that $(v_0,v_0):=1$. Denote by $(\cdot,\cdot)_\A$ and
$(\cdot,\cdot)_\e$ the induced scalar products on $F_\A$ and $F_\e$.


\begin{lemma}
  The triple $\structure{F, (\cdot,\cdot), \w}$ is a
  $\star$-representation of $\Uq(\H)$. The triple $\structure{F_\A,
    (\cdot,\cdot)_\A, \w}$ is a $\star$-representation of $\UA(\H)$.
  Let $\e\in\Ccross$. The triple $\structure{F_\e, (\cdot,\cdot), \w}$
  is a $\star$-representation of $\Ue(\H)$. If $\e\in\Rplus$ then
  $\structure{F_\e, (\cdot,\cdot), \w}$ is a unitary representation of
  $\Ue(\H)$.
\end{lemma}

\begin{proof}
  For $\e$ positive, it is easily checked that the sesquilinear scalar
  product is positive definite. Hence in this case the representation
  is unitary.
\end{proof}

\chapter{Quantum affine algebras at a root of unity}
\label{chap:qkm-root-of-1}

\section{Introduction}


\subsection{}
In this chapter the specialisation of the quantum affine Kac-Moody
algebra $\UA(\kmg)$ at an odd primitive root of unity is studied.
Using the action of Beck's extended braid group $\hat{B}$ introduced
in the previous chapter on some well-known central elements in
$\Ue(\kmg)$, I construct an infinite number of central elements in
$\Ue(\kmg)$ lying in $\Ue(\kmn_\pm)$. In fact I prove that, at an odd
$\l$-th root of unity, the $\l$-th power of every real root vector in
$\Ue(\kmg)$ is in the centre (proposition~\ref{qkm-1:real-root-centre}).

The Heisenberg algebra $\Ue(\H)$ is also studied at odd and even roots
of unity.  It is found that it also contains an infinite number of
central elements: the generators of the extended centre of $\Ue(\H)$
are in fact the generators that have mode number, which is an integer
(half integer in the even case) multiple of $\l$.  Further it turns
out that the extended central elements of $\Ue(\H)$ are also in the
centre of $\Ue(\kmg)$.

The centre $\Ze$ of $\Ue(\kmg)$ is infinite dimensional.  Nevertheless
$\Ue(\kmg)$ is still infinite dimensional over $\Ze$, in contrast with
the finite case.

\subsection{}
Until now, quantum affine algebras at a root of unity have only been
studied with zero central extension. In particular their (finite
dimensional) minimal cyclic representations has been
constructed~\cite{date-jimbo-miki-miwa:cyclic-kmsl3,%
  date-jimbo-miki-miwa:potts,arnaudon-chakrabarti:flat-periodic,%
  chari-pressley:min-cyclic}. In this chapter infinite dimensional
representations of $\Uq(\kmg)$ and $\Ue(\kmg)$ are also discussed. The
results on the representation theory are not definitive, but I am able
to construct some new representations of $\Ue(\kmg)$, which are
infinite dimensional and either nilpotent or semicyclic. It is also
possible to construct cyclic representations, on which all the real
root vectors act injectively.

The Fock modules over the Heisenberg subalgebra $\Ue(\H)$ become
reducible at the root of unity because of the new central elements.
However it is possible to quotient by a maximal submodule generated by
all the singular vectors giving an irreducible representation.  An
alternative approach is to take a Lusztig form of $\UA(\H)$, in this
case at the specialisation at a root of unity the algebra is still
isomorphic to the classical Heisenberg subalgebra.

\subsection{Historical note}
Let me mention how I came upon the results concerning the central
elements generated by the real root vectors in the algebra. Initially
in the Summer of 1993, I started by considering Drinfeld's
presentation of $\Uq(\sltwo)$ and made a conjecture that the
generators corresponding to real root vectors to the $\l$-th power
were central. For a long time I could only check the conjecture for
generators with small mode number by rather long tedious calculations
and with help from the symbolic manipulation programming language {\sc
  Form}~\cite{vermaseren:form}. Then at the end of April 1994 I
received the paper of Beck that proved the isomorphism between the
Drinfeld and Chevalley type presentations of $\Uq(\kmg)$. After I had
read this paper, I then realised that I could apply his extension of
the braid group action on $\Ue(\kmg)$ to prove my conjecture for
general $\kmg$.

\section{Heisenberg subalgebra at a root of unity}

\subsection{}
Let $\l$ be an integer such that $\l>2$ and $\l>d_i$
($i\in\range{0,r}$) and fix $\e$ to be an primitive $\l$-th root of
unity ($\e^\l=1$). As usual define
\begin{displaymath}
  \l'=
  \begin{cases}
    \l & \text{if $\l$ is odd},\\
    \frac{\l}{2} &\text{if $\l$ is even}.
  \end{cases}
\end{displaymath}

\begin{proposition}[centre] \label{qkm-1:H-centre}
  The elements $\set{H_{m\l'}^i \mid m\in\Zcross, i\in\range{1,r}}$
  are central in $\Ue(\H)$ at the root of unity $\e$. In fact these
  elements are also central in $\Ue(\kmg)$ at the root of unity.
\end{proposition}

\begin{proof}
  This follows immediately from the commutation relations
  in~\ref{km:drinfeld-presentation}, since $\brackets{m\l'}_\e= 0$ for
  all $m\in\Z$.
\end{proof}

\subsection{}
So at an $\l$-th root of unity the centre of the Heisenberg algebra
$\Ue(\H)$ is infinite dimensional. It is generated by
$\set{\gamma^{\pm\half}, K_i^{\pm1}, H_{m\l'}^i\mid m\in\Zcross,
  i\in\range{1,r}}$. The generators $H^i_m$, with mode number $m$
equaling a multiple of $\l'$, `decouple'
into the centre of the algebra.

\begin{corollary}
  The Fock module $F_\e$ of $\Ue(\H)$ at the root of unity $\e$ is
  reducible.
\end{corollary}

\begin{proof}
  This is clear since the central elements $H^i_{-m\l'}$ ($m\in\Zplus$,
  $i\in\range{1,r}$) generate an infinity of singular vectors in
  $F_\e$.
\end{proof}

\subsection{}
At the root of unity I define a new triangular decomposition of
$\Ue(\H)$. Let $\Ue(\H'_+)$, $\Ue(\H'_0)$ and $\Ue(\H'_-)$ be the
commutative algebras generated respectively by $\set{H_m^i|m\in
  \Zplus\setminus \l'\Zplus}$, $\set{\gamma^{\pm\half}, K_i^{\pm1},
  H_n^i| n\in \l'\Zcross}$ and $\set{H_m^i| m\in \Zminus\setminus
  \l'\Zminus}$.

\begin{lemma} \label{qkm-1:H-new-fock}
  Define $F'_\e$ to be a $\Ue(\H)$-module at the root of unity $\e$
  generated by a vector $v'_0$, such that $\Ue(\H'_+)\cdot v'_0=0$, $F'_\e=
  \Ue(\H'_-)\cdot v'_0$, and $\gamma\cdot v'_0=\e^c v'_0$ ($c\in\Ccross$).
  If $c\not\in \l'\Z$, then $F'_\e$ is an irreducible $\Ue(\H)$-module.
\end{lemma}

\begin{proof}
  At level $c$, $\comm{H^i_m, H^j_n}$ acts on $F'_\e$ as
  \begin{equation} \label{qkm-1:H-relation-in-module}
    \comm{H^i_m,H^j_n}= \delta_{m+n,0}
    \frac{\brackets{a_{ij}m}_{\e_i}}{m} \cdot \brackets{mc}_{\e_j}.
  \end{equation}
  The module is a highest weight module. If $c\in\l'\Z$, then the
  right-hand side of~\eqref{qkm-1:H-relation-in-module} is always
  zero. Therefore every element of $F_\e'$ is singular and the module
  is reducible. For $c\not\in \l'\Z$, if $i,j\in\range{1,r}$ are such
  that $a_{ij}\neq0$, then $\comm{H^i_m,H^j_{-m}}$ ($m\in\Zcross$)
  acts as a nonzero number on $F'_\e$ and the module $F'_\e$ is
  irreducible, since it contains no singular vectors.
\end{proof}

\begin{remark}
  Consider the $q_i$-boson field $\phi^i(z)$ (defined
  in~\ref{km:heisenberg-subalgebra}).  It is interesting to note that
  the terms in $\phi^i(z)$ that diverge at the root of unity $\e$ are
  exactly those in the central elements $H_{m\l'}^i\in\Ue(\H'_0)$
  (diverging by a factor $\frac{1}{\brackets{m\l'}_{\e_i}}$).  This
  fact may be useful for the construction the vertex operators and
  bosonisation of $\Ue(\kmg)$ at a root of unity.
\end{remark}

\subsection{}
Fix $c\in\Z$. Consider the algebra $\Uq(\tilde{\H}):= \Uq(\H)/
\ideal{\gamma- q^c}$ with generators $\tilde{K}^i:= K_i$ (central) and
$\tilde{H}_m^i:= \frac{H_m^i}{\brackets{m}_{q_i}}$, which satisfy
\begin{displaymath}
  \comm{\tilde{H}_m^i, \tilde{H}_n^j}= \delta_{m+n,0}
  \frac{\brackets{a_{ij}m}_{q_i}}{m\brackets{m}_{q_i}} \cdot
  \frac{\brackets{mc}_{q_j}}{\brackets{n}_{q_j}}
  \qquad (m,n\in\Zcross;i,j\in\range{1,r}).
\end{displaymath}
Note that the right hand side is well defined in $\C(q)$.  Let
$\Ue(\tilde{\H})$ be the specialisation of $\UA(\tilde{\H})$. It is
well-defined at the root of unity $\e$. The right hand side is also
well defined at the root of unity.  $\tilde{H}^i_{m\l'}$ are not
central in $\Ue(\tilde{\H})$: the centre of $\Ue(\tilde{\H})$ is {\em
  not\/} enlarged at the root of unity $\e$. I thank Tetsuji Miwa for
giving me this result, about this Lusztig-like form $\Ue(\tilde{\H})$
of $\Ue(\H)$.

\section{The centre of $\Ue(\kmg)$ at a root of unity}
\label{qkm-1:centre-of-Uqkmg}

Let $\l$ be an {\em odd\/} integer, such that $\l>d_i$ ($\forall
i\in\range{0,r}$) and fix $\e$ to be an odd primitive $\l$-th root of
unity ($\e=e^{2\pi i/\l}$).  Then the following fact about the Cartan
subalgebra generators is well known.

\begin{lemma}
  The elements $\set{K_i^{\pm \l}, {\deriv}^{\pm \l}\mid
    i\in\range{1,r}}$ are central in $\Ue(\kmg)$.
\end{lemma}

\subsection{}
Recall from chapter~\ref{chap:quantum-enveloping-algebras} the
following result concerning the centre of $\Ue(\g)$ at the root of
unity $\e$, which is a special case of
corollary~\ref{quea:centre-corol}.

\begin{lemma}
  The elements $e_i^\l$ and $f_i^\l$ ($i\in\range{1,r}$) lie in the
  centre of $\Ue(\g)$ at the root of unity $\e$.
\end{lemma}

\begin{proposition} \label{qkm-1:zero-modes^l-central}
  At the root of unity $\e$, the elements
  \begin{displaymath}
    \set{e_i^\l, f_i^\l\mid i\in \range{0,r}}
  \end{displaymath}
  lie in the centre of $\Ue(\kmg)$.
\end{proposition}

\begin{proof}
  The proposition follows from lemmas~\ref{quea:e^m-f-identity}
  and~\ref{quea:e^m-e-identity} with $\structure{\g, (a_{ij})_{i,j\in
      \range{1,r}}}$ replaced by $\structure{\kmg, (a_{ij})_{i,j\in
      \range{0,r}}}$.
\end{proof}

\begin{proposition} \label{qkm-1:real-root-centre}
  Let $\set{E_m^{\pm,\beta_k}\mid m\in \Z, k\in\range{1,n}}$ be the
  basis of the real root vectors of $\Ue(\kmg)$ introduced
  in~\ref{km:Uq(kmg)-basis}. The elements in the following set are in
  the centre of $\Ue(\kmg)$ at the root of unity $\e$
  \begin{displaymath}
    \set{(E_m^{+,\beta_k})^\l, (E_m^{-,\beta_k})^\l\mid m\in \Z,
      k\in\range{1,n}}.
  \end{displaymath}
\end{proposition}

\begin{proof}
  The case $m=0$ corresponds to
  proposition~\ref{qkm-1:zero-modes^l-central}. Consider now one of
  the generators $E_0^{+,i}$. By applying the translation automorphism
  $(\hat{T}_{\omega\czek_i})^m$ ($m\in\Zcross$) to $(E_0^{+,i})^\l$,
  it follows that $(E_{-m}^{+,i})^\l$ is in the centre of $\Ue(\kmg)$.
  Let $\set{\beta_k=s_{i_1}\cdots s_{i_{k-1}}\alpha_{i_k}\mid
    k\in\range{1,\ng}}$ be the set of roots $R_+$ of $\g$ ordered with
  respect to a reduced expression of $w_0$
  (see~\ref{km:Uq(kmg)-basis}).  Then $(E_m^{+,\beta_k})^\l=
  T_{i_1}\cdots T_{i_{k-1}} (E_m^{+,i_k})^\l$ lies in the centre of
  $\Ue(\kmg)$. The result for $E_m^{-,\beta_k}$ follows by applying
  the anti-automorphism $\w$ and the proposition is proved.
\end{proof}

\subsection{}
Let $\Ze$ denote the centre of $\Ue(\kmg)$.  Let $z_i^{\pm
  1}:= K_i^{\pm \l}$ and $z_D^{\pm1}:= \deriv^{\pm \l}$.  Denote by
$\Zo^0$ the subalgebra of $\Ue(\kmh)$ generated by $\set{z_i^{\pm1},
  z_D^{\pm1}\mid i\in\range{1,r}}$.

For $\alpha\in R_+$ and $m\in\Z$, define $x_{\alpha+m\delta}:=
(E_m^{+,\alpha})^\l$ and $x_{-\alpha+m\delta}:= (E_m^{-,\alpha})^\l$.
Let $\Zo^+$ be the subalgebra of $\Ue(\kmn_+)$ generated by
\begin{displaymath}
  \set{x_\alpha, x_{\beta+m\delta}, H^i_{m\l}\mid \alpha\in R_+,
    \beta\in R, m\in\Zplus, i\in\range{1,r}}.
\end{displaymath}
Let $\Zo^-$ be the subalgebra of $\Ue(\kmn_-)$ generated by
\begin{displaymath}
  \set{x_{-\alpha}, x_{\beta+m\delta}, H^i_{m\l}\mid \alpha\in R_+,
    \beta\in R, m\in\Zminus, i\in\range{1,r}}.
\end{displaymath}
Let $\Zo^{\pm,\text{re}}$ ($\Zo^{\pm,\text{im}}$) denote the
subalgebra of $\Zo^\pm$ generated by elements  constructed out
of real (respectively imaginary) root vectors.

Finally define $\Zo$ to be the subalgebra of $\Ze$ generated by $\Zo^+$,
$\Zo^0$ and $\Zo^-$. Then $\Zo= \Zo^-\tensor \Zo^0\tensor \Zo^+$.

\begin{lemma} \label{qkm-1:kmg-inf-over-Ze}
  $\Ue(\kmg)$ is infinite dimensional over $\Zo$ at the root of unity
  $\e$.
\end{lemma}

\begin{proof}
  The imaginary root vectors $H^i_m$ ($m\in\Z\setminus \l\Z$,
  $i\in\range{1,r}$) are not in the centre of $\Ue(\kmg)$ and do not
  generate central elements. Therefore as a free module over its
  centre, $\Ue(\kmg)$ has basis elements with arbitrarily high powers
  of $H^i_m$ and is infinite dimensional.
\end{proof}

\section{Representations}

\subsection{}
Let $\hat{P}$ and $\hat{Q}$ be the weight and root lattices of $\kmg$.
Let $(\cdot,\cdot):\hat{P}\times \hat{Q}\mapto \Z$ be the bilinear
form introduced in~\ref{km:symmetric-form}.

Let $\lambda=\sum_{i\in\range{0,r}} \lambda_i\omega_i\in \hat{P}$ be a
weight of $\kmg$ and $m\in\Z$. Denote by $M(\lambda,m)$ the Verma
module over $\Uq(\kmg)$ generated by a vector $v_\lambda$ such that
\begin{align*}
  \Uq(\kmn_+)\cdot v_\lambda &=0,\\ 
  \Uq(\kmn_-)\cdot v_\lambda& = M(\lambda,m),\\ 
  k_i\cdot v_\lambda &= q_i^{\lambda_i} v_\lambda \equiv
  q^{(\lambda,\alpha_i)} v_\lambda,\\ 
  \deriv \cdot v_\lambda &= q^m v_\lambda.
\end{align*}
Of course every highest weight $\Uq(\kmg)$-module with highest weight
$(\lambda,m)$, is a quotient of $M(\lambda,m)$. There exists a unique
maximal proper submodule $M'$ of $M(\lambda,m)$ and therefore the
quotient $L(\lambda,m):= M(\lambda,m)/ M'$ is the unique irreducible
highest weight $\Uq(\kmg)$-module with height weight $(\lambda,m)$.

\subsection{}
As in the classical case~\cite[9.2]{kac:book}, the Verma module
$M(\lambda,m)$ can be constructed as a quotient of $\Uq(\kmg)$ by an
left ideal $J(\lambda,m)$ in $\Uq(\kmg)$
\begin{displaymath}
  J(\lambda,m)= \Uq(\kmg) \left( \Uq(\kmn_+)+ \sum_{i\in\range{0,r}}
  \C(q)((k_i)^{\pm1}- q_i^{\pm\lambda_i})+ \C(q)(\deriv^{\pm1}- q^{\pm
    m}) \right).
\end{displaymath}
The quotient $\Uq(\kmg)/J(\lambda,m)\isomorphic M(\lambda,m)$.

\subsection{The level}
Let $C$ be the canonical central element of $\Uq(\kmg)$
(see~\ref{km:centre}). Define the level $c$ of the $\Uq(\kmg)$ Verma
module $M(\lambda,m)$ by $C\cdot v_\lambda= q^c v_\lambda$.

\begin{lemma}
  The level $c$ of the $\Uq(\kmg)$ Verma module $M(\lambda,m)$ is
  given by
  \begin{displaymath}
    c= \sum_{i\in\range{0,r}} a_i(\lambda,\alpha_i).
  \end{displaymath}
  Note that $c\in\Z$.
\end{lemma}

\begin{proof}
  \begin{align*}
    C\cdot v_\lambda &= \prod_{i=\range{0,r}} {k_i}^{a_i} v_\lambda\\
    &= \prod_{i\in\range{0,r}}
    q_i^{a_i\pairing{\lambda,\alpha\czek_i}} v_\lambda.
  \end{align*}
\end{proof}

\subsection{}
Let $V$ be a highest weight $\Uq(\kmg)$-module with highest weight
$(\lambda,m)$. Let $\mu\in\hat{Q}_+$. Define the subspace $V_\mu$ of
$V$ as
\begin{displaymath}
  V_\mu:=\set{v\in V\mid k_i\cdot v= q^{(\lambda-\mu,\alpha_i)} v}.
\end{displaymath}
Then $V$ admits the following weight space decomposition
($\hat{Q}_+$-gradation)
\begin{displaymath}
  V=\Directsum_{\mu\in \hat{Q}_+} V_\mu,
\end{displaymath}
and $V$ is said to be $\Uq(\kmh)$-diagonalisable.

\subsection{}
Integrable modules of $\Uq(\kmg)$ can be constructed in complete
analogy with integrable modules of $\Uq(\g)$
(see~\ref{quea:integrable-module-construction}) by quotienting
$\Uq(\kmg)$ by a suitable left ideal.

\section{Representations at a root of unity}

\subsection{}
Let $M(\lambda,m)$ be a $\Uq(\kmg)$ Verma module. Denote by
$M_\A(\lambda,m)$ the $\UA(\kmg)$-submodule of $M(\lambda,m)$. Denote
by $M_\e(\lambda,m)$ the $\Ue(\kmg)$ Verma module, which is the
specialisation of $M_\A(\lambda,m)$ at $q=\e$.

\subsection{Diagonal modules}
Consider the $\Ue(\kmg)$ Verma module $M_\e(\lambda,m)$ at the root of
unity $\e$, generated by $v_\lambda$. The action of the elements in
$\Zo^-$ on $v_\lambda$ generates singular vectors in $M_\e(\kmg)$.

Define the diagonal module $\bar{M}_\e(\lambda,m)$ of $\Ue(\kmg)$ to be
\begin{displaymath}
  \bar{M}_\e(\lambda,m):= M_\e(\lambda,m)/ (\Ue(\kmg) \Zo^-\cdot
  v_\lambda).
\end{displaymath}
The module $\bar{M}_\e(\lambda,m)$ is infinite dimensional. The
$\Ue(\g)$-submodule of $M_\e(\lambda,m)$ generated by its highest
weight vector $v_\lambda$ is finite dimensional and coincides with the
diagonal $\Ue(\g)$-module constructed in~\ref{quea:diagonal-module}.

\subsection{Triangular modules}
Let $\lambda\in \hat{P}$ and let $\nu$ be an algebra homomorphism
$\nu:\Zo^-\mapto \C$. Define the following left ideal in $\Ue(\kmg)$
\begin{displaymath}
  \cs{I}_{\text{tri}}(\lambda,m,\nu):= \Ue(\kmg)\left(
  \sum_{i\in\range{0,r}} e_i+ \sum_{i\in\range{0,r}} (k_i-
  q^{(\lambda,\alpha_i)})+ (\deriv- q^m)+ \sum_{y\in\Zo^-}(y- \nu(y))
\right).
\end{displaymath}
Then define the {\em triangular module\/} $\bar{M}_\e(\lambda,m,\nu)$
over $\Ue(\kmg)$ to be
\begin{displaymath}
  \bar{M}_\e(\lambda,m,\nu):=
  \Ue(\g)/\cs{I}_{\text{tri}}(\lambda,m,\nu).
\end{displaymath}

\subsection{}
The triangular module $\bar{M}_\e(\lambda,m,0)$ ($0:\Zo^-\mapto 0$)
coincides with the diagonal module $\bar{M}_\e(\lambda,m)$. This is a
nilpotent representation of $\Ue(\kmg)$ since the Chevalley generators
act on it nilpotently.

Let $\alpha\in R_+$. When $\nu(y_\alpha)\neq 0$, then $f_\alpha$ acts
cyclicly in $\bar{M}_\e(\lambda,m,\nu)$.  In the case that
$\nu(\Zo^{-,\text{re}})\neq 0$, the module is called semicyclic
(semiperiodic).

\subsection{Central characters}
Let $V$ be an irreducible $\Ue(\kmg)$-module at the root of unity
$\e$, then each central element $x\in\Ze$ acts on $V$ as a scalar
$\chi(x)$. The map $\chi:\Ze\mapto \C$ is the central character of the
representation. Note in particular that $\chi^\pi(z_i),
\chi^\pi(z_D)\in \Ccross$.

If a representation is a diagonal $\Ue(\g)$-module, then
$\chi:\Zo^{\pm,\text{re}}\mapto 0\in \C$. For a triangular module the
central character maps $\chi:\Zo^{+,\text{re}}\mapto 0$. In a
completely cyclic (periodic) module $\chi(x_\beta),\chi(y_\beta)\in
\Ccross$ ($\beta\in \hat{R}_+^{\text{re}}$).

\begin{proposition} \label{qkm-1:V-infinite-dim}
  Let $V$ be a indecomposable highest weight $\Ue(\kmg)$ module at the
  root of unity with nonzero level $c$ such that $\e^c\neq 1$. $V$ is
  infinite dimensional.
\end{proposition}

\begin{proof}
  This follows since every Heisenberg $\Ue(\H)$ submodule of $V$ will be
  infinite dimensional.
\end{proof}

\begin{lemma}
  Let $V$ be an indecomposable $\Ue(\kmg)$ module at the root of
  unity. $V$ has a finite number of weight spaces.
\end{lemma}

\subsection{}
Let $\nu:\Zo\mapto \C$, such that $\nu(z_i),\nu(z_D)\in \Ccross$. Define
$P_\e$ to be
\begin{displaymath}
  P_\e(\nu):=\Ue(\kmg)/ (\Ue(\kmg) \sum_{z\in\Zo} z- \nu(z)).
\end{displaymath}
$P_\e(\nu)$ has a natural $\Ue(\kmg)$ module structure.  Every
$\Ue(\kmg)$-module $V$ generated by a single vector, with central
character $\chi$, is equivalent to some quotient module of
$P_\e(\chi)$.

In the case $\nu(\Zo^\pm)=0$ the module $P_\e$ and its quotients are
called nilpotent. In the case that $\nu(\Zo^{+,\text{re}})=0$
($\nu(\Zo^{-,\text{re}})=0$) and $\nu(y_\beta)\neq0$
($\nu(x_\beta)\neq0$) ($\beta\in \hat{R}_+^{\text{re}}$) the module
$P_\e$ and its quotients are called completely semicyclic
(semiperiodic). In the case that $\nu(x_\beta)\neq0$ and
$\nu(y_\beta)\neq0$ ($\beta\in\hat{R}_+^{\text{re}}$) the module
$P_\e$ and its quotients are called completely cyclic (periodic).



\newcommand{\etalchar}[1]{$^{#1}$}
\ifx\undefined\bysame
\newcommand{\bysame}{\leavevmode\hbox to3em{\hrulefill}\,}
\fi

\end{document}